\documentclass[useAMS,usenatbib]{mn2e}

\usepackage{natbib}

\usepackage{latexsym,graphicx}
\usepackage{color}
\usepackage{fixltx2e}
\usepackage{verbatim}
\usepackage{float}
\usepackage{amsmath,amssymb}
\usepackage{times}
\usepackage{soul}
\usepackage{natbib}

\usepackage{setspace}

\def\apgt{\ {\raise-.5ex\hbox{$\buildrel>\over\sim$}}\ }
\def\aplt{\ {\raise-.5ex\hbox{$\buildrel<\over\sim$}}\ }

\title[Asymmetric, Galactic supernova remnants]{Asymmetric supernova remnants generated by Galactic, massive runaway stars}

\author[D. M.-A.~Meyer et al.]
       {D. M.-A.~Meyer,$^{1}$\thanks{E-mail: dmeyer@astro.uni-bonn.de}
       N.~Langer,$^{1}$
       J.~Mackey,$^{2,1}$ 
       P.~F.~Vel\'azquez$^{3}$
       and A.~Gusdorf$^{4}$\\
       $^{1}$Argelander-Institut f\"ur Astronomie der Universit\"at Bonn, Auf dem H\"ugel 71, 53121, Bonn, Germany \\
       $^{2}$\textcolor{black}{I. Physikalisches Institut, Universit\"at zu K\"oln, Z\"ulpicher Stra\ss e 77, 50937 K\"oln, Germany} \\
       $^{3}$Instituto de Ciencias Nucleares, UNAM, Apartado Postal 70-543, 04510 Mexico, DF, Mexico \\
       $^{4}$LERMA, UMR 8112 du CNRS, Observatoire de Paris, Ecole Normale Sup\' erieure, 24 rue Lhomond, 75231 Paris Cedex 05, France\\
       }

\begin{document}

\date{Received January 18  2015; accepted Month day, 2015}

\maketitle
   
\label{firstpage}

\begin{abstract}

After the death of a runaway massive star, its supernova shock wave interacts
with the bow shocks produced by its defunct progenitor, and may \textcolor{black}{lose 
energy, momentum, and its spherical} symmetry before expanding into the local interstellar medium (ISM). We
investigate whether the initial mass and space velocity of these progenitors can
be associated with asymmetric supernova remnants. We run hydrodynamical models
of supernovae exploding in the pre-shaped medium of moving Galactic
core-collapse progenitors. We find that bow shocks that accumulate more than
about $1.5\, \rm M_{\odot}$ generate asymmetric remnants. The shock wave first
collides with these bow shocks $160-750\, \rm yr$ after the supernova, \textcolor{black}{and the
collision lasts until $830-4900\, \rm yr$. The shock wave is then} located
$1.35-5\, \rm pc$ from the center of the explosion, and it expands freely into
the ISM, whereas in the opposite direction it is channelled into the region of
undisturbed wind material. This applies to an initially $20\, \rm M_{\odot}$
progenitor moving with velocity $20\, \rm km\, \rm s^{-1}$ and to our initially
$40\, \rm M_{\odot}$ progenitor. These remnants generate mixing of ISM gas,
stellar wind and supernova ejecta that is particularly important upstream from
the center of the explosion. Their lightcurves are dominated by emission from
optically-thin cooling and by X-ray emission of the shocked ISM gas. We find
that these remnants are  likely to be observed in the [O{\sc iii}] $\lambda \,
5007$ spectral line emission or in the soft energy-band of X-rays. Finally, we
discuss our results in the context of observed Galactic supernova remnants such
as 3C391 and the Cygnus Loop.

\end{abstract}



\section{Introduction}
\label{sect:introduction}

Massive stars are rare but crucial to understand the cycle of
matter in the interstellar medium (ISM) of galaxies~\citep{langer_araa_50_2012}.
Significantly influenced by their rotation~\citep{langer_ApJ_520_1999,
vanmarle_aa_478_2008, chita_aa_488_2008}, bulk
motion~\citep{brighenti_mnras_277_1995, brighenti_mnras_273_1995} or by the
presence of a companion~\citep{stevens_apj_386_1992}, their strong winds shape
their surroundings and chemically augment their ambient
medium~\citep{vink_asp_353_2006}. Some of these stars explode as
luminous supernovae which release ejecta interacting with their pre-shaped
environment~\citep{borkowski_apj_400_1992,vink_aa_307_1996,vink_aa_328_1997,
vanveelen_aa_50._2009}. This event gives birth to supernova remnants replenishing the
ISM with momentum and kinetic energy up to about $120\, \rm pc$ from the center
of the explosion~\citep{badenes_mnras_407_2010}.

Supernovae have been noticed in ancient Asia with the naked eye, e.g. the
guest-star recorded in AD185 by chinese astronomers~\citep{green_lnp_598_2003},
whereas the first supernova remnant has been identified spectroscopically almost
two millennia later~\citep{baade_apj_88_1938}. Nowadays, surveys provide us with
observations of Galactic supernova remnants, e.g. in gamma-rays~\citep{abdo_apj_722_2010}, 
X-ray~\citep{pannuti_147_aj_2014}, the infrared~\citep{reach_aj_131_2006} or in 
\textcolor{black}{submillimeter~\citep{vandishoeck_aa_279_1993}}. Catalogues of remnants visible in the
radio waveband in the northern and southern hemisphere are available
in~\citet{kothes_aa_457_2006} and~\citet{whiteoak_aas_118_1996}, respectively.
An exhaustive catalogue of the known Galactic supernova remnants was compiled
by~\citet{green_cat_2009}. These abundant observations reveal a diversity of
complex morphologies such as shells, annuli, cylinders, rings and bipolar
structures~\citep[cf.][]{gaensler_phd_1999}.

The shape of young supernova remnants depends (i) on the geometry of the
supernova explosion and (ii) on the (an)isotropy of their ambient
medium~\citep{vink_aarv_20_2012}. They therefore exhibit a wide range of
morphologies that can be used to constrain their progenitors and/or ambient
medium properties. The distribution of circumstellar matter depends on the
progenitor properties~\citep{bedogni_190_aa_1988,ciotti_aa_215_1989,
dwarkadas_apj_630_2005,dwarkadas_apj_667_2007} and the presence of ISM
structures, e.g. borders of neighbouring diffuse nebulae or filaments that
affect the propagation of the supernova ejecta. Models of remnants developing in
a pre-existing wind cavity are shown
in~\citet{tenoriotagle_mnras_244_1990,tenoriotagle_mnras_251_1991}, and
demonstrate that mixing of material happens in the former wind bubble.
Multi-dimensional models of the formation of knots by wind-wind collision around
Cassiopeia A are shown in~\citet{peresrendon_aa_506_2009} and the effects of this
fragmented Wolf-Rayet shell on the rebrightening of young remnants is explored
in~\citet{vanveelen_aa_50._2009}.  Supernova remnants developing through an edge
of a dense region, e.g. a molecular cloud, give rise to champagne 
flows~\citep{tenoriotagle_aa_148_1985}. If
the supernova happens near a denser region, the reverse shock is reflected
towards the center of the explosion and a hot region of shocked material
forms~\citep{ferreira_478_aa_2008}. A strong magnetization of the ISM can induce
a collimation of the supernova ejecta engendering elongated
remnants~\citep{rozyczka_274_MNRAS_1995}.

Particularly, the bow shocks produced by runaway massive stars are an ideal site
for the generation of an anisotropic circumstellar distribution. This is likely to
happen in the Galactic plane, where most of the massive stars which are both in
the field and classified as runaway are found~\citep{gies_apjs_64_1987,
blau1993ASPC...35..207B, huthoff_aa_383_2002}. A few of them are identified as
evolved massive stars and three of them are red supergiants with detected bow
shocks, i.e. Betelgeuse~\citep{noriegacrespo_aj_114_1997, decin_aa_548_2012},
$\mu$ Cep~\citep{cox_aa_537_2012} and IRC$-$10414~\citep{Gvaramadze_2013}.
Consequently, and because these stars will explode as core-collapse supernova,
their circumstellar medium is of prime interest in the study of aspherical
supernova remnants.

The circumstellar medium of Galactic runaway red supergiant stars is numerically
studied in~\citet{brighenti_mnras_277_1995,brighenti_mnras_273_1995} as an attempt 
to explain non-spherical supernova remnants. The works
by~\citet{vanmarle_apj_734_2011} and~\citet{decin_aa_548_2012} tailor models to
Betelgeuse's bow shock and estimate in the context of recent
observations~\citep{cox_aa_537_2012} how the drag force on dust grains modifies
the evolution of its contact discontinuity. The effects of the mass loss and
space velocity on the shape and luminosity of bow shocks around red supergiant
stars is investigated in~\citet[][hereafter Paper~\,I]{meyer}. The repercussions
of a weak ISM magnetic field on the damping of instabilities in the bow shocks
of Betelgeuse is presented in~\citet{vanmarle_aa_561_2014}. The stabilizing role
of photoionization by an external source of radiation on the bow shock of
IRC$-$10414 is shown in~\citet{meyer_mnras_2013}. The above cited models can be
understood as investigations of the circumstellar medium of Galactic
runaway core-collapse progenitors near their pre-supernova phase.

After the supernova explosion, the forward shock of the blastwave interacts with
the free-streaming stellar
wind~\citep{chevalier_apj_344_1989,chevalier_apj_258_1982}. Supernovae showing
evidence of interaction with circumstellar structures are commonly denoted as
Type IIn and their corresponding lightcurves provide information on the
progenitor and properties of its close
surroundings~\citep{schlegel_mnras_244_1990,filippenko_araa_35_1997,
vanmarle_mnras_407_2010}. About $10$$-$$100\,\rm yr$
after the explosion, the shock wave collides with the bow shock along the
direction of motion of its progenitor, whereas it expands in a cavity of wind
material in the opposite direction~\citep{borkowski_apj_400_1992}.

In this spirit,~\citet{rozyczka_mnras_261_1993} model supernovae in oval bubbles
generated by moving progenitors. They neglect the progenitor stellar evolution
but demonstrate that elongated jet-like structures of size of about $10\, \rm
pc$ form when the shock wave expands into the wind bubble. A model interpreting
the cool jet-like \textcolor{black}{[O{\sc iii}] $\lambda \, 5007$} feature found in the Crab
nebula~\citep{blandford_301_natur_1983} as a shock wave channelled into the trail
produced by its progenitor's motion is presented in~\citet{cox_mnras_250_1991}.
~\citet{brighenti_mnras_270_1994} show that if the runaway progenitor evolves
beyond the main-sequence phase, the supernova explosion happens out of the
main-sequence wind bubble, and the subsequent remnant develops as an outflow
upstream from the direction of motion of the progenitor.

In this work, we aim to determine the degree of anisotropy of supernova remnants
generated by runaway core-collapse progenitors moving through the Galactic
plane. We model the circumstellar medium from near the pre-supernova phase of
a representative sample of the most common runaway massive 
stars~\citep{eldridge_mnras_414_2011}. We
calculate one-dimensional hydrodynamic models of the supernova ejecta
interacting with the stellar wind and use them as initial conditions for
two-dimensional simulations of the supernova remnants. Finally, we discuss the
emitting properties of the most aspherical of these remnants.

This project is different from previous studies~\citep{brighenti_mnras_270_1994}
because (i) we use self-consistent stellar evolution models, (ii)
we consider both optically-thin cooling and heating along with thermal
conduction, (iii) we trace the mixing between ISM, stellar wind and supernovae ejecta
inside the remnants and (iv) our grid of models explores a broader space of
parameters than works tailored to a particular supernova remnant, e.g. 
Kepler's supernova remnant~\citep{borkowski_apj_400_1992,
velazquez_apj_649_2006,chiotellis_aa_537_2012,toledo_mnras_442_2014} or 
Tycho's supernova remnant~\citep{vigh_apj_727_2011}. We neglect the magnetization, inhomogenity and
turbulence of the ISM and ignore the cooling in the shock wave induced by the
production of Galactic cosmic
rays~\citep{orlando_apj_749_2012,schure_mnras_435_2013}. We assume that the
supernova explosions do not have any intrinsic anisotropy. Furthermore, we
assume than no pulsar wind nebula remains inside the supernova remnants and
modifies the reflection of the reverse shock towards the center of the
explosion~\citep{bucciantini_aa_405_2003}.

This paper is structured as follows. We begin Section~\ref{sect:method} by
discussing our numerical methods and initial parameters. The modelling of the
circumstellar medium of our progenitors is shown in
Section~\ref{sect:result_sncsm}. We describe the calculations of supernova
remnants developing inside and beyond their progenitors' bow shock in
Sections~\ref{sect:young_remnant} and~\ref{subsect:large_scale_remnant},
respectively. Section~\ref{sect:discussion} discusses and compares our models of
aspherical remnants with observations. We conclude in Section~\ref{section:cc}.


\section{Method and initial parameters}
\label{sect:method}

In this section, we review the \textcolor{black}{numerical methods used} to 
model the circumstellar medium of our progenitors and we present 
the procedure to set up supernova blastwaves.

\subsection{Modelling the circumstellar medium}
\label{subsect:method_wind}

We perfom two-dimensional simulations using the code {\sc
pluto}~\citep{mignone_apj_170_2007,migmone_apjs_198_2012} to model the
circumstellar medium of moving core-collapse supernova progenitors. We solve the
equations of hydrodynamics in a cylindrical computational domain $(O; R, z)$ of
origin $O$, which is coincident with the location of the runaway star and has rotational
symmetry about $R=0$. A uniform grid of $N_{\rm R} \times N_{\rm z}$ grid cells
is mapped onto a domain of size $[0,R_{\rm max}] \times [z_{\rm min},z_{\rm
max}]$, respectively. We define $\bmath{\hat{R}}$ and $\bmath{\hat{z}}$ as the
unit vectors of the axis $OR$ and $Oz$, respectively. The grid spatial
resolution is $\itl{\Delta}=R_{\rm max}/N_{\rm R}$. Following the method
of~\citet{comeron_aa_338_1998}, we release the stellar wind on a circle of
radius $20$ grid cells centered on the origin and compute the wind-ISM
interaction in the frame of reference of the moving progenitor.

We model the circumstellar medium of initially $10$, $20$ and $40\, \rm
M_{\odot}$ stars moving with space velocity ranging from $v_{\star}=20$ to $70\,
\rm km\, \rm s^{-1}$. The (time-dependent) stellar wind properties are taken from stellar
evolution models~\citep{brott_aa_530_2011a}. We consider a homogeneous ISM with
a hydrogen number density $n_{\rm H}=0.57\, \rm
cm^{-3}$~\citep{wolfire_apj_587_2003}, i.e. we assume that the stars are exiled
from their parent cluster and move in the low-density ISM. We set the ISM
temperature to $T_{\rm ISM} \approx\, 8000\, \rm K$ and include gain/losses by
optically-thin radiative cooling assuming that the gas has solar metallicity
(sections 2.3 and 2.4 of Paper~I). All our bow shock models include electronic
thermal conduction~\citep{spitzer_1962, cowie_apj_211_1977}.

We start our models at $t_{\rm start} \approx t_{\rm psn} - 32\, t_{\rm cross}$,
where $t_{\rm psn}$ is the time at the end of the stellar evolution model. 
The time interval $32\, t_{\rm cross}$ is sufficient to simulate 
these bow shocks getting rid of any switch-on effects arising during their development, where
$t_{\rm cross} = R(0)/v_{\star}$ is the crossing-time and $R(0)$ is the
stand-off distance of the bow shock~\citep*{baranov_sphd_15_1971}. The
calculation of each bow shock model is followed until the end of the stellar
evolution model, at $t_{\rm end}$. Note that our initially $10$ and $20\, \rm M_{\odot}$
progenitors explode as a red supergiant (Paper~I). We run these models with the same
underlying assumptions as in Paper~I, especially considering that the stellar
radiation field is not trapped by the bow shock~\citep{weaver_apj_218_1977}.
Moreover, we assume that the evolutionary model of our initially $40\,
M_{\odot}$ progenitor, that does not go all the way up to the pre-supernova
phase, is sufficient to approximate the mass
distribution at the time of the supernova explosion.  Our stellar evolution
models are described in section 2.2 of Paper~I and the wind properties at
$t_{\rm psn}$ are shown in Table~\ref{tab:wind_prop}.

\begin{table}
	\centering
	\caption{ Wind properties at the end of the used stellar evolution models, at $t_{\rm psn}$.
	$M_{\star}\, (\rm M_{\odot})$ is the initial mass of each star, $\dot{M}\, (\rm M_{\odot}\, \rm yr^{-1})$
	their mass loss and $v_{\rm wind}\, (\mathrm{km}\, \mathrm{s}^{-1})$ their wind velocity at 
	a distance of $0.01\, \rm pc$ from the star, respectively. $T_{\rm eff}\, (\rm K)$ is the effective 
	temperature of the stars~\textcolor{black}{\citep{brott_aa_530_2011a}}. }
	\begin{tabular}{cccc}
	\hline
	\hline
	$M_{\star}\, (\rm M_{\odot})$ 
	                        &   $\dot{M}\, (\rm M_{\odot}\, \rm yr^{-1})$                           
 			        &   $v_{\rm wind}\, (\mathrm{km}\, \mathrm{s}^{-1})$ 
 	                        &   $T_{\rm eff}\, (\rm K)$  \\ \hline   
	$20$   &  $10^{-6.11}$  &  $16$  &  $3.2 \times 10^{3}$ \\             
	$40$   &  $10^{-4.79}$  &  $11$  &  $3.2 \times 10^{3}$ \\        
	$70$   &  $10^{-4.48}$  &  $50$  &  $5.8 \times 10^{3}$ \\ 	
	\hline 
	\end{tabular}
\label{tab:wind_prop}
\end{table}

Our models of the circumstellar medium from near the pre-supernova phase are
named with the prefix PSN followed by the initial mass of the progenitor
$M_{\star}$ (first two digits, in $\mathrm{M}_{\odot}$) and its space velocity
$v_{\star}$ (two last digits, in $\mathrm{km}\, \mathrm{s}^{-1}$). We adopt grid
dimensions such that it includes the wake of the bow shocks produced during
about $v_{\star} 10 t_{\rm cross}$ (our Table~\ref{tab:psn}) in order to
properly model the expansion of the shock wave through the tail of the bow shock
\textcolor{black}{up to times of the order of $10^{4}\, \rm yr$. } The stellar
wind is distinguished from the ISM material using a scalar $Q_{1}(\bmath{r})$
passively advected with the gas, where $\bmath{r}$ is the position vector of a
grid cell. Its value is set to $Q_{1}(\bmath{r})=1$ for the wind material and to
$Q_{1}(\bmath{r})=0$ for the ISM gas. Our cooling curves and numerical methods
are extensively detailed in section 2 of Paper~\,I. Results are presented in
Section~\ref{sect:result_sncsm}.

\begin{table*}
	\centering
	\caption{Input parameters used in our simulations of the bow shocks generated by supernova progenitors.  
	 Parameters $z_{\mathrm{min}}$ and $R_{\mathrm{max}}$ are the limits of the domain along the $x$-axis 
	 and $y$-axis (in $\mathrm{pc}$), respectively. $N_{\mathrm{R}}$ and $N_{\mathrm{z}}$ are the number 
	 of cells discretising the corresponding directions and $\it \Delta$ is the grid resolution 
	 (in $\mathrm{pc}\, \mathrm{{cell}^{-1}}$). 
	 The simulations start at a time $t_{\rm start}$ (in $\mathrm{Myr}$) after the star's birth and 
	 are run until the end of the used stellar evolution models, at $\, t_{\rm psn}$ (in $\mathrm{Myr}$).}
	\begin{tabular}{cccccccccc}
	\hline
	\hline
	${\rm {Model}}$ &   $M_{\star}\, (\rm M_{\odot})$                           
  
 			&   $v_{\star}\, (\mathrm{km}\, \mathrm{s}^{-1})$
			&   $z_{\mathrm{min}}\, (\mathrm{pc})$ 
			&   $R_{\mathrm{max}}\, (\mathrm{pc})$ 
			&   $N_{\rm R}$  
       	        	&   $N_{\rm z}$
			&   $\itl{ \Delta}\, (\mathrm{pc}\, \mathrm{cell}^{-1})$
			&   $t_{\mathrm{start}}\, (\mathrm{Myr})$
			&   $t_{\mathrm{psn}}\, (\mathrm{Myr})$ 
			\\ \hline   
	PSN1020   &  $10$  &  $20$  &  $-6.0$   &  $7.0$   &  $1225$   & $1400$ &  $5.71\times 10^{-3}$  	  & $23.7$   &  $24.7$      \\             
	PSN1040   &  $10$  &  $40$  &  $-2.1$   &  $2.0$   &  $1143$   & $1600$ &  $1.75\times 10^{-3}$ 	  & $24.5$   &  $24.7$      \\        
	PSN1070   &  $10$  &  $70$  &  $-0.9$   &  $0.8$   &  $950$    & $1425$ &  $8.42\times 10^{-4}$	  & $24.6$   &  $24.7$      \\ 
	\hline 
	PSN2020   &  $20$  &  $20$  &  $-30.0$  &  $35.0$  &  $2333$   & $2333$ &  $1.49\times 10^{-2}$        & $6.63$    &  $9.05$      \\             
	PSN2040   &  $20$  &  $40$  &  $-9.0$   &  $8.0$   &  $600$    & $1200$ &  $1.33\times 10^{-2}$ 	  & $8.65$    &  $9.05$      \\        
	PSN2070   &  $20$  &  $70$  &  $-4.5$   &  $4.0$   &  $1067$   & $1600$ &  $3.75\times 10^{-3}$	  & $8.70$    &  $9.05$      \\ 
	\hline 
	PSN4020   &  $40$  &  $20$  &  $-90.0$  &  $100.0$ &  $1500$   & $1575$ &  $6.00\times 10^{-2}$         & $0.0$     &  $4.5$      \\             
	PSN4040   &  $40$  &  $40$  &  $-36.0$  &  $30.0$  &  $1500$   & $2100$ &  $2.00\times 10^{-2}$ 	  & $2.00$    &  $4.5$      \\        
	PSN4070   &  $40$  &  $70$  &  $-30.0$  &  $20.0$  &  $1000$   & $2000$ &  $2.00\times 10^{-2}$	  & $3.50$    &  $4.5$      \\ 	
	\hline 
	\end{tabular}
\label{tab:psn}
\end{table*}


\subsection{Setting up the supernova shock wave}
\label{subsect:method_sn}

We perform one-dimensional hydrodynamical simulations of the shock wave
expanding into the stellar wind. The blastwave is characterized by its energy
fixed to $E_{\rm ej}=10^{51}\, \rm erg$ and by the mass of the ejecta 
$M_{\rm ej}$. The latter is estimated as,
\begin{equation}
   M_{\rm ej} =  M_{\star} - \int_{t_{0}}^{t_{\rm psn}} \dot{M}(t) dt -
M_{\rm co},
   \label{eq:co}
\end{equation}
where $t_{0}$ and $t_{\rm psn}$ are the time at the beginning and the end of the
used stellar evolution model, respectively. Note that we assume $M_{\rm ej}$ for our
$40\, \rm M_{\odot}$ progenitor, because we ignore its post-main-sequence 
evolution. The quantity $M_{\rm co}=2\, \rm M_{\odot}$ in Eq.~(\ref{eq:co})
is the assumed mass of the residual compact object left after the supernova (our
Table~\ref{tab:sncsm}).

\begin{table}
	\centering
	\caption{Simulations parameters used in our simulations of supernovae
interacting with the unperturbed stellar wind. Parameter $M_{\rm ej}$ is the
mass of the ejecta (in $\mathrm{M}_{\odot}$) and $r_{\mathrm{max}}$ is the size
of the one-dimensional spherically symmetric domain (in $\rm pc$). The
simulations are started at $t=0.04\, \mathrm{yr}$. The last column indicates the
time at the end of our simulations, $t_{\mathrm{sncsm}}$ (in $\mathrm{yr}$).}
	\begin{tabular}{ccccc}
	\hline
	\hline
	${\rm {Model}}$  &   $M_{\mathrm{ej}}\, (\mathrm{M}_{\odot})$
			 &   $r_{\mathrm{max}}\, (\mathrm{pc})$ 
			 &   $t_{\mathrm{sncsm}}\, (\mathrm{yr})$
			\\ \hline   
	SNCSM 1020   &  $7.7$   & $0.30$  & $40$  \\           
	SNCSM 1040   &  $7.7$   & $0.20$  & $13.5$  \\     
	SNCSM 1070   &  $7.7$   & $0.13$  & $15$  \\      
	\hline 
	SNCSM 2020   &  $17.7$  & $0.90$  & $40$  \\
	SNCSM 2040   &  $17.7$  & $0.50$  & $25$  \\       
	SNCSM 2070   &  $17.7$  & $0.25$  & $20$  \\     
	\hline 
	SNCSM 4020   &  $20$    & $3.00$  & $400$  \\
	SNCSM 4040   &  $20$    & $1.50$  & $200$  \\       
	SNCSM 4070   &  $20$    & $0.90$  & $180$  \\ 	
	\hline 
	\end{tabular}
\label{tab:sncsm}
\end{table}

We set up the supernova using the method detailed in~\citet{whalen_apj_682_2008}
and in~\citet{vanveelen_aa_50._2009}. It assumes that the blastwave density
profile $\rho(r)$ is a radial piece-wise function of the distance $r$ to the
center of the explosion in the region $[0,r_{\rm max}]$, where $r_{\rm max}$ is
the radius of the shock wave when we start the simulations. Under these
assumption, the ejecta density profile is,
\begin{equation}
\rho(r) = \begin{cases}
        \rho_{\rm core}(r) & \text{if $r \le r_{\rm core}$ },               \\
        \rho_{\rm max}(r)  & \text{if $r_{\rm core} < r < r_{\rm max}$},    \\
        \rho_{\rm csm}(r)  & \text{if $r \ge r_{\rm max}$},                 \\
        \end{cases}
	\label{cases}
\end{equation}
where,
\begin{equation}
   \rho_{\rm core}(r) =  \frac{1}{ 4 \pi n } \frac{ (10 E_{\rm ej}^{n-5})^{-3/2}
 }{  (3 M_{\rm ej}^{n-3})^{-5/2}  }  t_{\rm max}^{-3},
   \label{sn:density_1}
\end{equation}
is constant up to the inner core of radius $r_{\rm core}$ and,
\begin{equation}
   \rho_{\rm max}(r) =  \frac{1}{ 4 \pi n } \frac{ (10 E_{\rm
ej}^{n-5})^{(n-3)/2}  }{  (3 M_{\rm ej}^{n-3})^{(n-5)/2}  } 
\bigg(\frac{r}{t_{\rm max}}\bigg)^{-n},
   \label{sn:density_2}
\end{equation}
is a steeply decreasing function of inner radius
$r_{\rm core}$ and external radius $r_{\rm max}$~\citep{truelove_apjs_120_1999}.
The power law index $n$ of Eq.~(\ref{sn:density_1})$-$(\ref{sn:density_2}) is
set to the usual value $n=11$ for core-collapse
supernovae~\citep{chevalier_apj_258_1982}. In relation~(\ref{cases}), $\rho_{\rm
cms}$ is the freely-expanding wind profile measured from the simulations along
the symmetry axis $Oz$, in the direction of motion of the progenitor ($z\ge0$). 
We use it as initial condition in the [$r_{\rm max}$,$r_{\rm sncsm}$] of the domain,
where $r_{\rm sncsm} < R(0)$ is outer border of the domain.

The ejecta obey a homologous expansion, i.e. the velocity profile $v(r)$ is,
\begin{equation}
   v(r) =  \frac{r}{t},\, \mathrm{if}\, t>\textcolor{black}{t_{\rm max}},
   \label{sn:vel}
\end{equation}
where $t$ is the time after the supernova explosion. The ejecta velocity at
$r_{\rm core}$ is therefore,
\begin{equation}
   v_{\rm core} = \bigg(  \frac{ 10(n-5)E_{\rm ej} }{ 3(n-3)M_{\rm ej} } 
\bigg)^{1/2},
   \label{sn:vcore}
\end{equation}
~\citep{truelove_apjs_120_1999}. The choice of $r_{\rm max}$ is free, as long as
a mass of stellar wind smaller than $M_{\rm ej}$ is enclosed in $[r_{\rm
max},r_{\rm sncsm}]$. We determined its $r_{\rm max}$ using the numerical
procedure described in~\citet{whalen_apj_682_2008}.
We start the simulation at $t_{\rm max}=r_{\rm max}/v_{\rm
max}$, where $v_{\rm max}$ is set to $30000\, \rm km\, \rm
s^{-1}$~\citep{vanveelen_aa_50._2009}.

We choose \textcolor{black}{a} uniform grid of resolution $\Delta \le 10^{-4}\, \rm {pc}\, \rm
{cell}^{-1}$ and follow the expansion of the shock wave until
slightly before it reaches the reverse shock of the bow shocks produced by our
progenitors. These models are labelled with the prefix SNCSM (our
Table~\ref{tab:sncsm}). Additionally, we use a second passive scalar
$Q_{2}(\bmath{r})$ to distinguish the ejecta from the stellar wind. We carry out
these one-dimensional calculations using a uniform spherically symmetric grid.
We use a finite volume method with the Harten-Lax-van Leer approximate Riemann
solver, and integrate the Euler equations with a second order, unsplit,
time-marching algorithm. Dissipative processes are computed using our cooling
curve for fully ionized gas. Results are presented in
Section~\ref{sect:young_remnant}.


\subsection{Modelling the supernova remnants}
\label{subsect:method_remnant}

In order to resolve both the early interaction between the
blastwave interacting with the circumstellar medium and the old supernova remnant, 
we adopt a mapping strategy. We run two-dimensional hydrodynamical
simulations of the shock waves interacting with the bow shocks using a squared
computational domain of size about $4R(0)$ which is supplied with a uniform
rectilinear grid. These models are labelled with the prefix YSNR. The above-described 
one-dimensional simulations of the ejecta interacting with the stellar
wind are mapped into a circle of radius $r_{\rm max}<R(0)$ centered on the origin
$O$ of the domain. We run these simulations starting at $t_{\rm sncsm}$ until
the shock wave has passed through the forward shock of the bow shock and reaches
a distance of about $2R(0)$ in the direction of motion of the progenitor, at
$t_{\rm ysnr}$ (our Table~\ref{tab:ysnr}).

The remnants at $t_{\rm ysnr}$ are mapped a second time onto a larger
computational domain which includes both the entire pre-calculated 
circumstellar medium and the calculations of the young supernova remnants (our
Tables~\ref{tab:psn} and~\ref{tab:ysnr}). The regions of this domain which 
overlap neither the bow shock nor the remnant are filled with
unperturbed ISM gas. We start the simulations at time $t_{\rm ysnr}$ and follow
edge of the domain in the $-\bmath{\hat{z}}$ direction, at $t_{\rm osnr}$. These simulations
are labelled with the prefix OSNR (our Table~\ref{tab:osnr}). Results are presented in
Section~\ref{subsect:large_scale_remnant}.

\begin{table*}
	\centering
	\caption{Input parameters used in our simulations of the
supernova blastwaves interacting with the bow shocks of our progenitors.  As
input we use the solution of the shock waves interacting with the stellar winds
(our Table~\ref{tab:sncsm}). The grid parameters are similar as in our
Table~\ref{tab:psn}. Our simulations start at $t_{\mathrm{sncsm}}$ (our
Table~\ref{tab:psn}) shortly before that the shock wave interacts with the 
bow shock and the models are run until $t_{\mathrm{ysnr}}$ (in
$\mathrm{yr})$ once the shock wave has gone through it. }
	\begin{tabular}{ccccccccc}
	\hline
	\hline
	${\rm {Model}}$ &   ${\rm {Input}}$
			&   $z_{\mathrm{min}}\, (\mathrm{pc})$ 
			&   $R_{\mathrm{max}}\, (\mathrm{pc})$ 
			&   $N_{\rm R}$  
		        &   $N_{\rm z}$
			&   $\it \Delta\, (\mathrm{pc}\, \mathrm{cell}^{-1})$
			&   $t_{\mathrm{sncsm}}\, (\mathrm{yr})$
			&   $t_{\mathrm{ysnr}}\, (\mathrm{yr})$ 
			\\ \hline  
	YSNR1020   & SNCSM1020  &     $-2.0$ & $2.0$  &  $1000$  &  $2000$   & $2.0\times 10^{-3}$    & $40$     &  $264$     \\  
          
	YSNR1040   & SNCSM1040  &     $-1.3$ & $1.3$  &  $1000$  &  $2000$   & $1.3\times 10^{-3}$    & $13.5$   &  $150$     \\  
     
	YSNR1070   & SNCSM1070  &     $-0.7$ & $0.7$  &  $1000$  &  $2000$   & $4.0\times 10^{-4}$    & $15$     &  $60$     \\ 
	\hline 
	YSNR2020   & SNCSM2020  &     $-8.0$ & $8.0$  &  $1000$  &  $2000$   & $8.0\times 10^{-3}$    & $40$   &  $2400$      \\ 
           
	YSNR2040   & SNCSM2040  &     $-2.0$ & $2.0$  &  $1000$  &  $2000$   & $2.0\times 10^{-3}$    & $20$   &  $450$      \\ 
      
	YSNR2070   & SNCSM2070  &     $-1.0$ & $1.0$  &  $1000$  &  $2000$   & $1.0\times 10^{-3}$    & $25$   &  $170$      \\
 
	\hline 
	YSNR4020   & SNCSM4020  &     $-25$  & $25$   &  $1000$  &  $2000$   & $2.5\times 10^{-2}$   & $400$   &  $4900$      \\ 
           
	YSNR4040   & SNCSM4040  &     $-7.0$ & $7.0$  &  $1500$  &  $3000$   & $4.7\times 10^{-3}$   & $200$   &  $1360$      \\ 
      
	YSNR4070   & SNCSM4070  &     $-4.0$ & $4.0$  &  $1500$  &  $3000$   & $2.7\times 10^{-3}$   & $180$   &  $830$      \\
 
	\hline 	
	\end{tabular}
\label{tab:ysnr}
\end{table*}

\begin{table*}
	\centering
	\caption{
	Input parameters used in our simulations of the
supernova blastwaves interacting with the tails of the bow shocks generated by
our progenitors.  As input we use the solution of the shock waves interacting
with the bow shocks (our Table~\ref{tab:ysnr}). The grid parameters are similar
as in our Table~\ref{tab:psn}. Our simulations start at $t_{\mathrm{ysnr}}$ (our
Table~\ref{tab:ysnr}) and the models are run until $t_{\mathrm{osnr}}$ (in
$\mathrm{yr})$.
	}
	\begin{tabular}{ccccccccc}
	\hline
	\hline
	${\rm {Model}}$ &   ${\rm {Input}}$
			&   $z_{\mathrm{min}}\, (\mathrm{pc})$ 
			&   $R_{\mathrm{max}}\, (\mathrm{pc})$ 
			&   $N_{\rm R}$  
		        &   $N_{\rm z}$
			&   $\it \Delta\, (\mathrm{pc}\, \mathrm{cell}^{-1})$
			&   $t_{\mathrm{ysnr}}\, (\mathrm{yr})$
			&   $t_{\mathrm{osnr}}\, (\mathrm{yr})$ 
			\\ \hline  
	OSNR1020   & YSNR1020  &     $-6.0$  & $6.0$   &  $500$   &  $1000$   & $1.2\times 10^{-2}$    & $264$   &  $1500$      \\  
          
	OSNR1040   & YSNR1040  &     $-2.1$  & $2.1$   &  $500$   &  $1000$   & $4.2\times 10^{-3}$    & $150$   &  $1300$      \\  
     
	OSNR1070   & YSNR1070  &     $-0.9$  & $0.9$   &  $500$   &  $1000$   & $4.0\times 10^{-4}$    & $60$   &  $1300$      \\ 
	\hline 
	OSNR2020   & YSNR2020  &     $-30.0$ & $25.0$  &  $1000$  &  $2000$   & $9.0\times 10^{-2}$    & $2400$   &  $21100$      \\ 
           
	OSNR2040   & YSNR2040  &     $-9.0$  & $9.0$   &  $1000$  &  $2000$   & $2.5\times 10^{-3}$    & $450$   &  $15000$      \\ 
      
	OSNR2070   & YSNR2070  &     $-4.5$  & $4.5$   &  $1000$  &  $2000$   & $9.0\times 10^{-3}$    & $170$   &  $10000$      \\
 
	\hline 
	OSNR4020   & YSNR4020  &     $-90.0$ & $70.$   &  $1000$  &  $1714$   & $7.0\times 10^{-2}$    & $4900$   &  $49500$      \\ 
           
	OSNR4040   & YSNR4040  &     $-35.0$ & $25.0$  &  $1000$  &  $2200$   & $2.5\times 10^{-2}$    & $1360$   &  $14000$      \\ 
   
	OSNR4070   & YSNR4070  &     $-30.0$ & $15.0$  &  $700$   &  $2333$   & $2.1\times10^{-2}$    & $830$   &  $10500$      \\ 
 
	\hline 	
	\end{tabular}
\label{tab:osnr}
\end{table*}


\section{The pre-supernova phase}
\label{sect:result_sncsm}

\begin{figure}
	\centering
	\begin{minipage}[b]{ 0.45\textwidth}
\includegraphics[width=1.0\textwidth]{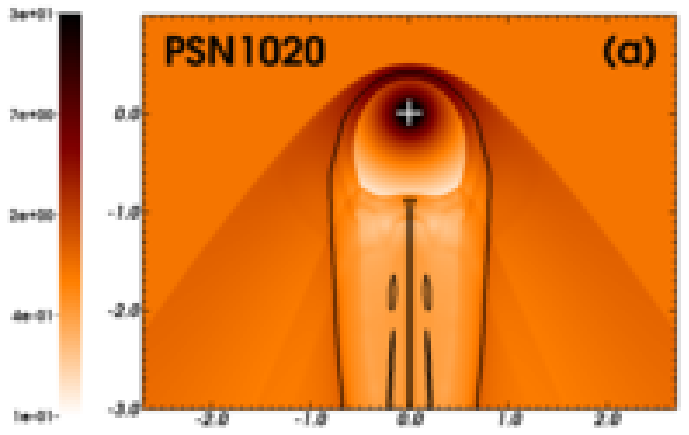}
	\end{minipage}\\
	\begin{minipage}[b]{ 0.45\textwidth}
\includegraphics[width=1.0\textwidth]{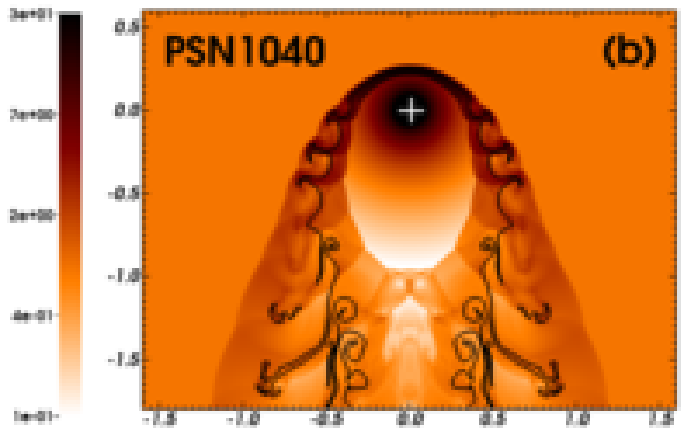}
	\end{minipage}\\
	\begin{minipage}[b]{ 0.45\textwidth}
\includegraphics[width=1.0\textwidth]{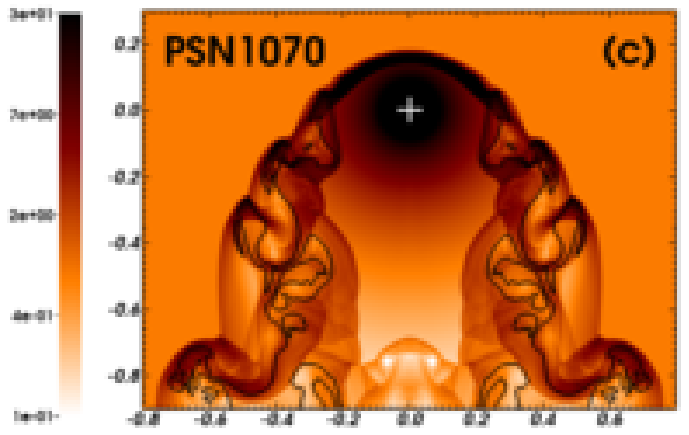}
	\end{minipage} 
        \\	
	\caption{ 
		Grid of stellar wind bow shocks from the pre-supernova phase of
		our initially $10\, \rm M_{\odot}$ progenitor as a function of its space velocity
		with respect to the ISM, with velocity $v_{\star}=20$ (a), $40$ (b) and $70\,
		\mathrm{km}\, \mathrm{s}^{-1}$ (c). The nomenclature of the models follows Table~\ref{tab:psn}.
		The bow shocks are shown at $t_{\rm psn}$. The gas number density is shown with a
		density range from $10^{-1}$ to $30.0\, \mathrm{cm}^{-3}$ in the logarithmic scale. 
		\textcolor{black}{The white cross marks the position of the star.} The solid 
		black contour traces the boundary between wind and ISM material $Q_{1}(\bmath{r})=1/2$. 
		The $x$-axis represents the radial direction and the $y$-axis the direction of stellar motion (in
		$\mathrm{pc}$). Only part of the computational domain is shown in the figures.  
		}
	\label{fig:progenitor1}  
\end{figure}

\begin{figure}
	\centering
        \begin{minipage}[b]{ 0.45\textwidth}
\includegraphics[width=1.0\textwidth]{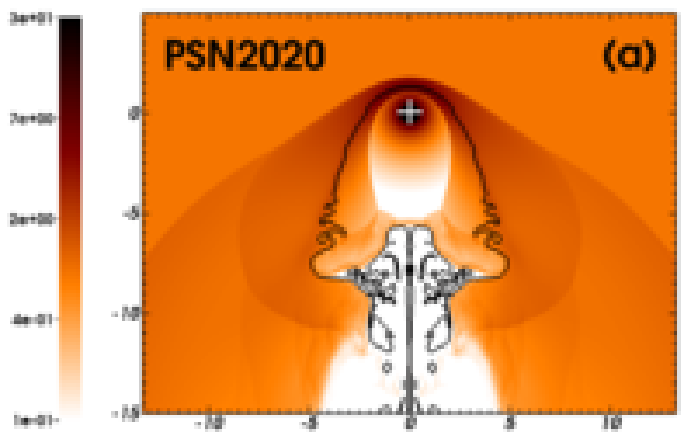}
	\end{minipage}
	\begin{minipage}[b]{ 0.45\textwidth}
\includegraphics[width=1.0\textwidth]{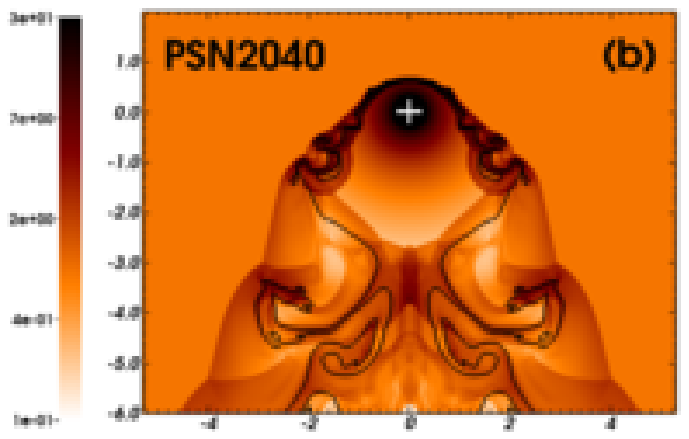}
	\end{minipage}
	\begin{minipage}[b]{ 0.45\textwidth}
\includegraphics[width=1.0\textwidth]{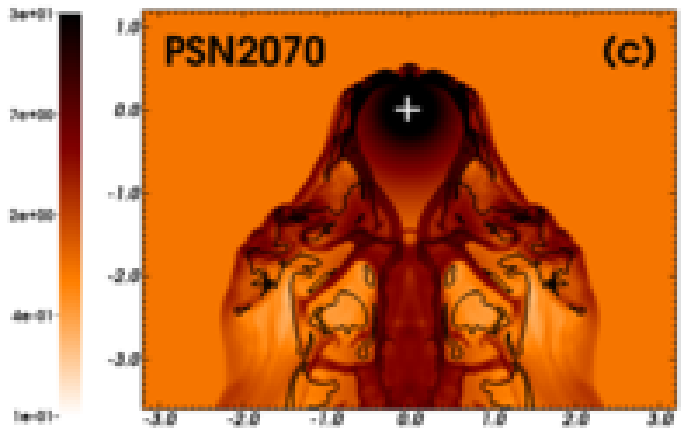}
	\end{minipage} 
        \\	
	\caption{ 
                 As Fig.~\ref{fig:progenitor1}, with our initially $20\, \rm
M_{\odot}$ progenitor. \textcolor{black}{Note the Napoleon's hat structure of the bow shock in panel (a).}    
		}
	\label{fig:progenitor2}  
\end{figure}

\begin{figure}
	\centering
	\begin{minipage}[b]{ 0.45\textwidth}
\includegraphics[width=1.0\textwidth]{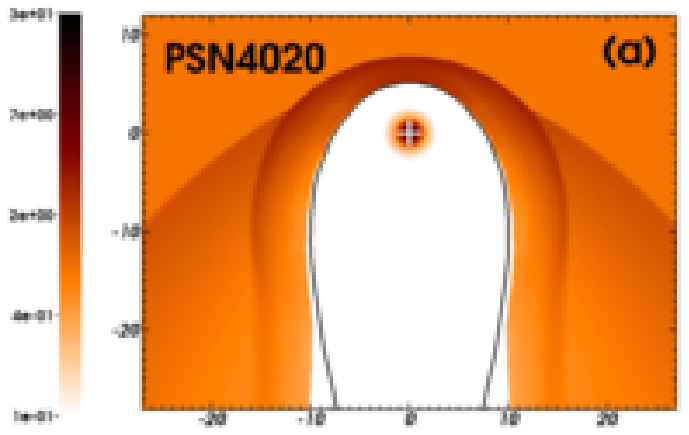}
	\end{minipage}
	\begin{minipage}[b]{ 0.45\textwidth}
\includegraphics[width=1.0\textwidth]{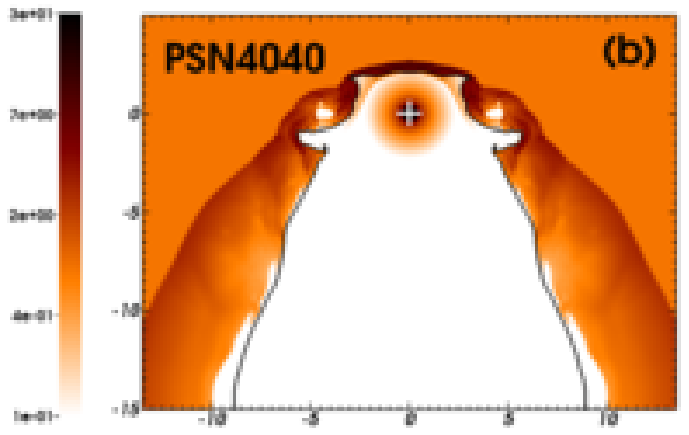}
	\end{minipage}
	\begin{minipage}[b]{ 0.45\textwidth}
\includegraphics[width=1.0\textwidth]{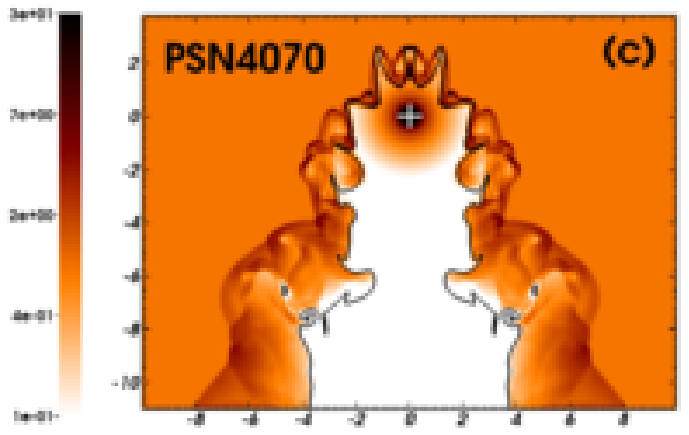}
	\end{minipage} 
        \\	
	\caption{ 
                 As Fig.~\ref{fig:progenitor1}, with our initially $40\, \rm
M_{\odot}$ progenitor.    
                 }
	\label{fig:progenitor3}  
\end{figure}

In Fig.~\ref{fig:progenitor1} we show the gas density fields in our bow shocks
from near the pre-supernova phase in the models PSN1020 (initially $10\, \rm
M_{\odot}$ star, $v_{\star}=20\, \rm km\, \rm s^{-1}$,
Fig.~\ref{fig:progenitor1}a), PSN1040 (initially $10\, \rm M_{\odot}$ star,
$v_{\star}=40\, \rm km\, \rm s^{-1}$, Fig.~\ref{fig:progenitor1}b) and PSN1070
(initially $10\, \rm M_{\odot}$ star, $v_{\star}=70\, \rm km\, \rm s^{-1}$,
Fig.~\ref{fig:progenitor1}c). Figs.~\ref{fig:progenitor2}
and~\ref{fig:progenitor3} show the same, but for our $20$ and $40\,
M_{\odot}$ stars. The figures plot the density at a time $t_{\rm psn}$ and do
not show all of the computational domain. In
Figs.~\ref{fig:progenitor1},~\ref{fig:progenitor2} and~\ref{fig:progenitor3} the
overplotted solid black line is the contact discontinuity, i.e. the border
between the wind and ISM gas where the value of the passive scalar
$Q_{1}(\bmath{r})=1/2$.

The bow shocks of our $10$ and $20\, \rm M_{\odot}$ stars have morphologies
consistent with previous
studies~\citep[][Paper~\,I]{vanmarle_apj_734_2011,mohamed_aa_541_2012}. Their
overall shape is rather stable if $v_{\star} \le v_{\rm
w}$~\citep[][]{dgani_apj_461_1996} and the flow inside the bow shocks is laminar
(Fig.~\ref{fig:progenitor1}a). The bow shocks are unstable and exhibit
Rayleigh-Taylor and/or Kelvin-Helmholtz instabilities for $v_{\star} \ge v_{\rm
w}$ because of the large density difference between the dense red supergiant
wind and the ISM gas (Figs.~\ref{fig:progenitor1}b,c
and~\ref{fig:progenitor2}a,b). For high-mass stars moving with large space
velocities, e.g. the models with $M_{\star} \ge 20\, \rm M_{\odot}$ and $v_{\star}
\ge 40\, \rm km\, \rm s^{-1}$, the shocked layers develop non-linear thin-shell
instabilities~\citep{vishniac_apj_428_1994,garciasegura_1996_aa_316,
blondin_na_57_1998} and induce a strong mixing in the wakes of the bow shocks
(Fig.~\ref{fig:progenitor2}c).

The stellar motion displaces the position of the star from the center of the
cavity of unshocked wind material~\citep{brighenti_mnras_277_1995,
brighenti_mnras_273_1995}, and this displacement is larger for velocities
$v_{\star} \ge 20\, \rm km\, \rm s^{-1}$ (Fig.~\ref{fig:progenitor2}a). The bow
shocks which have the most pronounced tunnels of low-density gas are produced
either by our initially $20\, \rm M_{\odot}$ star moving with $20\, \rm km\, \rm
s^{-1}$ or by our initially $40\, \rm M_{\odot}$ star (Fig.~\ref{fig:progenitor2}a
and~\ref{fig:progenitor3}a-c). In the region downstream from the progenitor, the
reverse shock, which forms the walls of the cavity, has a rather smooth
appearance for $v_{\star} \le 20\, \rm km\, \rm s^{-1}$
(Fig.~\ref{fig:progenitor2}a) but it is ragged for $v_{\star} \ge 40\, \rm
km\, \rm s^{-1}$ (Fig.~\ref{fig:progenitor3}c). Finally, note that the model
PSN2020 has a double bow shock due to the final increase of the mass loss that
ends the red supergiant phase. This structure is called a Napoleon's hat and it
develops when the bow shock from a new mass-loss event goes through the one
generated by the previous evolutionary
phase~\citep{wang_MNRAS_261_1993,mackey_apjlett_751_2012}.

\begin{figure} 
	\centering
	\begin{minipage}[b]{ 0.45\textwidth}
\includegraphics[width=1.0\textwidth,angle=0]{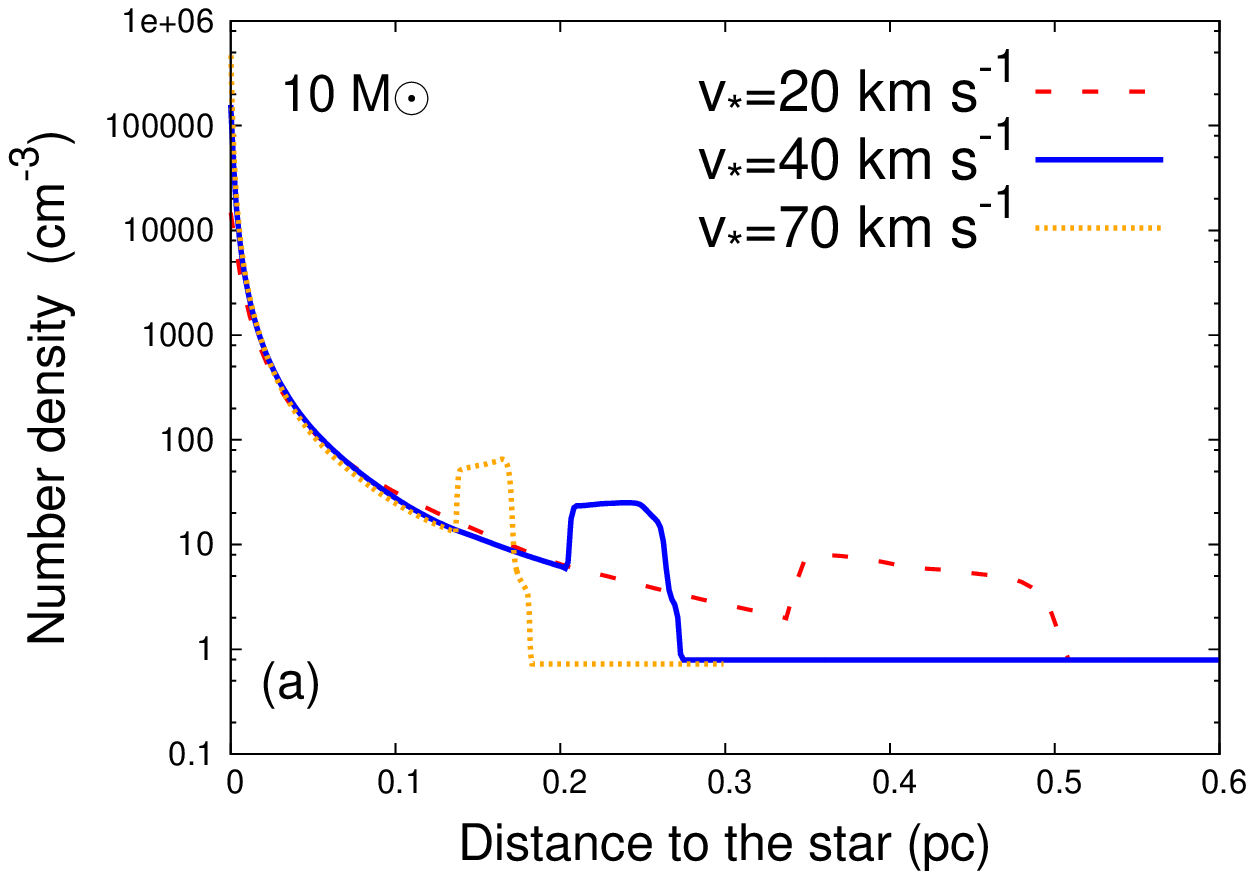}
	\end{minipage} 
	\begin{minipage}[b]{ 0.45\textwidth}
\includegraphics[width=1.0\textwidth,angle=0]{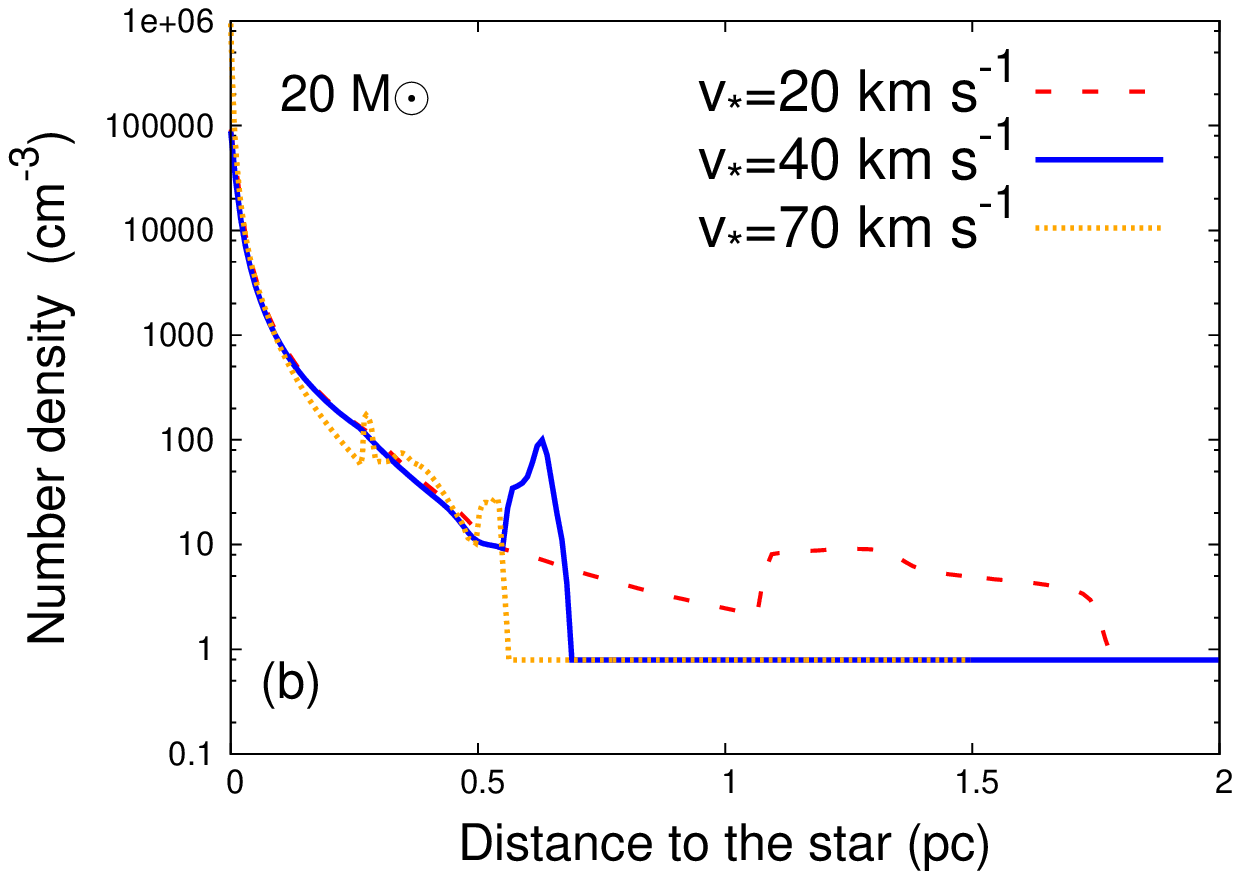}
	\end{minipage}
	\begin{minipage}[b]{ 0.45\textwidth}
\includegraphics[width=1.0\textwidth,angle=0]{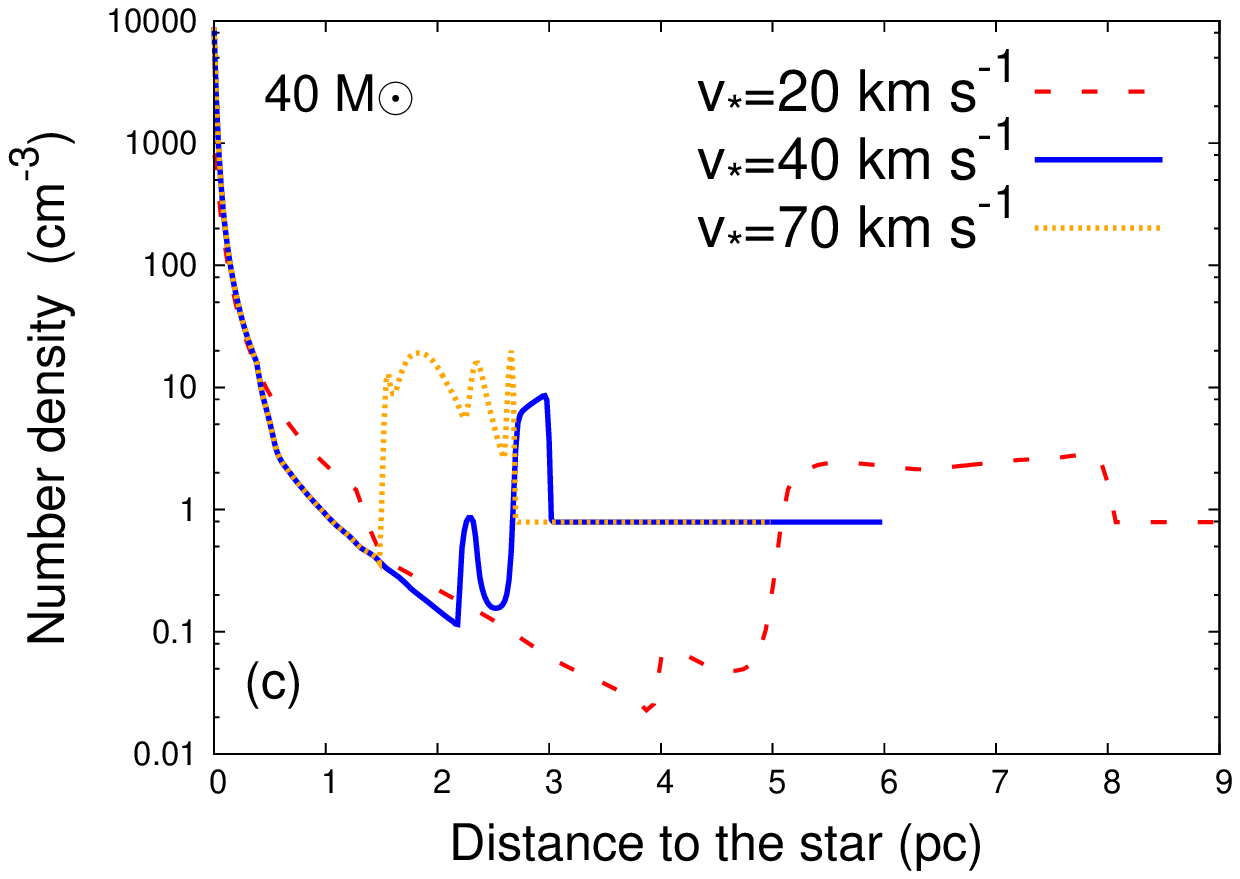}
	\end{minipage}
	\caption{ 
                Gas number density profiles (in $\rm cm^{-3}$) taken along the direction of motion of our 
                initially $10$ (a), $20$ (b) and $40\, \rm M_{\odot}$ (c) progenitors at $t_{\rm psn}$. 
		}
	\label{fig:cuts_profiles_psn}  
\end{figure}

The stand-off distance $R(0)$ and the mass $M$ trapped in the bow shocks
upstream from the star ($z\ge0$) are summarised in Table~\ref{tab:psn1}. The
more massive bow shocks are the biggest ones, e.g. our bow shock model PSN4020
has the largest stand-off distance $R(0)\approx5\, \rm pc$ and has accumulated
about $116\, \rm M_{\odot}$ of shocked gas. They are generated by high-mass stars 
moving with small space velocities, i.e. $M_{\star} \ge 20\, \rm M_{\odot}$ and
$v_{\star} \le 40\, \rm km\, \rm s^{-1}$
(Figs.~\ref{fig:progenitor2}a-\ref{fig:progenitor3}ab).  In
Fig.~\ref{fig:cuts_profiles_psn} we show the average density
profiles in our simulations of our $10$ (Fig.~\ref{fig:cuts_profiles_psn}a), $20$
(Fig.~\ref{fig:cuts_profiles_psn}b) and $40\, \rm M_{\odot}$ models
(Fig.~\ref{fig:cuts_profiles_psn}c), that we use as initial conditions
for our one-dimensional simulations of the supernova shock waves interacting
with their surroundings, see Eqs.~(\ref{cases})$-$(\ref{sn:density_1}). 
Fig.~\ref{fig:cuts_profiles_psn} illustrates that the most massive bow shocks have
the largest $R(0)$, i.e. they are the most voluminous and are reached by the
shock wave about $R(0)/v(r) \approx 10^{3} \rm yr$ after the explosion
(Section~\ref{sect:young_remnant}).

\begin{table}
	\centering
	\caption{Gas mass $M\, (\rm M_{\odot})$ in the region of the bow shocks that is upstream from the progenitor ($z\ge0$)
		 and stand-off distance at the contact discontinuity (in $\mathrm{pc}$)
		 measured in our simulations of the circumstellar medium from near 
		 the pre-supernova phase, at a time $t_{\rm psn}$. 
		 }
	\begin{tabular}{ccc}
	\hline
	\hline
	${\rm {Model}}$  & $M\, (\rm M_{\odot})$ & $R(0)\, (\rm pc)$    
	\\ \hline   
	PSN1020   &  $0.06$   &  $0.42$ \\             
	PSN1040   &  $0.03$   &  $0.25$ \\        
	PSN1070   &  $0.01$   &  $0.17$ \\ 
	\hline 
	PSN2020   &  $3.87$   &  $1.35$ \\             
	PSN2040   &  $1.10$   &  $0.64$ \\        
	PSN2070   &  $0.75$   &  $0.55$ \\  
	\hline 
	PSN4020   &  $116.00$ &  $5.00$ \\             
	PSN4040   &  $9.40$   &  $2.70$ \\        
	PSN4070   &  $1.65$   &  $1.55$ \\  
	\hline 
	\end{tabular}
\label{tab:psn1}
\end{table}


\section{The young supernova remnant}
\label{sect:young_remnant}

This section presents our simulations of the supernova blastwaves interacting
with their circumstellar medium until the shock wave reaches the outer layer of
its surrounding bow shock. Spherical remnants are distinguished from
asymmetric remnants as a function of their progenitors' properties.

\subsection{The ejecta-stellar-wind interaction}
\label{subsect:sn_wind}

\begin{figure*} 
        \centering
	\begin{minipage}[b]{ 0.55\textwidth}
	\centering
\includegraphics[width=1.0\textwidth,angle=0]{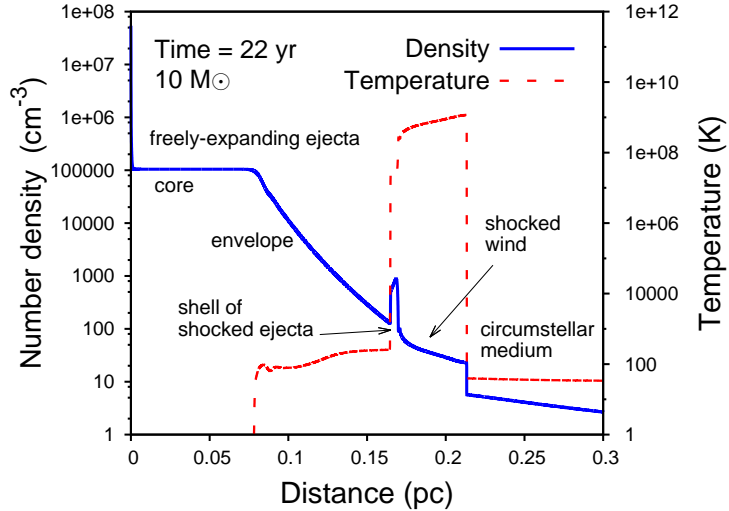}
	\end{minipage} 
	\caption{ 
                Total number density (solid blue line, in $\rm cm^{-3}$) and temperature 
                (dashed red line, in $\rm km\, \rm s^{-1}$) profiles of the 
                supernova ejecta interacting with the circumstellar medium of 
                our initially $10\, \rm M_{\odot}$ progenitor. 
	        Profiles are shown at about $22\, \rm yr$ after the supernova 
	        explosion. The distinct regions composing 
	        the supernova remnant are indicated on the figure. The substructure 
	        of the shocked ejecta follows the nomenclature in~\citet{truelove_apjs_120_1999}.
		}
	\label{fig:sncsm}  
\end{figure*}

In Fig.~\ref{fig:sncsm} we show a typical interaction between a supernova
shock wave and the surrounding stellar wind, before the shock wave collides
with the bow shock. We plot the gas number density (solid blue line) and
temperature (dashed red line) profiles in our model SNCSM1020 at a time about
$22\, \rm yr$ after the supernova explosion. It assumes a release of $M_{\rm ej}=7.7\,
M_{\odot}$ of ejecta together with a kinetic energy of $E_{\rm ej}=10^{51}\, \rm erg$ (our
Table~\ref{tab:sncsm}). The structure is composed of 4 distinct regions: the
expanding ejecta profile, itself made of two regions, the core
and the envelope~\citep{truelove_apjs_120_1999}, the shell of swept up shocked
ejecta and shocked wind material, and finally the undisturbed circumstellar
material~\citep{chevalier_apj_258_1982}.

The shell is bordered by two shocks, a reverse shock that is the interface
between ejecta and shocked ejecta, and a forward shock constituting the border
between shocked wind and undisturbed freely-expanding stellar
wind~\citep{chiotellis_aa_537_2012}. The core of the ejecta ($r<0.08\, \rm pc$)
has a very low temperature because its thermal pressure is initially
null~\citep{whalen_apj_682_2008,vanveelen_aa_50._2009}. The temperature slightly
increases up to a few tens of degrees in the envelope of ejecta ($0.08\le r \le
0.16\, \rm pc$) because (i) we use a homologous velocity profile which results
in increasing the thermal pressure close to the high-velocity shock wave and
(ii) the decreasing density $\rho_{\rm max}\propto r^{-11}$ increases the
temperature $T \propto p/\rho$. At radii $r\approx 0.16\, \rm pc$ is the hot
($T\approx 10^{6}\, \rm K$) and dense ($n\gg 10^{3}\, \rm cm^{-3}$) gas. This
region between the shell and the shock wave is hot because it is shock-heated 
by the blastwave and it has not yet cooled. All our models have a similar 
behaviour.

\subsection{The shock wave interacting with the bow shock}
\label{subsect:sn_wind}

\begin{figure*}
	\begin{minipage}[b]{ 0.33\textwidth}
\includegraphics[width=1.0\textwidth]{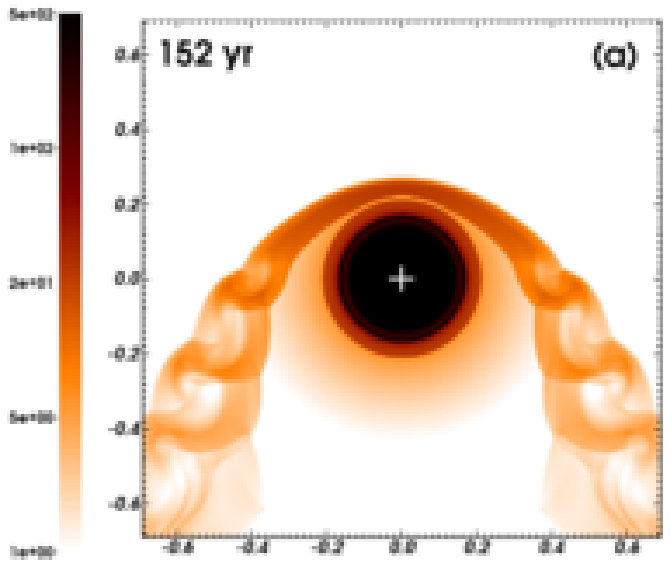}
	\end{minipage} 
	\begin{minipage}[b]{ 0.33\textwidth}
\includegraphics[width=1.0\textwidth]{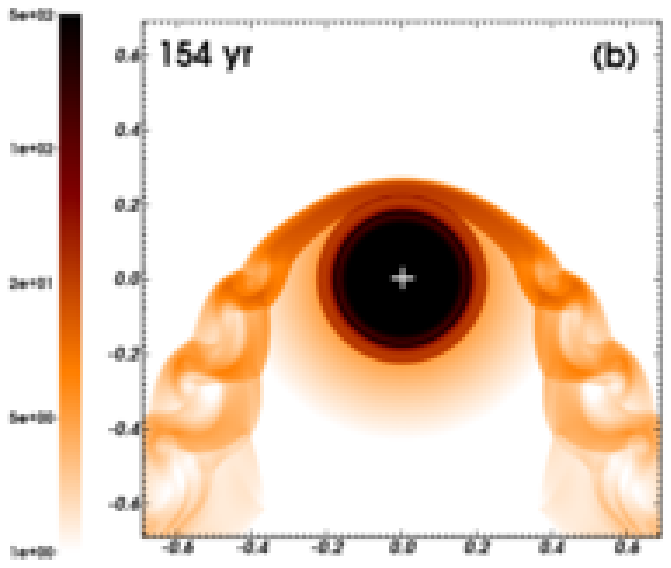}
	\end{minipage} 
	\begin{minipage}[b]{ 0.33\textwidth}
\includegraphics[width=1.0\textwidth]{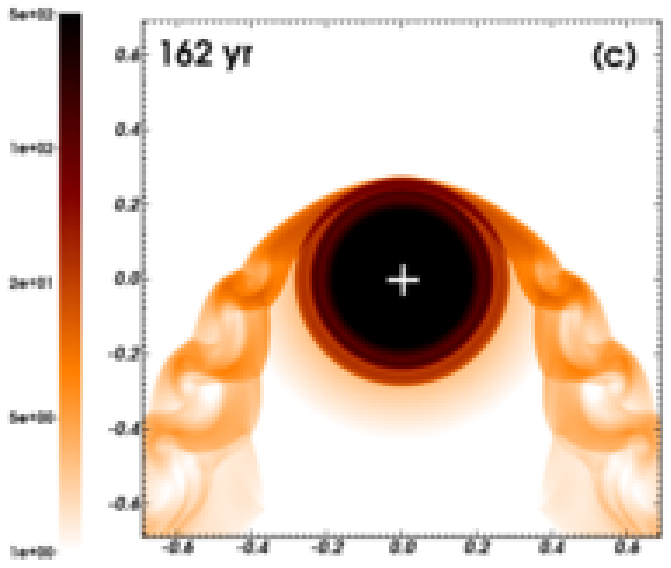}
	\end{minipage} \\
	\begin{minipage}[b]{ 0.33\textwidth}
\includegraphics[width=1.0\textwidth]{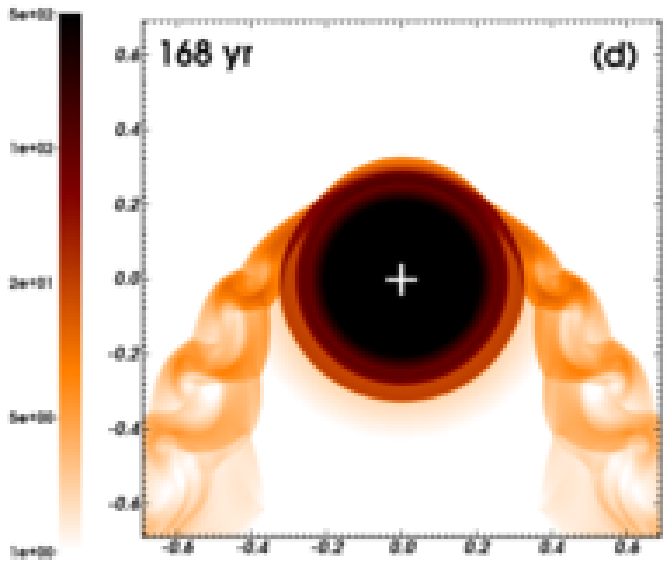}
	\end{minipage} 
	\begin{minipage}[b]{ 0.33\textwidth}
\includegraphics[width=1.0\textwidth]{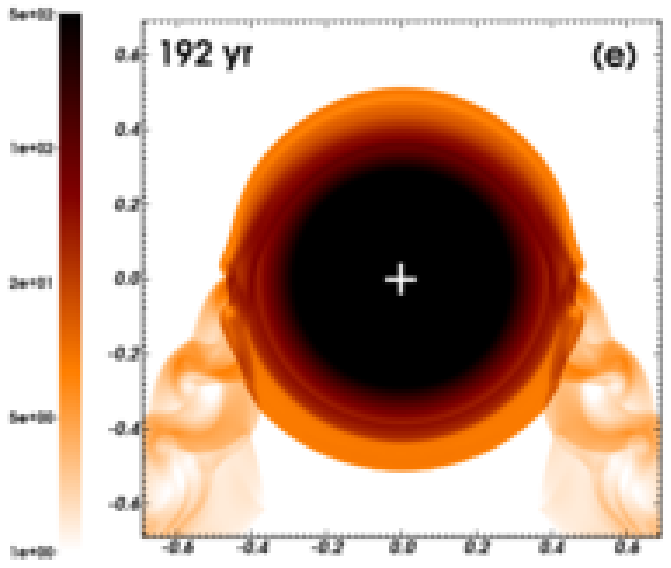}
	\end{minipage} 
	\begin{minipage}[b]{ 0.33\textwidth}
\includegraphics[width=1.0\textwidth]{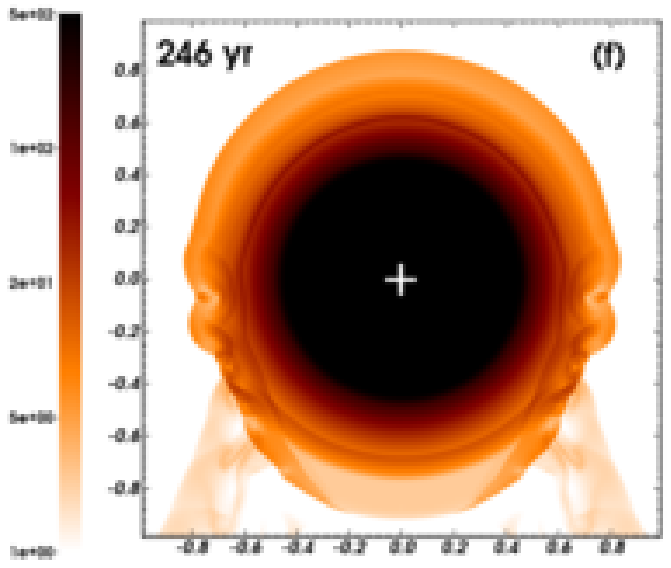}
	\end{minipage} 	\\
	\caption{ 
                 Time sequence of a supernova interacting with
the bow shock generated by our initially $10\, \rm M_{\odot}$ progenitor moving with
velocity $40\, \rm km\, \rm s^{-1}$. The figures correspond to times $t_{\rm sncsm}$
(a) up to about $t_{\rm ysnr}$ (f). The gas number density is shown with a
density range from $1.0$ to $5\times 10^{2}\, \mathrm{cm}^{-3}$ on a 
logarithmic scale. The white cross marks the center of the explosion.
The $x$-axis represents the radial direction and the $y$-axis
the direction of stellar motion (in $\mathrm{pc}$). Only part of the
computational domain is shown in the figures. Note that the panel (f) shows the supernova
remnant at larger scale than in panels (a-e).
		}
	\label{fig:snr_interacting}  
\end{figure*}

\begin{figure} 
        \centering
	\begin{minipage}[b]{ 0.43\textwidth}
        \includegraphics[width=1.0\textwidth,angle=0]{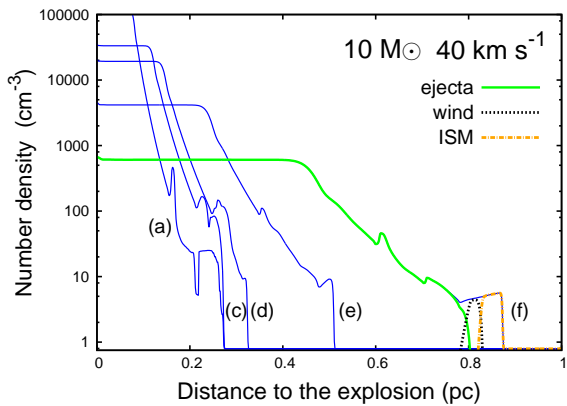}
	\end{minipage} 
	\caption{ 
	        Cross-sections taken along the direction of motion of the progenitor 
	        through the number density field of the supernova blastwave interacting 
	        with the $z\ge0$ region of the bow shock produced by our initially $10\, \rm M_{\odot}$ 
	        progenitor moving with velocity $40\, \rm km\, \rm s^{-1}$ (Fig.~\ref{fig:snr_interacting}a-f).
	        \textcolor{black}{Note that the cross-section from Fig.~\ref{fig:snr_interacting}b is omitted in the 
	        figure because it is very similar the curve (a). }
	        The respective proportion of the supernova ejecta (solid green), stellar wind (dotted \textcolor{black}{black}) 
	        and ISM material (dashed orange) are indicated for the curve (f).
		}
	\label{fig:profiles_interaction}  
\end{figure}

The supernova interacting with the bow shock generated by the $10\, \rm M_{\odot}$
star moving with $v_{\star}=40\, \rm km\, \rm s^{-1}$ is illustrated in
Fig.~\ref{fig:snr_interacting}. We show the density field in our simulation
YSNR1040, composed of a shock wave interacting with its circumstellar medium
(our model SNCSM1040). The density stratification is shown at times $t_{\rm sncsm} \approx
152\, \rm yr$ (Fig.~\ref{fig:snr_interacting}a), $154$, $162$, $168$, $192$ and
$t_{\rm ysnr} \approx 246\, \rm yr$ (Fig.~\ref{fig:snr_interacting}f) after the
explosion, respectively. Note that the part of the computational domain plotted
in Fig.~\ref{fig:snr_interacting}f is larger than in
Figs.~\ref{fig:snr_interacting}a-e. Corresponding cross-sections measured in
these density fields along the $Oz$ axis are shown in
Fig.~\ref{fig:profiles_interaction}.

At a time $t_{\rm sncsm}\approx152\, \rm yr$ the solution SNCSM1040 is mapped on
the model PSN1040 shortly before the supernova remnant shock wave reaches the
reverse shock of the bow shock (Fig.~\ref{fig:snr_interacting}a).
\textcolor{black}{The organisation of the remnant is similar to the situation
depicted in fig.~5 of~\citet{velazquez_apj_649_2006}.} From the origin to larger
radius, the curve (a) of Fig.~\ref{fig:profiles_interaction} plots the initial
plateau of density $\rho_{\rm core}(r)\approx 10^{5}\rm cm^{-3}$, the steep
profile $\rho_{\rm max}(r)\propto r^{-n}$, the shell of swept-up ejecta at about
$0.16\, \rm pc$ (cf. Fig.~\ref{fig:sncsm}), the shock wave progressing in the
freely-expanding wind, the red-supergiant bow shock from about $0.21$ to about
$0.27\, \rm pc$, and the unperturbed ISM.

In Fig.~\ref{fig:snr_interacting}b the shock wave collides with the reverse
shock of the bow shock and begins to interact with the dense ($n\approx\, 25\rm
cm^{-3}$) shocked wind. The interaction starts at the stagnation point because 
this is the part of the bow shock with the smallest radius~\citep{borkowski_apj_400_1992}. 
The blastwave decelerates and loses its
spherical symmetry, the shock wave penetrates the reverse shock of the bow shock
at a time $154\, \rm yr$ with velocity $v\approx 6700\, \rm km\, \rm s^{-1}$
whereas it hits its forward shock $8\, \rm yr$ later with velocity $v\approx
4500\, \rm km\, \rm s^{-1}$. At $162\, \rm yr$, the shell of shocked ejecta
merges with the former post-shock region at the reverse shock of the bow shock,
and its material is compressed to $n\approx \, 85\, \rm cm^{-3}$ (curve (c)
of Fig.~\ref{fig:profiles_interaction}).

In Fig.~\ref{fig:snr_interacting}c the shock wave has totally penetrated the bow
shock, both a reflected and a transmitted shock wave form at both the ends of the bow
shock. In Fig.~\ref{fig:snr_interacting}d the transmitted shock at the former
forward shock starts expanding into the undisturbed ISM. As sketched in fig.~3
of~\citet{borkowski_apj_400_1992}, a bump emerges beyond the bow shock and it
reaches about $0.32\, \rm pc$ at a time $168\, \rm yr$ after the explosion
(curve (d) of Fig.~\ref{fig:profiles_interaction}).  It expands and enlarges
laterally as the shock wave, that is no longer restrained by the material of the
bow shock, penetrates the undisturbed ISM, accelerates and progressively
recovers its spherical symmetry (Fig.~\ref{fig:snr_interacting}e, \textcolor{black}{see also 
Fig.~1 of~\citealp{brighenti_mnras_270_1994}}).

At a time $246\, \rm yr$, the shock wave has recovered its sphericity
(Fig.~\ref{fig:snr_interacting}f). Note that the shock wave is slightly
constricted in the cavity of unshocked wind as it expands downstream from the
direction of motion of the progenitor. This anisotropy is a function of the
circumstellar density distribution at the pre-supernova phase and 
governs the long term evolution of the supernova
remnant (Section~\ref{subsect:sphericity}). The curve (f) of
Fig.~\ref{fig:profiles_interaction} shows the density structure composed of the
plateau whose density has diminished to $\rho_{\rm core}\approx600\, \rm
cm^{-3}$, the twice shocked ejecta, the twice shocked stellar wind,
twice shocked ISM, shocked ISM and finally the unperturbed
ISM~\citep[c.f.][]{borkowski_apj_400_1992}. These regions are not clearly
discernible because of the mixing at work provoked by the multiple reflections
and refractions proliferating through the remnant (curve (f) of
Fig.~\ref{fig:profiles_interaction}). 
\textcolor{black}{ 
We underline that our one-dimensional, spherically-symmetric 
calculations do not allow us to model the Rayleigh-Taylor instabilities triggered at 
the interface between supernova ejecta and uniformly-expanding stellar 
wind~\citep{chevelier_apj_392_1992}. These instabilities may affect 
the propagation of the shock wave through the bow shock. 
}

\subsection{Spherical and aspherical supernova remnants}
\label{subsect:sphericity}

\begin{figure}

	\centering
	\begin{minipage}[b]{ 0.41\textwidth}
	
\includegraphics[width=1.0\textwidth,angle=0]{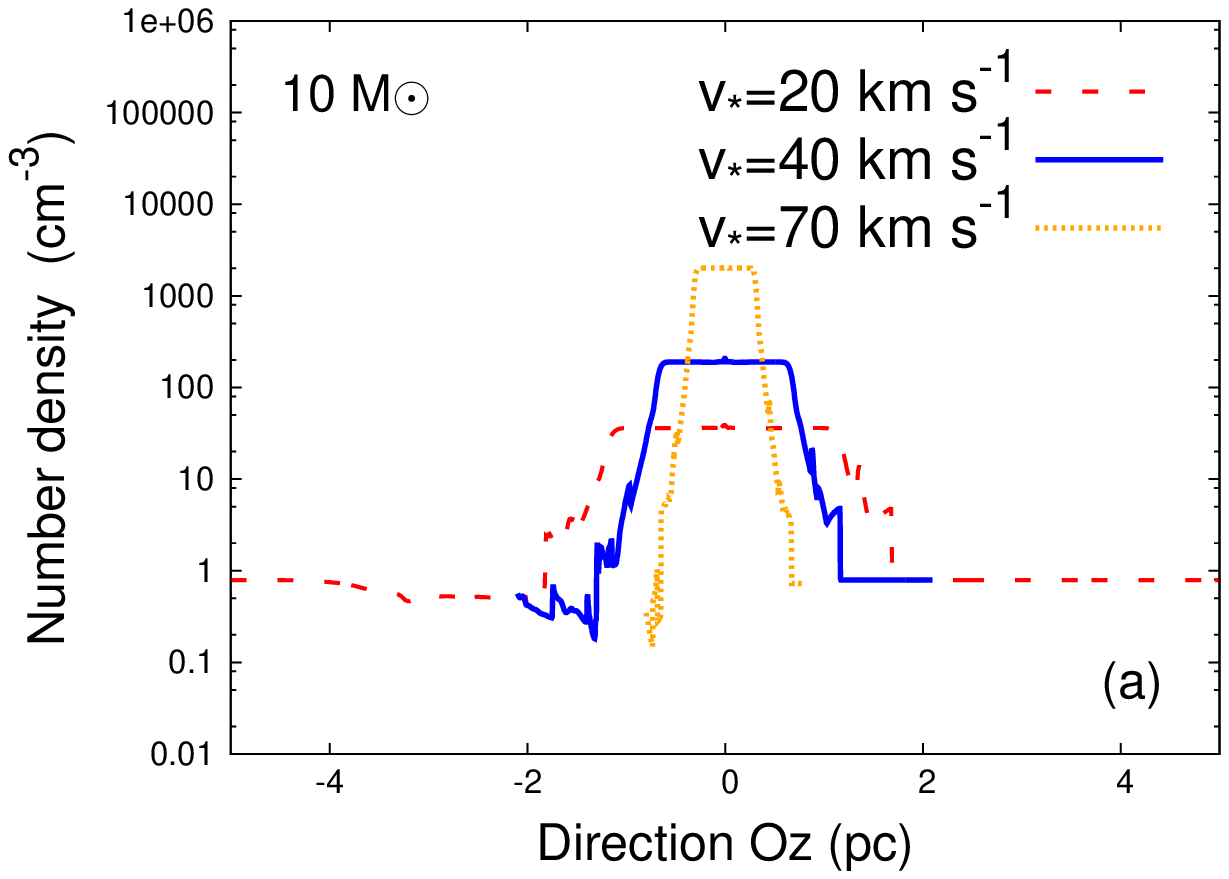}
	\end{minipage}   
	\begin{minipage}[b]{ 0.41\textwidth}
	
\includegraphics[width=1.0\textwidth,angle=0]{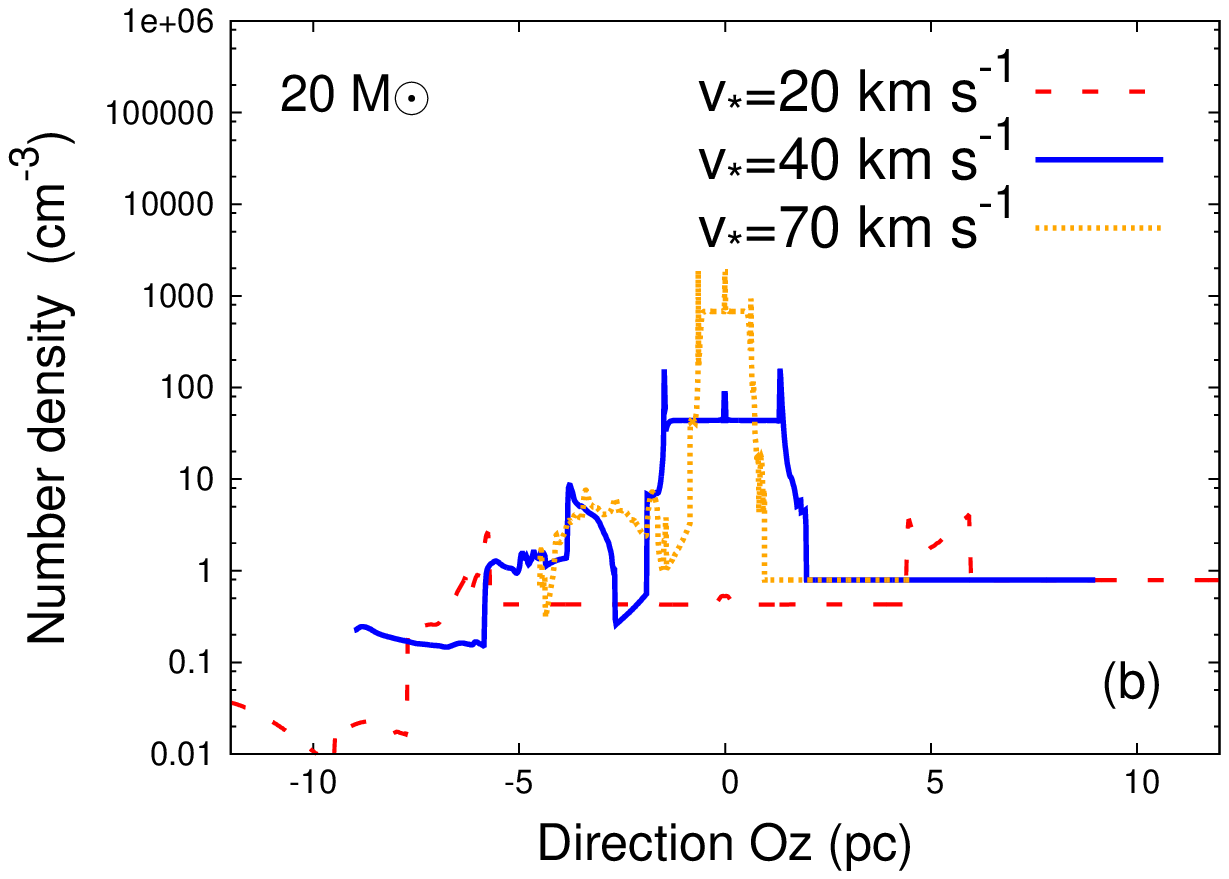}
	\end{minipage}  
	\begin{minipage}[b]{ 0.41\textwidth}
	
\includegraphics[width=1.0\textwidth,angle=0]{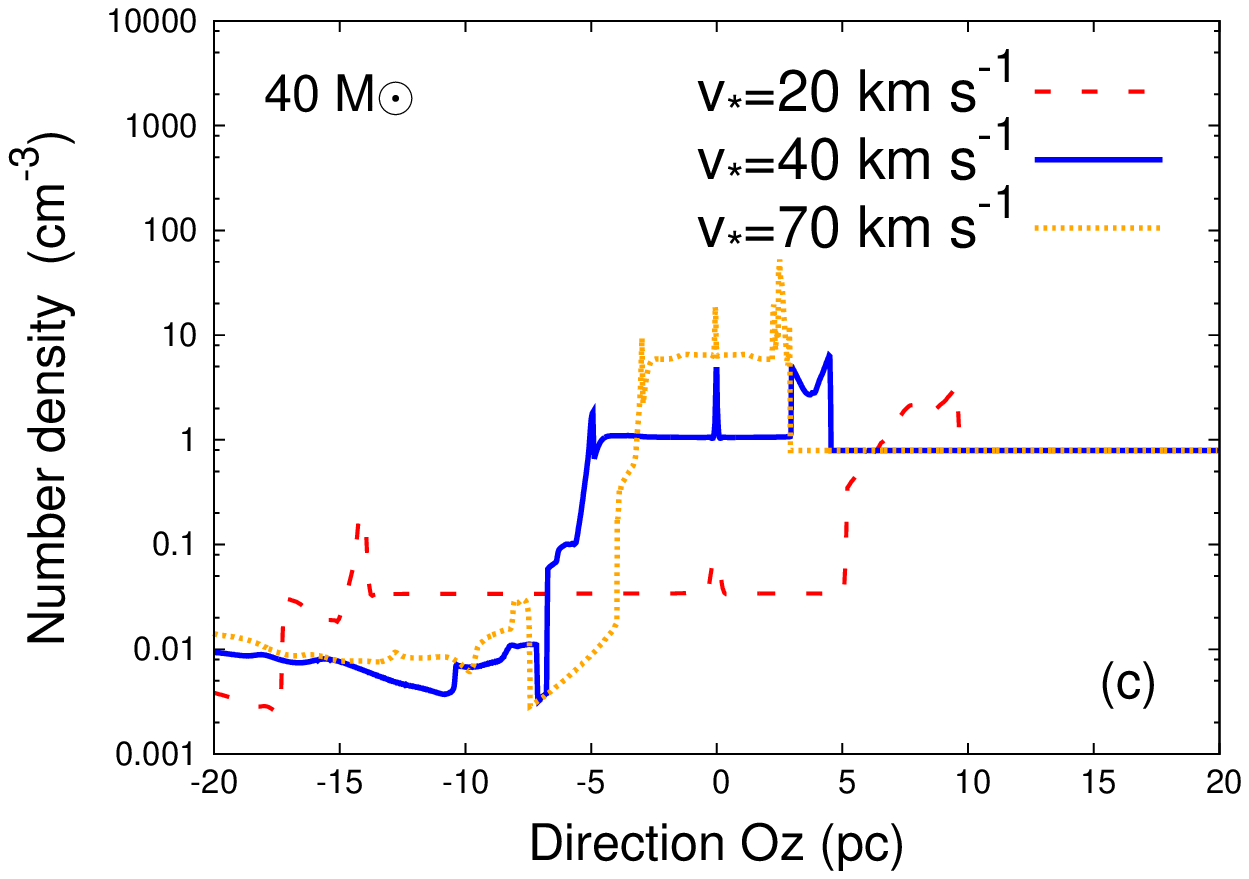}
	\end{minipage}
	\caption{                  
		Density profiles (in $\rm cm^{-3}$) of our supernova remnants taken 
		along the direction of motion of our initially $10$ (a), $20$ (b) and 
		$40\, \rm M_{\odot}$ (c) progenitors, taken when the supernova shock wave 
		has reached a distance of about $2R(0)$ in the direction of motion of 
		the progenitors, at $t_{\rm ysnr}$. 
		}
	\label{fig:cuts_for_asphericity}  
\end{figure}

In Fig.~\ref{fig:cuts_for_asphericity} we represent the density profiles of our
models of supernova remnants at a time $t_{\rm ysnr}$, taken along the direction
of motion of the progenitors. The models produced by our $10\,
M_{\odot}$ progenitor conserve their symmetry after the shock waves collide with
the bow shocks, e.g. our model YSNR1020 has a plateau of density
$n\approx10^{2}\, \rm cm^{-3}$ at $|z|\le\, 1.7\, \rm pc$, whereas its density
distribution in both directions beyond the shock wave is about the ISM ambient
medium density. Here, the shock wave expansion is barely disturbed by the light
bow shocks (our Table~\ref{tab:psn1}) and the remnants grow quasi-spherically
(Fig.~\ref{fig:snr_interacting}).

The remnant generated by our $20\, \rm M_{\odot}$ star moving with $v_{\star}=20\,
\rm km\, \rm s^{-1}$ is aspherical, in that it has a cavity of $n \approx
10^{-2}\, \rm cm^{-3}$ at $z<-8\, \rm pc$ (red curve in
Fig.~\ref{fig:cuts_for_asphericity}b). This is much less pronounced for models
with a larger $v_{\star}$, i.e. our models YSNR2040 and YSNR2070, see blue and
yellow curves in Fig.~\ref{fig:cuts_for_asphericity}b. They have a rather
spherical density distribution, and only a bulge of swept-up wind material in
the direction $+\bmath{\hat{z}}$ distinguishes them from a spherically symmetric
structure. A small accumulation of swept-up gas that could slightly decelerate the
shock wave forms in the direction $-\bmath{\hat{z}}$, i.e., these models do form
a pronounced wind cavity (Fig.~\ref{fig:cuts_for_asphericity}bc).

The supernova remnants of our $40\, \rm M_{\odot}$ progenitor are all strongly
anisotropic, e.g. our model with velocity $v_{\star}=70\, \rm km\, \rm s^{-1}$
has a dense shell of density $n\approx 10^{2}\, \rm cm^{-3}$ along the direction 
$+\bmath{\hat{z}}$ and a cavity of density $n\approx10^{-2}\, \rm cm^{-3}$ in the
opposite direction (Fig.~\ref{fig:cuts_for_asphericity}c). On the basis of
Table~\ref{tab:psn1} and according to the above discussion, we find that the bow
shocks of runaway stars that we simulate and which accumulate more than about
$1.5\, \rm M_{\odot}$ generate asymmetric supernova remnants. \textcolor{black}{We measure from our simulations that the collision 
between the shock waves and these dense bow shocks, located at $R(0)\approx\, 1.35-5\, \rm pc$ 
from the center of the explosion, begins about $160-750\, \rm yr$ and ends 
about $830-4900\, \rm yr$ after the supernova, respectively.} In the next section,
we continue our study focusing on the asymmetric models only, i.e. generated
either by our $20\, \rm M_{\odot}$ progenitor moving with $v_{\star}=20\,
\rm km\, \rm s^{-1}$ or produced by our $40\, \rm M_{\odot}$ star.

\begin{figure}
	\centering
	\begin{minipage}[b]{ 0.39\textwidth}
	
\includegraphics[width=1.0\textwidth]{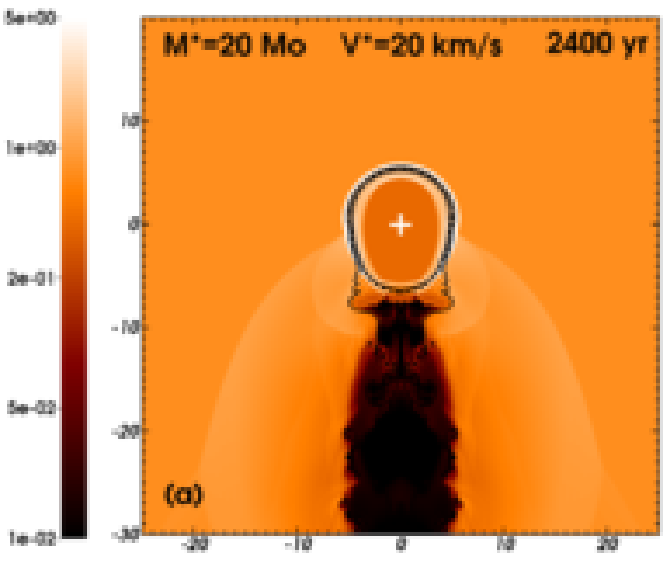}
	\end{minipage} \\
	\begin{minipage}[b]{ 0.39\textwidth}
\includegraphics[width=1.0\textwidth]{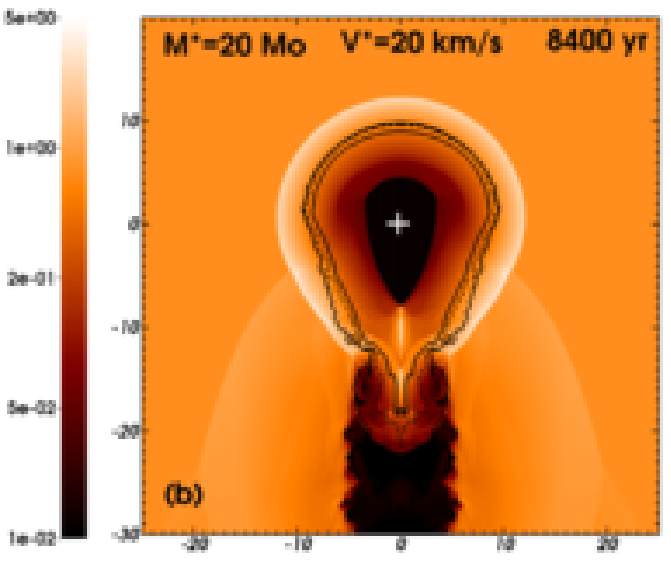}
	\end{minipage} \\
	\begin{minipage}[b]{ 0.39\textwidth}
\includegraphics[width=1.0\textwidth]{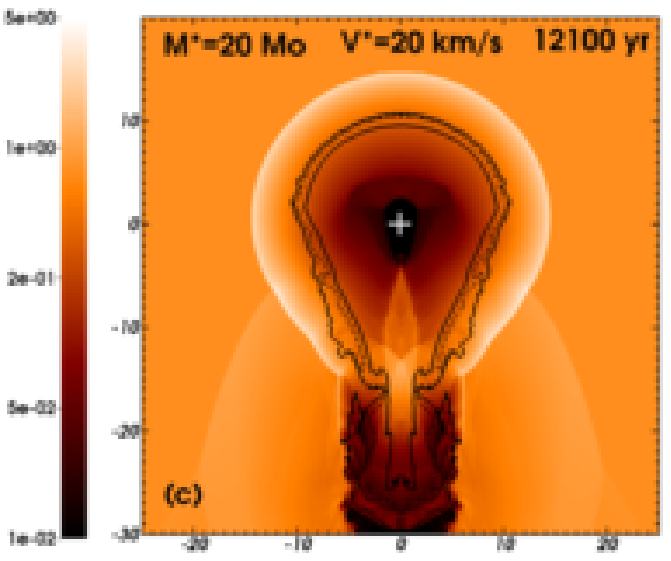}
	\end{minipage} 
	\caption{ 
                 Time sequence of the evolution of the supernova remnant
generated by our initially $20\, \rm M_{\odot}$ progenitor moving with velocity
$20\, \rm km\, \rm s^{-1}$. The figures are shown at a time $t_{\rm ysnr}$ (a), at
an intermediate time (b) and at the end of the simulation, at a time $t_{\rm osnr}$ (c). 
The gas number density is shown with a density
range from $10^{-2}$ to $5.0\, \mathrm{cm}^{-3}$ on the logarithmic scale. 
The color scale is reversed compare to Fig.~\ref{fig:progenitor1} to~\ref{fig:progenitor3}. 
The cross marks the center of the explosion. 
\textcolor{black}{Note the Napoleon's hat structure of the unperturbed circumstellar medium in panel (a).} 
The solid black contours trace the boundary between stellar wind material and
either the supernova ejecta or ISM gas $Q_{1}(\bmath{r})=1/2$. The dotted black contour is
the discontinuity between supernova ejecta and wind material
$Q_{2}(\bmath{r})=1/2$. The nomenclature of the models follows our
Table~\ref{tab:osnr}. The $x$-axis represents the radial direction and the
$y$-axis the direction of stellar motion (in $\mathrm{pc}$). Only part of the
computational domain is shown in the figures. 
		}
	\label{fig:snr_large_scale_2020}  
\end{figure}


\section{The old supernova remnant phase}
\label{subsect:large_scale_remnant}

In this section we detail the interaction between supernova ejecta and pre-shaped 
circumstellar medium once the expanding front has passed through the bow shock. 
We focus on our four models of supernova remnants whose solutions strongly deviate 
from sphericity.

\subsection{Physical characteristics of the remnants}
\label{subsect:morpho}

\subsubsection{Asymmetric structures...}
\label{subsect:phys_1}

We show the gas density fields in the old supernova remnant produced by our
$20\, \rm M_{\odot}$ progenitor moving with velocity $v_{\star}=20\, \rm km\, \rm
s^{-1}$ in Fig.~\ref{fig:snr_large_scale_2020}.
Figs.~\ref{fig:snr_large_scale_4020},~\ref{fig:snr_large_scale_4040}
and~\ref{fig:snr_large_scale_4070} are similar, but showing our $40\, \rm M_{\odot}$
progenitor moving with velocity $v_{\star}=20$, $40$ and $v_{\star}=70\, \rm
km\, \rm s^{-1}$, respectively. In each of these figures, panel (a)
corresponds to a time $t_{\rm ysnr}$ (our Table~\ref{tab:ysnr}), panel (b)
shows the shock wave expanding into the wind cavity and panel (c) shows the
remnant at $t_{\rm osnr}$ when the simulation ends (our Table~\ref{tab:osnr}).
The figures do not show the full computational domain. In
Figs.~\ref{fig:snr_large_scale_2020} to~\ref{fig:snr_large_scale_4070}, the
overplotted solid black line is the border between the wind and ISM gas where
the value of the passive scalar $Q_{1}(\bmath{r})=1/2$, and the dashed black
line is the discontinuity between the ejecta and the other materials where
$Q_{2}(\bmath{r})=1/2$.

At a time $t_{\rm ysnr}$, the shock wave is already asymmetric because its
unusually dense surroundings (see our Table~\ref{tab:psn1}) strongly restrain it
from expanding into the direction normal to the direction of motion of the
progenitor (panels (a) of Figs.~\ref{fig:snr_large_scale_2020}
to~\ref{fig:snr_large_scale_4070}). As an example, the forward shock in our
model OSNR2020 at a time $2400\, \rm yr$ has reached about $5.8\, \rm pc$ and $6\,
\rm pc$ along the $+\bmath{\hat{R}}$ and $+\bmath{\hat{z}}$ directions,
respectively, whereas it expands to about $-7.9\, \rm pc$ along the
$-\bmath{\hat{z}}$ direction. This asymmetry of the shock wave is particularly
pronounced in our simulation OSNR4020 whose pre-shaped circumstellar medium contains
the most massive bow shock with a mass of $116\, \rm M_{\odot}$
(Fig.~\ref{fig:snr_large_scale_4020}ab).

At times larger than $t_{\rm ysnr}$, the shock wave freely expands into the
undisturbed ISM \textcolor{black}{in the $+\bmath{\hat{R}}$ and $+\bmath{\hat{z}}$ directions}  
because it has entirely overtaken its circumstellar
structure~\citep{brighenti_mnras_270_1994}, see
Figs.~\ref{fig:snr_large_scale_4020}-\ref{fig:snr_large_scale_4070}bc. 
It progressively recovers its sphericity, but this takes longer in our simulations
with slowly moving progenitors because the mass in their bow shock is 
larger (Fig.~\ref{fig:snr_large_scale_4020}c and~\ref{fig:snr_large_scale_4070}c). 
The penetration of the shock wave through
the wake of the bow shock results in its chanelling into the tubular cavity of
unshocked stellar wind~\citep{blandford_301_natur_1983}. The constant
cross-sectional area of the cavity continues to impose large temperature and
pressure jumps at the post-shock region of the shock wave, which prevents it
from decelerating and which collimates the ejecta as a tubular/jet-like extension to
the spherical region of shocked ejecta~\citep{cox_mnras_250_1991}.

\subsubsection{... of \textcolor{black}{angle-dependent physical properties}}
\label{subsect:phys_2}

\begin{figure}
	\centering
	\begin{minipage}[b]{ 0.39\textwidth}
	
\includegraphics[width=1.0\textwidth]{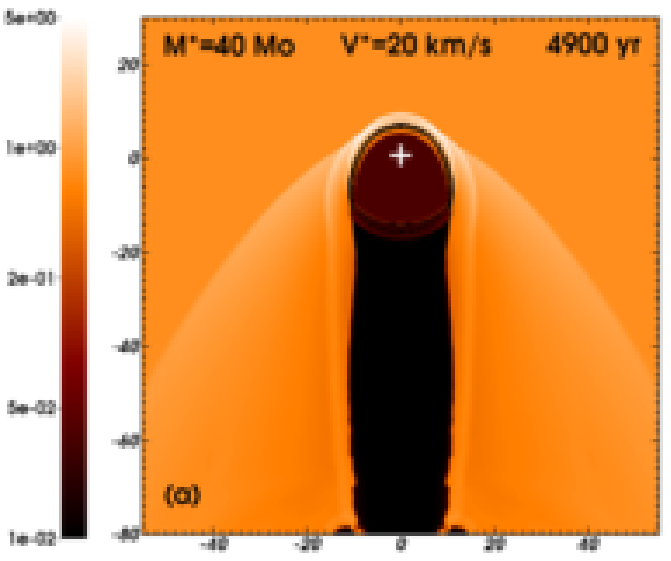}
	\end{minipage} \\
	\begin{minipage}[b]{ 0.39\textwidth}
	
\includegraphics[width=1.0\textwidth]{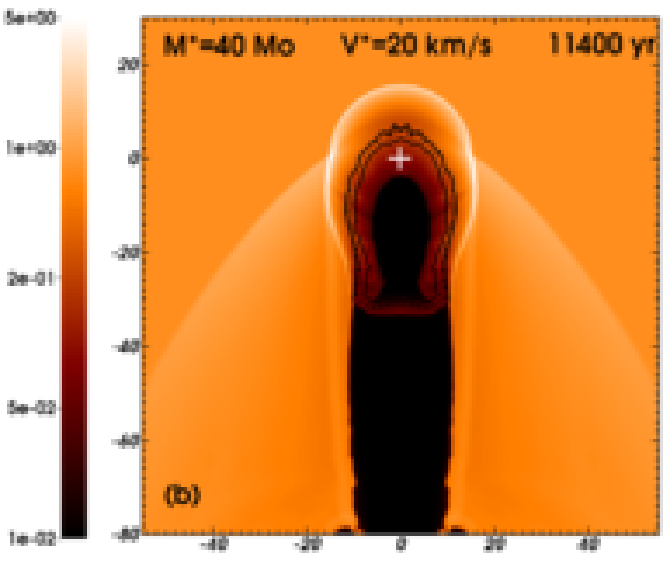}
	\end{minipage} \\
	\begin{minipage}[b]{ 0.39\textwidth}
	
\includegraphics[width=1.0\textwidth]{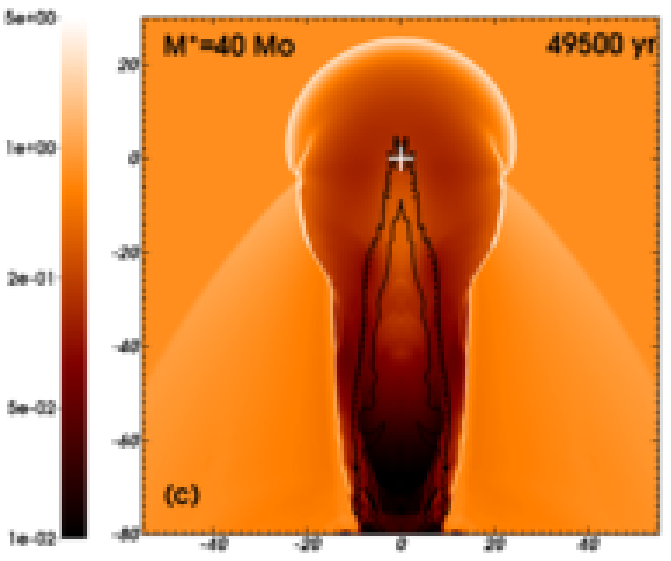}
	\end{minipage} 	
	\caption{ 
                 As Fig.~\ref{fig:snr_large_scale_2020}, with our initially $40\, \rm M_{\odot}$ progenitor.
		}
	\label{fig:snr_large_scale_4020}  
\end{figure}

Fig.~\ref{fig:expansion} plots the spatio-temporal evolution of the shock front
measured in our apsherical remnants along the $+\bmath{\hat{z}}$ and
$-\bmath{\hat{z}}$ directions. The shock wave typically expands into the ISM at
velocities of the order of a few hundred \textcolor{black}{$\rm km\, \rm s^{-1}$} whereas it
propagates inside the trail of the bow shock with a velocity of the order of a
few thousands \textcolor{black}{$\rm km\, \rm s^{-1}$}. E.g. at a time $4500\, \rm yr$ after the
explosion, the model OSNR2020 has a shock wave velocity of $564$ and $1578\, \rm
km\, \rm s^{-1}$ at $8.73$ and $13.03\, \rm pc$ from the center of the explosion
in the direction along and opposite of the progenitor's motion, respectively.
The \textcolor{black}{deceleration} of the shock wave is more important for our slowly
moving progenitors which induce the strongest anisotropy in their circumstellar
medium (Fig.~\ref{fig:expansion}a-b), \textcolor{black}{whereas the ejecta velocity is larger after 
the collision with the lighter bow shocks of our fast-moving progenitors 
(Fig.~\ref{fig:expansion}d and our Tab.~\ref{tab:psn1})}.

Because the blastwave expand in non-uniform medium~\citep{ferreira_478_aa_2008}, a
wave created during the collision with the dense bow shock is reflected towards
the center of the explosion and shocks back the unperturbed supernova ejecta
(see Fig~\ref{fig:snr_large_scale_2020}a-c). This mechanism generates a hot  
region of ejecta which progressively fills the entire cavity of the
remnants, e.g. the shocked ejecta of our model OSNR2020 has $n\approx 3\, \rm
cm^{-3}$ and $T\approx10^{6}\, \rm K$ (Fig.~\ref{fig:snr_large_scale_2020}bc).
Simultaneously, the collimated shock wave continues expanding downstream from
the center of the explosion. It hits the tunnel's side,~\textcolor{black}{i.e., 
the walls of the wind cavity,} which post-shock density
increases up to about $30\, \rm cm^{-3}$ in model OSNR4020 and cools to 
less than $10^{5}\, \rm K$. The shocked walls produce strong optical line
emission~\citep[][see also Section~\ref{subsect:maps}]{cox_mnras_250_1991}.

\begin{figure}
	\centering
	\begin{minipage}[b]{ 0.39\textwidth}
	
\includegraphics[width=1.0\textwidth]{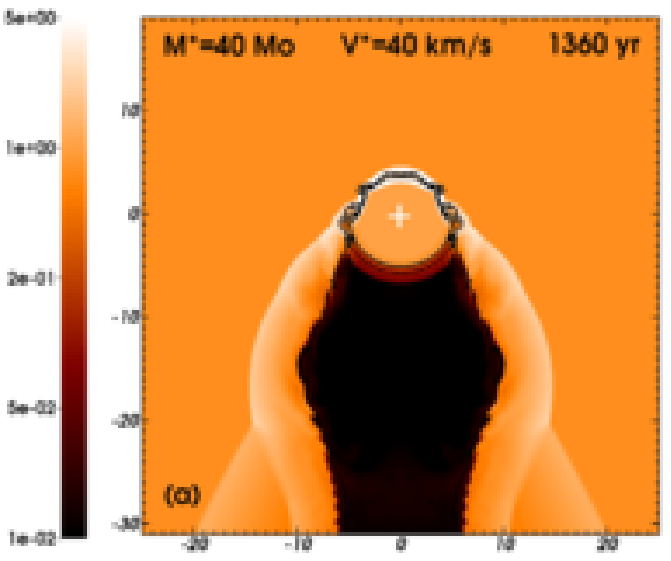}
	\end{minipage} \\
	\begin{minipage}[b]{ 0.39\textwidth}
	
\includegraphics[width=1.0\textwidth]{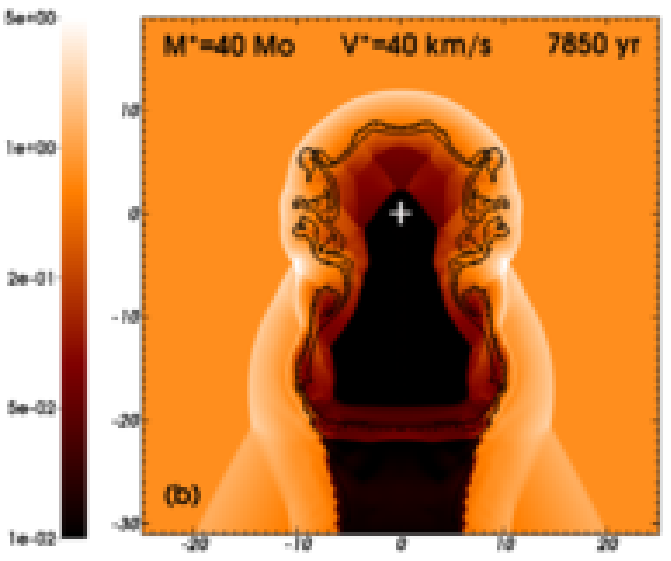}
	\end{minipage} \\
	\begin{minipage}[b]{ 0.39\textwidth}
	
\includegraphics[width=1.0\textwidth]{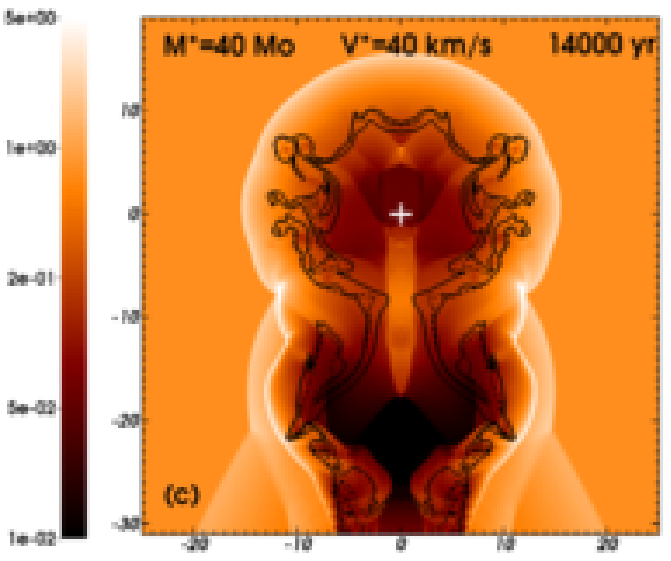}
	\end{minipage} 	
	\caption{ 
                 As Fig.~\ref{fig:snr_large_scale_4020}, with space velocity $v_{\star}=40\, \rm km\, \rm s^{-1}$.
		}
	\label{fig:snr_large_scale_4040}  
\end{figure}

\begin{figure*}
	\begin{minipage}[b]{ 0.31\textwidth}
	
\includegraphics[width=1.0\textwidth]{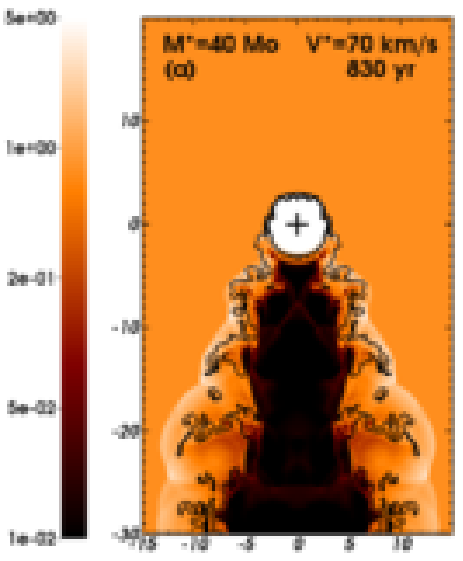}
	\end{minipage} 
	\begin{minipage}[b]{ 0.31\textwidth}
	
\includegraphics[width=1.0\textwidth]{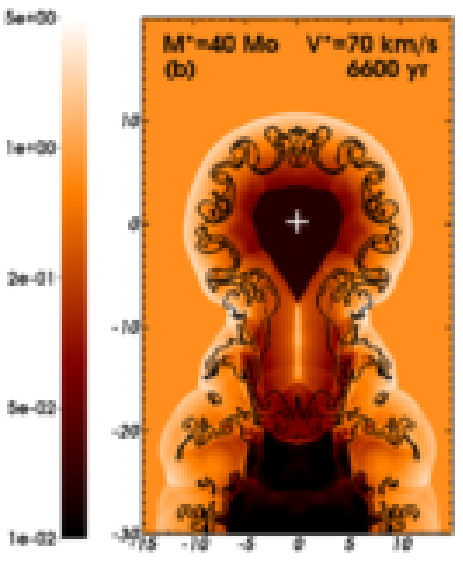}
	\end{minipage} 
	\begin{minipage}[b]{ 0.31\textwidth}
	
\includegraphics[width=1.0\textwidth]{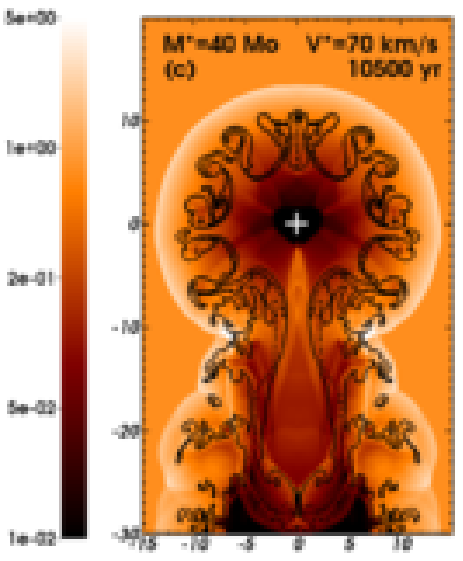}
	\end{minipage} 	
	\caption{ 
                 As Fig.~\ref{fig:snr_large_scale_4020}, with space velocity $v_{\star}=70\, \rm km\, \rm s^{-1}$.
		}
	\label{fig:snr_large_scale_4070}  
\end{figure*}

A transmitted wave penetrates the shocked wind material in the bow
shock~\citep{brighenti_mnras_270_1994} while a \textcolor{black}{second} wave is reflected towards the
center of the tunnel. After the passage of the shock wave through the interface
that separates the tunnel from the bow shock, its cross-sectional area is
locally constricted and accelerates the flow. It is about $1500\, \rm km\, \rm
s^{-1}$ at a time $4200\, \rm yr$, about $2750\, \rm km\, \rm s^{-1}$ at a time
\textcolor{black}{$8000\, \rm yr$} and decelerates down to $1450\, \rm km\, \rm s^{-1}$ when the
shock wave expands further in the tunnel at a time $12000\, \rm yr$ in our
simulation OSNR2020 (Fig.~\ref{fig:expansion}). The same happens when the
reflected waves collide at the center of the tunnel
(Fig.~\ref{fig:expansion}a-b). At a time $t_{\rm ysnr}$ almost the whole
interior of the remnant is shocked by the reflected shock wave, and these
multiple reflections induce a strong mixing of ejecta, wind and ISM (see the
overlapping of the lines where $Q_{1}(\bmath{r})=Q_{2}(\bmath{r})=1/2$). 
\textcolor{black}{The Rayleigh-Taylor instabilities developing upstream from 
the center of the explosion (see, e.g. Fig.~\ref{fig:snr_large_scale_4070}c) 
have an origin similar to the ones described in~\citet{chevelier_apj_392_1992}, 
but take place at the interface with the shocked ISM gas. 
Additionally, the interaction with the thin layer of stellar wind material
elongates them up to close to the forward shock~\citep{kane_apj_511_1999}. 
}

\begin{figure*} 
	\centering
	\begin{minipage}[b]{ 0.45\textwidth}
\includegraphics[width=1.0\textwidth,angle=0]{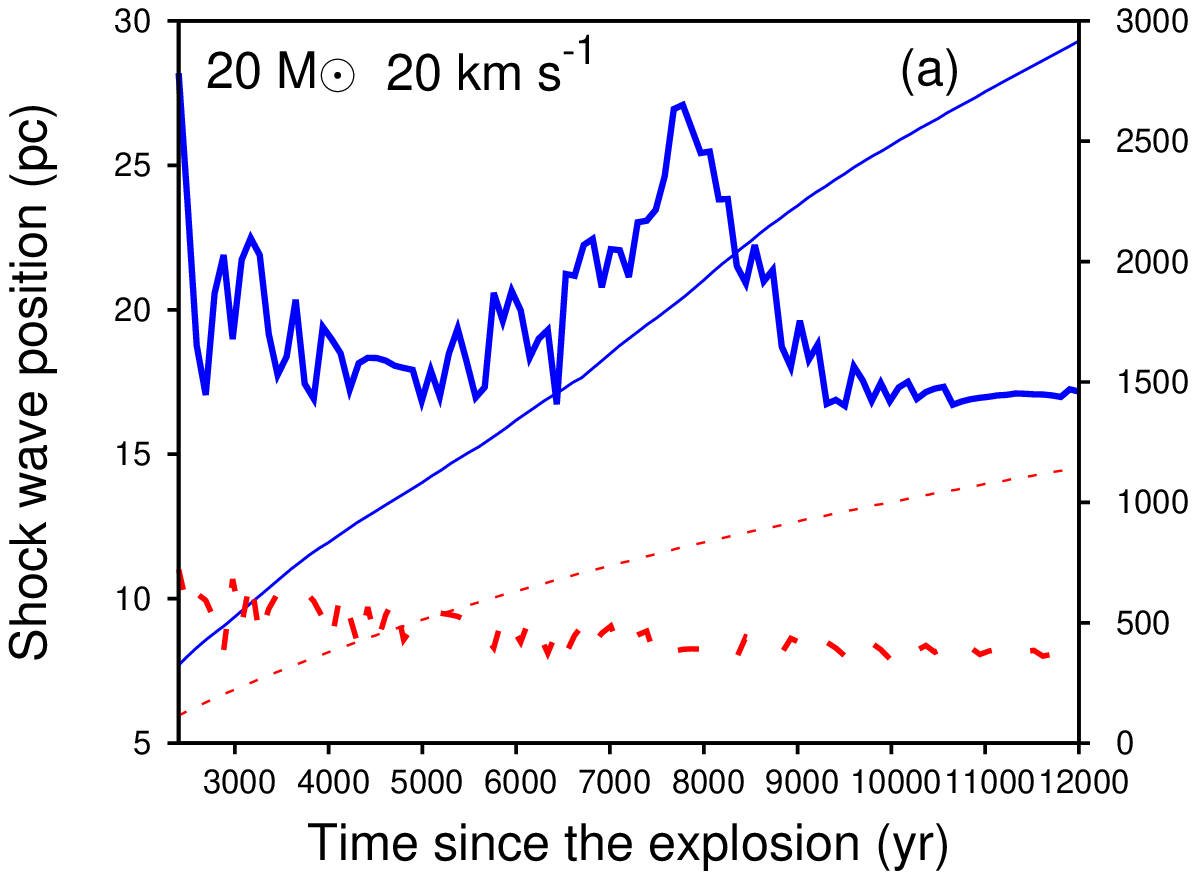}
	\end{minipage} 
	\begin{minipage}[b]{ 0.45\textwidth}
\includegraphics[width=1.0\textwidth,angle=0]{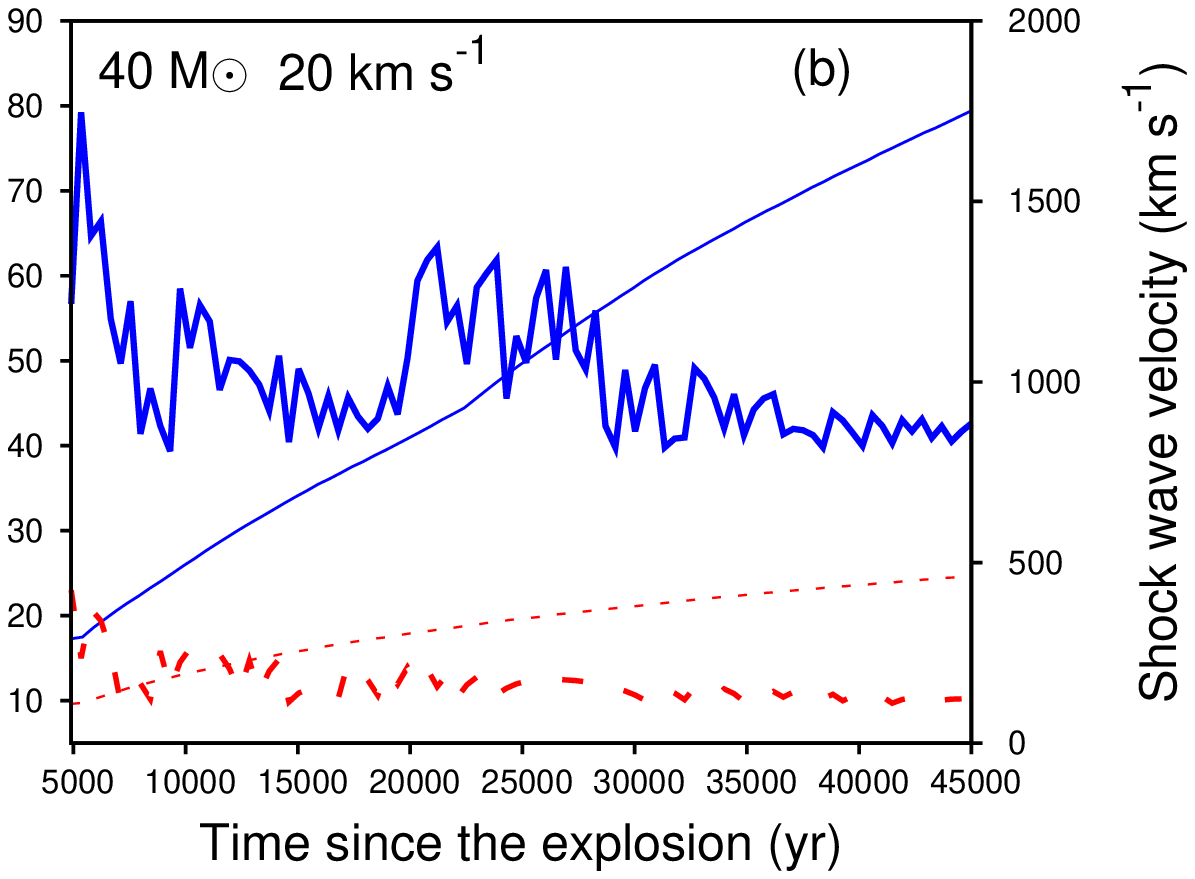}
	\end{minipage} \\
	\begin{minipage}[b]{ 0.45\textwidth}
\includegraphics[width=1.0\textwidth,angle=0]{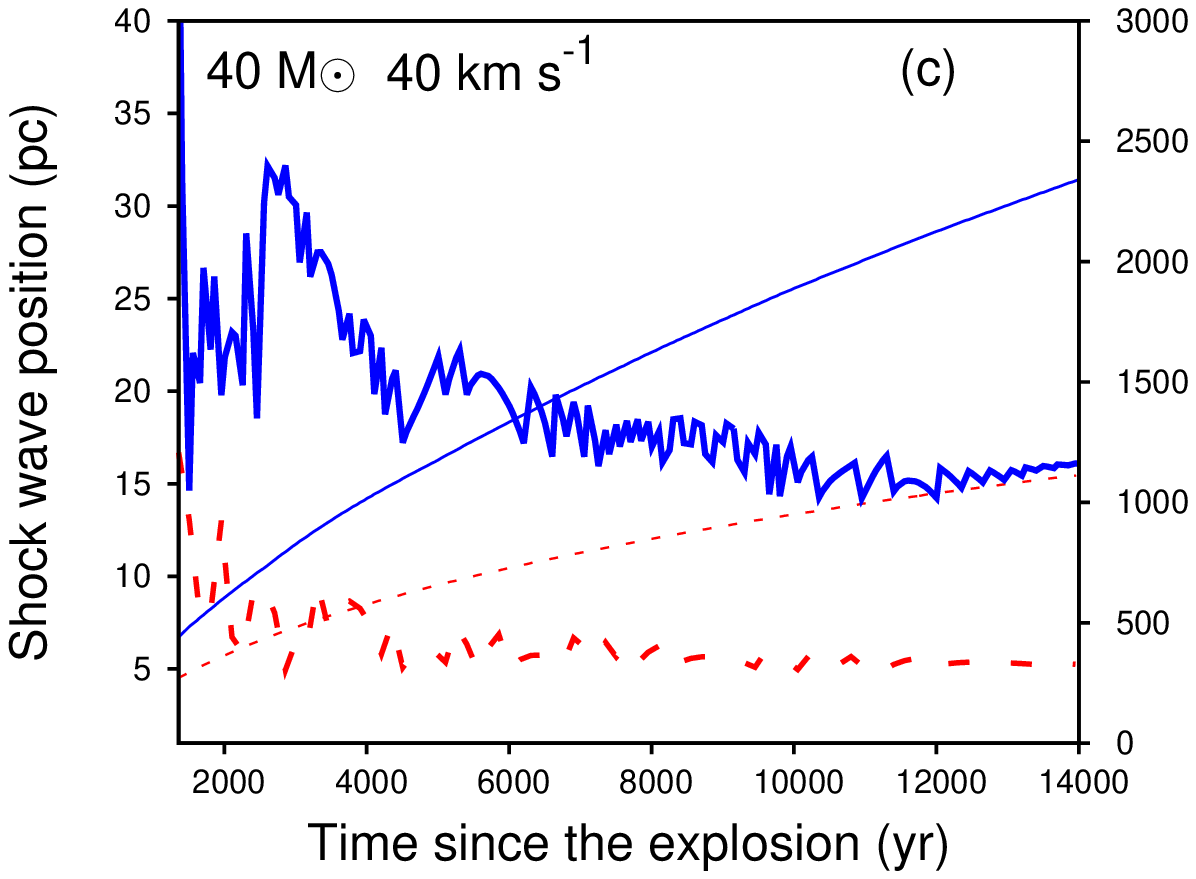}
	\end{minipage} 
	\begin{minipage}[b]{ 0.45\textwidth}
\includegraphics[width=1.0\textwidth,angle=0]{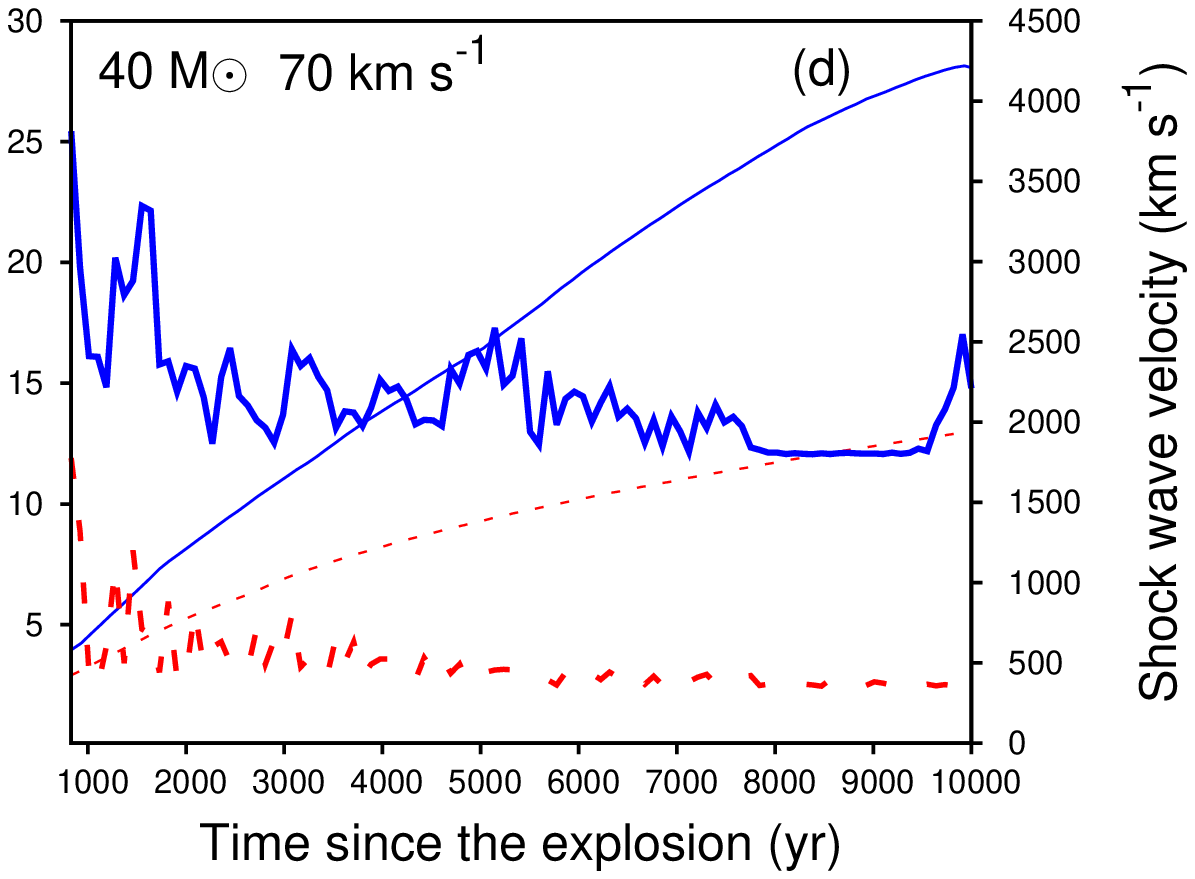}
	\end{minipage} 	
	\caption{                  
Time evolution of the shock waves of our aspherical supernova remnant models.
The figure plots the shock wave position (thin lines, in $\rm pc$) and velocity
(thick lines, in $\rm km\, \rm s^{-1}$), \textcolor{black}{measured along the 
direction of motion of the progenitor, both} upstream (dashed red lines,
$+\bmath{\hat{z}}$ direction) and downstream (solid blue lines,
$-\bmath{\hat{z}}$ direction) \textcolor{black}{from} the center of the explosion, respectively. The
figure represents the expansion from $t_{\rm ysnr}$ up to the end of the
simulation at $t_{\rm osnr}$ (in $\rm yr$) of the remnants generated by our
initially $20\, \rm M_{\odot}$ progenitor moving with velocity $20\, \rm km\,
\rm s^{-1}$ (a), our initially $40\, \rm M_{\odot}$ progenitor moving with
velocity $20$ (b), $40$ (c) and $70\, \rm km\, \rm s^{-1}$ (d), respectively.
		}
	\label{fig:expansion}  
\end{figure*}

The stars end their lives as core-collapse supernovae and the explosion can
produce a runaway neutron star~\citep{lyne_natur_369_1994}. Assuming a typical
velocity of the compact object of about $400\, \rm km\, \rm
s^{-1}$~\citep{hibbs_360_mnras_2005}, one finds that it could not be further than about
$4.84,\, 19.8,\, 5.6$ and $4.2\, \rm pc$ from the center of the explosion in our
simulations at $t_{\rm osnr}$
(Figs.~\ref{fig:snr_large_scale_2020}c,~\ref{fig:snr_large_scale_4020}c,
~\ref{fig:snr_large_scale_4040}c and~\ref{fig:snr_large_scale_4070}c).
Consequently, we suggest that our remnants host a neutron star of mass $M_{\rm
co}\approx2\, \rm M_{\odot}$ in the region close to the center of the explosion,
i.e. out of the chimney-like extension of channelled ejecta.


\subsection{Remnants luminosity}
\label{subsect:luminosity}

\begin{figure*}
	\centering	
	\begin{minipage}[b]{ 0.45\textwidth}
\includegraphics[width=1.0\textwidth,angle=0]{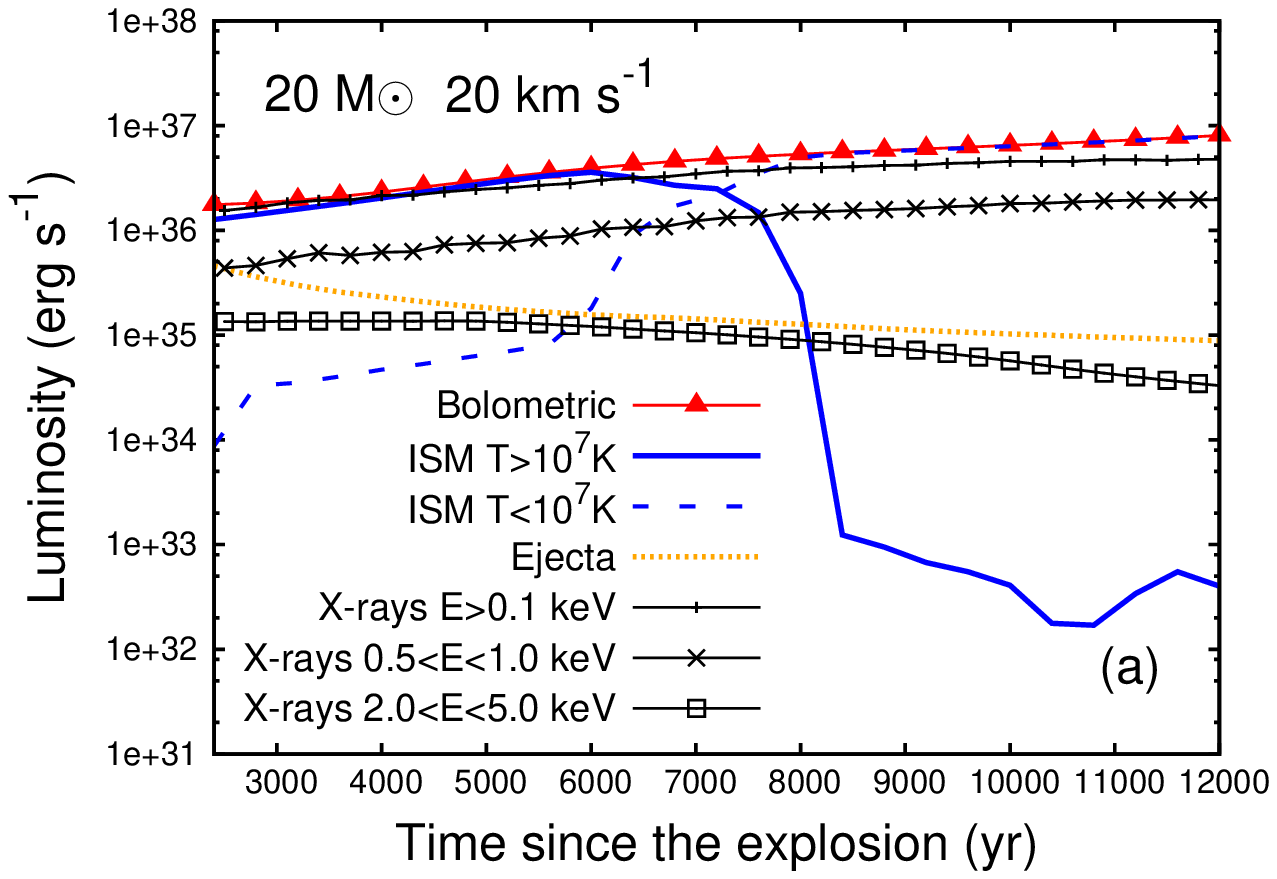}
	\end{minipage} 
	\begin{minipage}[b]{ 0.45\textwidth}
\includegraphics[width=1.0\textwidth,angle=0]{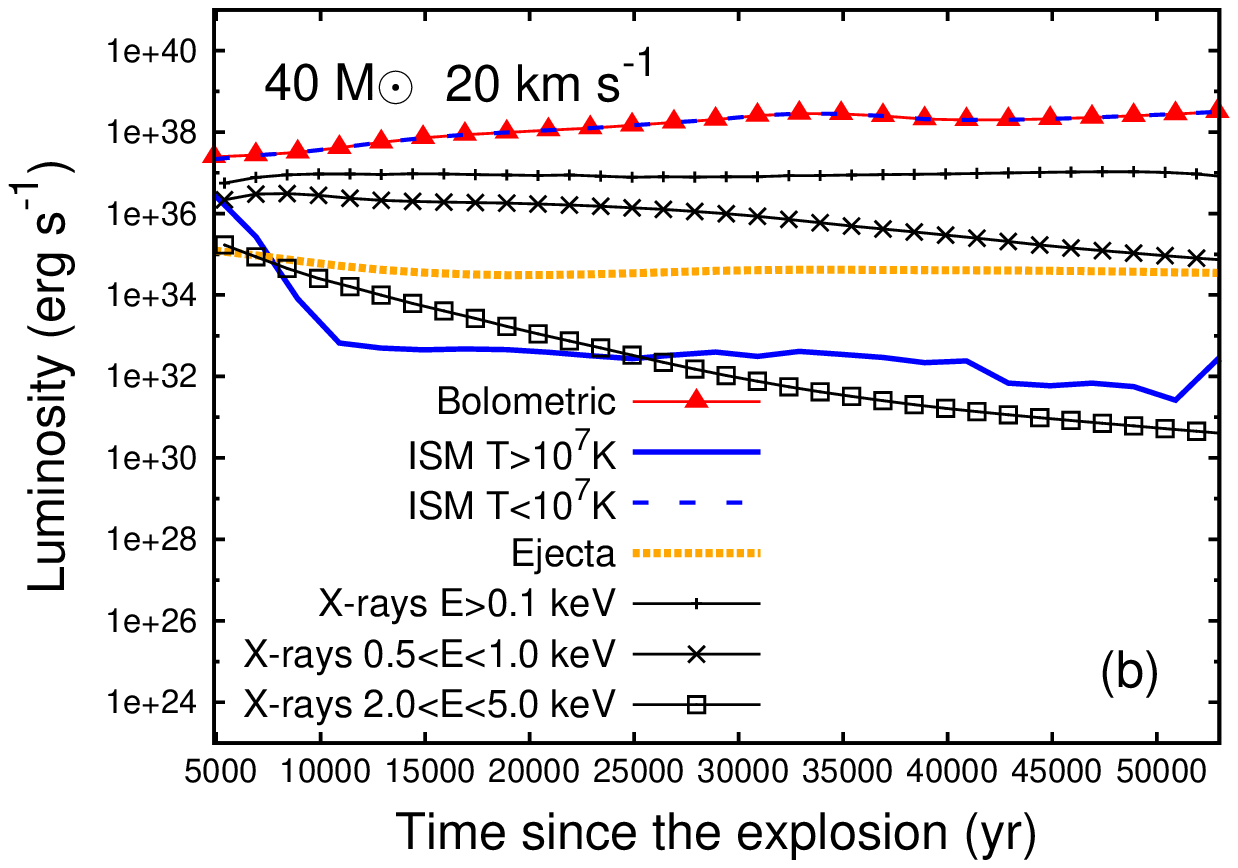}
	\end{minipage} 	\\	
	\begin{minipage}[b]{ 0.45\textwidth}
\includegraphics[width=1.0\textwidth,angle=0]{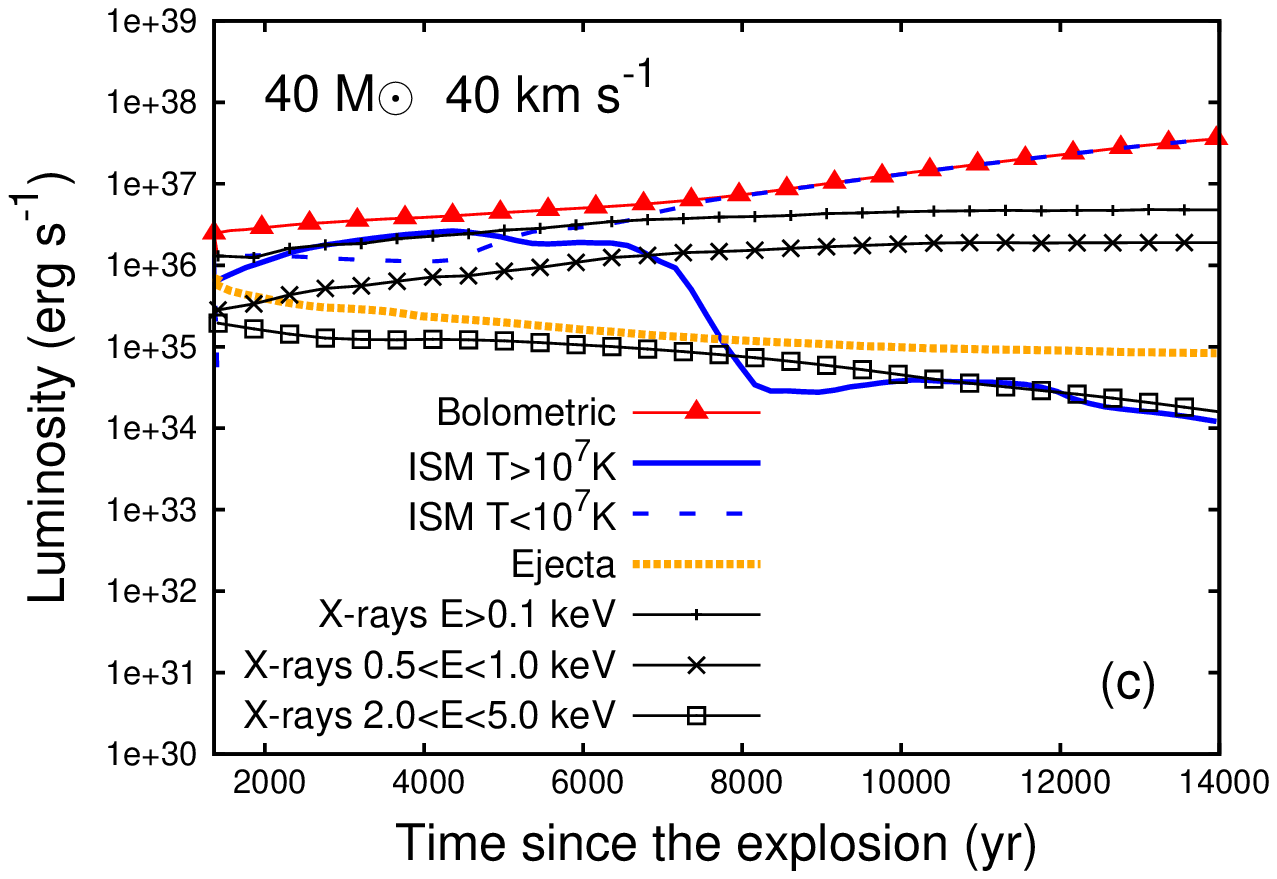}
	\end{minipage} 		
	\begin{minipage}[b]{ 0.45\textwidth}
\includegraphics[width=1.0\textwidth,angle=0]{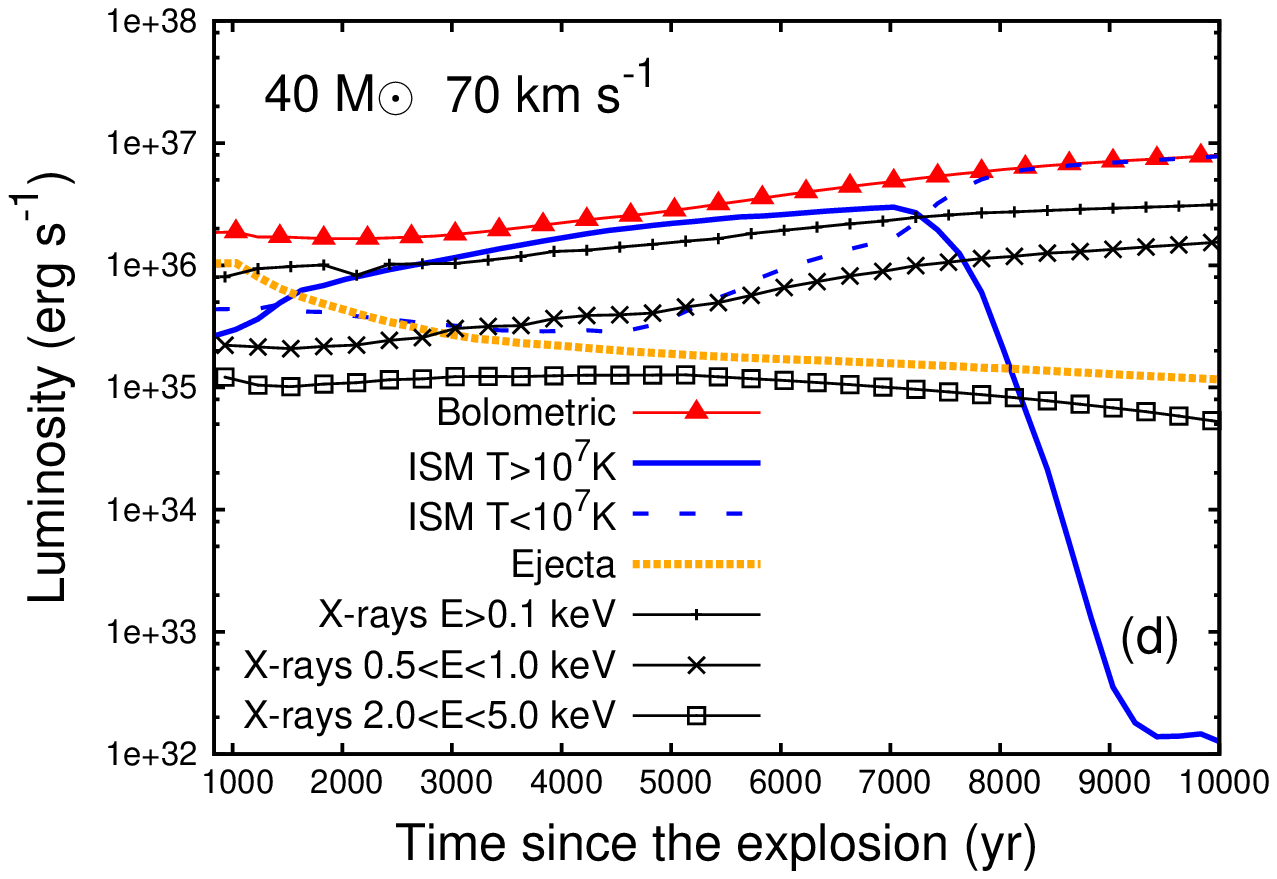}
	\end{minipage} 
	\caption{ 
Lightcurves of our old apherical supernova remnants. We show the luminosities
(in $\rm erg\, \rm s^{-1}$) during the time interval [$t_{\rm ysnr}$,$t_{\rm
osnr}$] (in $\rm yr$) of our aspherical supernova remnants generated by our
initially $20\, \rm M_{\odot}$ progenitor moving with velocity $20\, \rm km\,
\rm s^{-1}$ (a), our initially $40\, \rm M_{\odot}$ progenitor moving with
velocity $20$ (b), $40$ (c) and $70\, \rm km\, \rm s^{-1}$ (d), respectively.
The figures distinguish between the bolometric (total) luminosity from the gas
(red thin line with triangles), the contribution from the hot ISM gas
($T>10^{7}\, \rm K$, blue solid line), from the warm ISM gas ($T\le10^{7}\,
\rm K$, blue thin dashed lines) and the contribution from the ejecta (orange dotted
line). The black lines are the X-ray luminosity calculated at photons energies
$E>0.1\, \rm keV$ (\textcolor{black}{$T>1.2 \times 10^{6}\, \rm K$}, dashed line), in the soft energy 
band $0.5$-$1.0\, \rm keV$ (\textcolor{black}{$5.8 \times 10^{6}<T<1.2 \times 10^{7}\, \rm K$}, crossed line) 
and hard energy band $2.0$-$5.0\, \rm keV$ (\textcolor{black}{$2.3 \times 10^{7}<T<5.8 \times 10^{7}\, \rm K$}, 
squared line), respectively. }
	\label{fig:curves}  
\end{figure*}

We plot the luminosities (in $\rm erg\, \rm s^{-1}$) of the aspherical supernova
remmants as a function of time (in $\rm yr$) in Fig.~\ref{fig:curves}. The
bolometric luminosity by optically-thin radiation, $\it L$ (red thin line with
triangles), is estimated with the used cooling curve, integrating the energy
emitted per unit time and per unit volume over the whole remnant. Similarly, we
plot the contribution $L_{\rm ISM}$ from the hot ISM gas ($T>10^{7}\, \rm K$,
blue solid line) \textcolor{black}{where radiative losses are mostly due to Bremsstrahlung}
and from the warm ISM gas ($T\le10^{7}\, \rm K$, blue thin dashed lines) 
\textcolor{black}{where the emission are principally caused by cooling 
from Helium and metals together with O oxygen forbidden lines emission (see Section 2.4 of Paper~I)}
and the contribution $L_{\rm ej}$ from the ejecta (orange dotted
line). \textcolor{black}{We discriminate the various contributions of the 
luminosity $L$ on the basis of the passive scalars $Q_{1}(\bmath{r})$ and 
$Q_{2}(\bmath{r})$ advected with the fluid. They allow us to time-dependently 
trace the proportions of ISM gas, wind and supernova ejecta 
in the remnants (see section 3.4.1 and Eq.~(18) of Paper~I).} 
The X-ray luminosity, $L_{\rm X}$, is calculated for gas temperatures
$10^{5} \le T \le 1.58\times10^{8}\, \rm K$, with emission coefficients
interpolated from tables generated with the {\sc xspec}
software~\citep{arnaud_aspc_101_1996} with solar metalicity and chemical
abundances from~\citet{asplund_araa_47_2009}, as in~\citet{mackey_sept_2014}.

The total luminosity $L$ is first dominated by the hot
($T\ge10^{7}\, \rm K$) \textcolor{black}{emission from the shocked gas} before becoming
dominated by emission from the warm ISM gas of temperature $T\le10^{7}\, \rm K$
at a time about $8000\, \rm yr$ (see thick solid blue line in
Fig.~\ref{fig:curves}a-d) because the post-shock temperature at the shock
wave decreases during the adiabatic expansion of the blastwave. The cooling 
from the metals is stronger at $T\approx10^{6}\, \rm K$ than at $T\ge 10^{7}\, \rm K$
(Fig.4a of Paper~I), so $L$ increases \textcolor{black}{at larger times}.

$L$ increases as a function of time from $t_{\rm ysnr}$ up to the end of our
simulation. The reflection of the shock wave towards the center of the explosion
produces a growing hot, dense region that is upstream from the center of the
explosion and augments the luminosity. \textcolor{black}{The emission is
influenced by the size of the bow shock at the pre-supernova phase which
decreases with $v_{\star}$ (Paper~I) and governs the reflection of
the shock wave towards the center of the explosion.} $L$ monotonically increases
by less than one order of magnitude over a timescale of about $10^{4}\, \rm yr$,
e.g. in our model OSNR4070 $L\approx 1.2\times 10^{36}\, \rm erg\, \rm s^{-1}$
at $t\approx 10^{3}\, \rm yr$ and $L\approx 9.5\times 10^{36}\, \rm erg\, \rm
s^{-1}$ at a time $10^{4}\, \rm yr$ (Fig.~\ref{fig:curves}d).


$L_{\rm ej}$ is smaller than $L_{\rm ISM}$ by \textcolor{black}{a factor of a few} 
at $t_{\rm ysnr}$, e.g. just after the end of the shock
wave-bow shock collision, $L\approx 8.0\times 10^{35}$ and $L_{\rm ej}\approx
3.0\times 10^{35}\, \rm erg\, \rm s^{-1}$ in model OSNR4070
(Fig.~\ref{fig:curves}d). It monotonically decreases with time as the shocked
ejecta \textcolor{black}{expands and its density decreases such that its emission
} finally becomes a negligible fraction of $L$ (Fig.~\ref{fig:curves}a-d). The
contribution from the wind material is negligible compared to the bolometric
luminosity. It is not shown in Fig.~\ref{fig:curves} since it does not
influence $L$ at all.

The total X-ray emission, $L_{\rm X}$, is calculated as the emission from
photons at energies $0.1$-$50\, \rm keV$. Not surprisingly, this is slightly
smaller than the component of $L_{\rm ISM}$ from the gas with $T\le 10^{7}\, \rm
K$ and follows the same trend as $L$ (Fig.~\ref{fig:curves}a-d). The soft X-ray
emission, i.e. from photons in the $0.5$-$1.0\, \rm keV$ energy band, is fainter
than $L_{\rm X}$ by less than an order of magnitude and has a similar behaviour
as a function of time except for our older and larger supernova remnant OSNR4040
(Fig.~\ref{fig:curves}b). The hard X-ray emission in the $2.0$-$5.0\, \rm keV$
energy band is fainter than $L_{\rm X}$ by about an order of magnitude at
$t_{\rm ysnr}$. It decreases as a function of
time~\citep{brighenti_mnras_270_1994} because the emission of very energetic
X-ray photons ceases as the remnant expands and cools. Consequently, our old
remnants are more likely to be observed in the soft energy band of X-ray
emission.

\subsection{Emission maps}
\label{subsect:maps}

\begin{figure*}
	\begin{minipage}[b]{ 0.37\textwidth}
\includegraphics[width=1.0\textwidth]{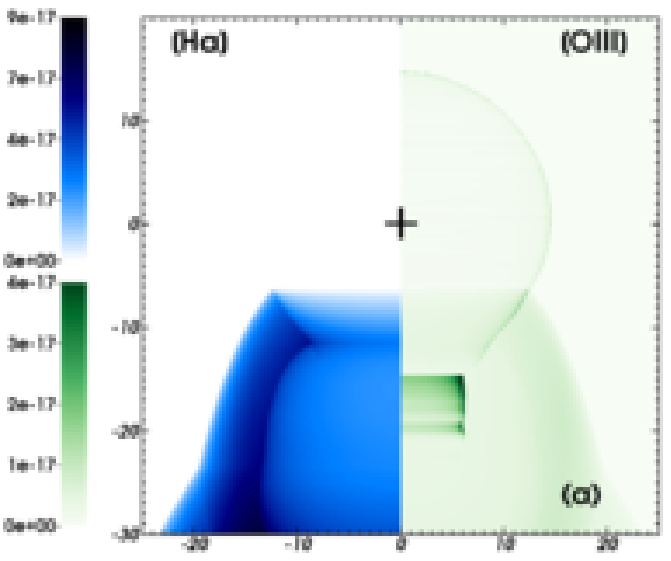}
	\end{minipage} 
	\begin{minipage}[b]{ 0.37\textwidth}
	\includegraphics[width=1.0\textwidth]{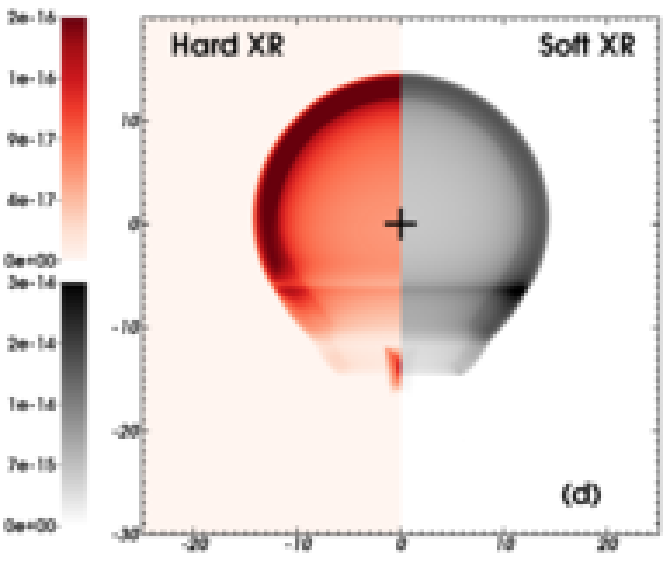}	
	\end{minipage} \\
	\begin{minipage}[b]{ 0.37\textwidth}
	\includegraphics[width=1.0\textwidth]{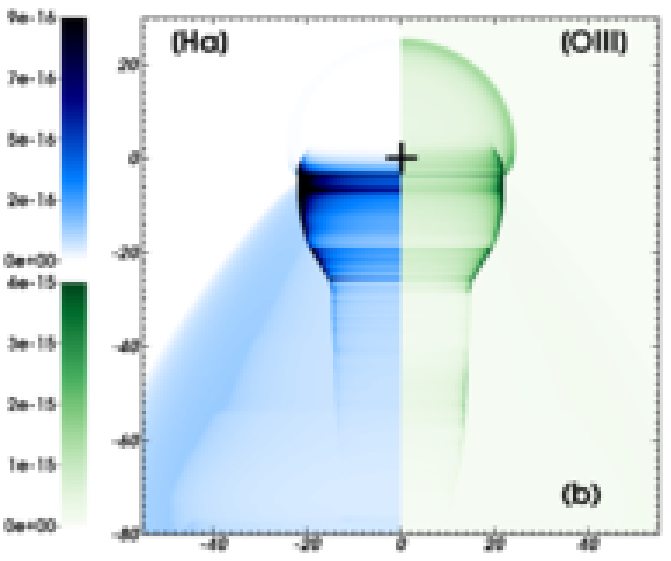}
	\end{minipage} 		
	\begin{minipage}[b]{ 0.37\textwidth}
	\includegraphics[width=1.0\textwidth]{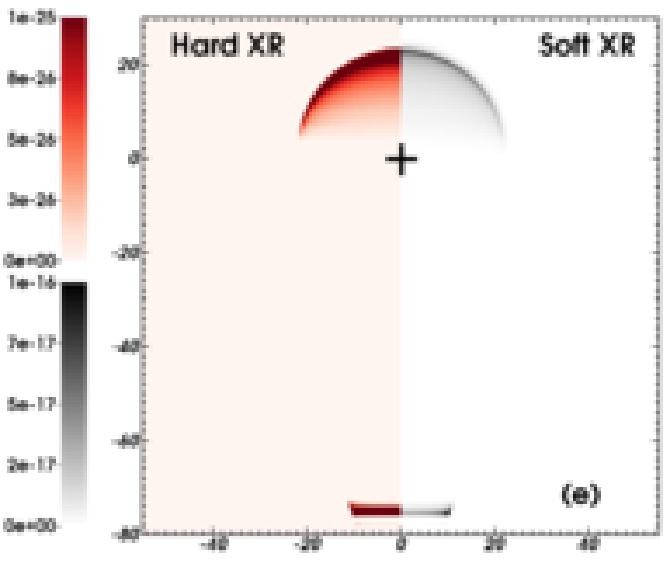}
	\end{minipage} 
	\begin{minipage}[b]{ 0.37\textwidth}
	\includegraphics[width=1.0\textwidth]{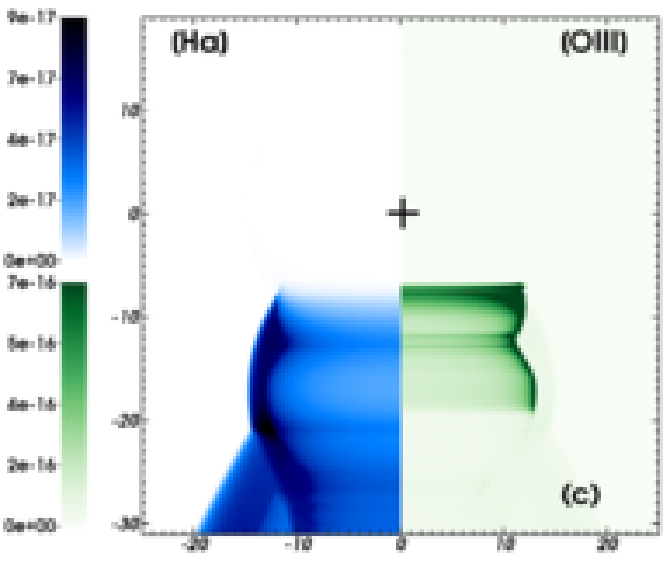}
	\end{minipage} 
	\begin{minipage}[b]{ 0.37\textwidth}
\includegraphics[width=1.0\textwidth]{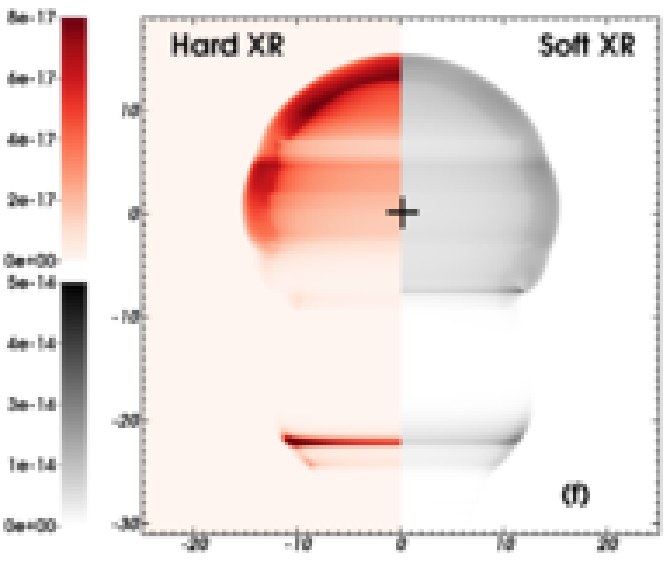}
	\end{minipage}	
	\caption{ 
                 Synthetic emission maps of our models OSNR2020 at a time
$12100\, \rm yr$ (a,d), OSNR4020 at time $49500\, \rm yr$ (b,e) and OSNR4040 at a
time $14000\, \rm yr$ (c,f), plotted on the linear scale in units of $\rm erg\,
\rm s^{-1}\, \rm cm^{-2}\, \rm arcsec^{-2}$. The left-hand part of the figures
show the H$\alpha$ surface brightness (blue) and the right-hand part the [O{\sc
iii}] surface brightness (green), respectively. The black crosses mark the
center of the explosion. The panels d-f
show the projected X-ray emission maps in the hard ($2.0$-$5.0\, \rm keV$~
\textcolor{black}{corresponding to $2.3 \times 10^{7}<T<5.8 \times 10^{7}\, \rm K$}, left-hand part) and soft 
($0.5$-$1.0\, \rm keV$~\textcolor{black}{corresponding to $5.8 \times 10^{6}<T<1.2 \times 10^{7}\, \rm K$}, right-hand part) X-ray bands 
for the same models. The $x$-axis represents the radial direction and the $y$-axis
the direction of stellar motion (in $\mathrm{pc}$). Only part of the
computational domain is shown in the figures. 
		}
	\label{fig:snr_maps_4020}  
\end{figure*}

\begin{figure*}
	\centering
	\begin{minipage}[b]{ 0.3\textwidth}
\includegraphics[width=1.0\textwidth]{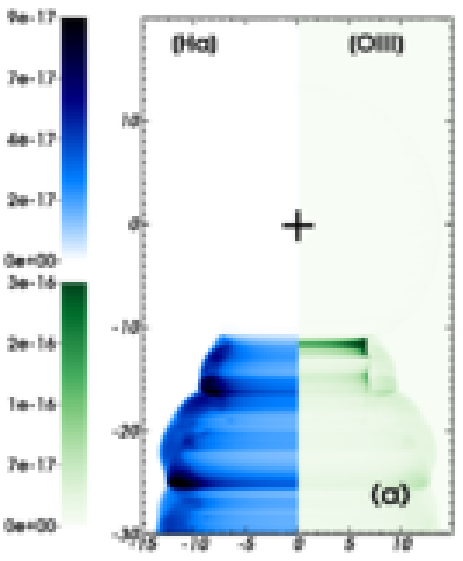}
	\end{minipage} 	
	\begin{minipage}[b]{ 0.3\textwidth}
\includegraphics[width=1.0\textwidth]{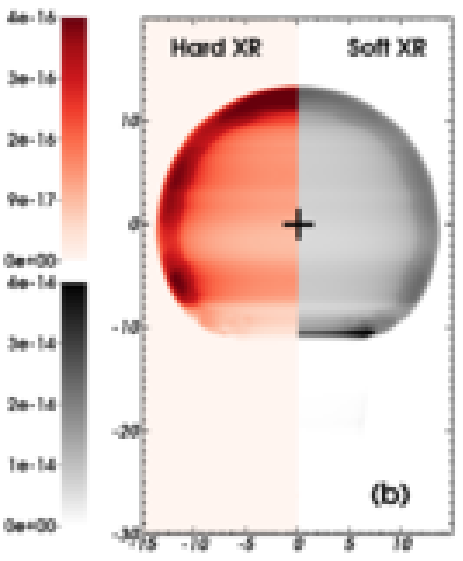}
	\end{minipage}	
	\caption{ 
                 As Fig.~\ref{fig:snr_maps_4020}, with an initially $40\, \rm M_{\odot}$ 
                 progenitor moving with space velocity $v_{\star}=70\, \rm km\, \rm s^{-1}$. 
                 The H$\alpha$ and \textcolor{black}{[O{\sc iii}] $\lambda \, 5007$} surface 
                 brightnesses are plotted in the left-hand panel, 
                 the projected hard and soft X-ray emissivity in the right-hand panel. 
		}
	\label{fig:snr_maps_4070}  
\end{figure*}

In Fig.~\ref{fig:snr_maps_4020} we show synthetic emission maps corresponding to
the time $t_{\rm osnr}$ of the supernova remnants generated by our $20\,
M_{\odot}$ progenitor moving with $v_{\star}=20\, \rm km\, \rm s^{-1}$ (\textcolor{black}{panels} a,d) and
by our $40\, \rm M_{\odot}$ progenitor moving with $v_{\star}=20$ (\textcolor{black}{panels} b,e) and $40\,
\rm km\, \rm s^{-1}$ (\textcolor{black}{panels} c,f), respectively. Fig.~\ref{fig:snr_maps_4070} is
similar for our $40\, \rm M_{\odot}$ progenitor moving with $v_{\star}=70\, \rm km\,
\rm s^{-1}$. The left-hand side of each panel plots the H$\alpha$ surface
brightness (blue) whereas the right-hand side shows the [O{\sc iii}] $\lambda \,
5007$ surface brightness (green). The left-hand side of each bottom panel plots
the hard X-ray surface brightness (red) and the right-hand side shows the soft
X-ray surface brightness (grey). The spectral line emission coefficients are
taken from the prescriptions by~\citet{osterbrock_1989}
and~\citet{dopita_aa_29_1973} for H$\alpha$ and \textcolor{black}{[O{\sc iii}]
$\lambda \, 5007$}, respectively, with solar oxygen
abundance~\citep{lodders_apj_591_2003} \textcolor{black}{and imposing a cut-off
temperature at $10^{6}\, \rm K$~\citep[cf.][]{cox_mnras_250_1991} when oxygen
becomes further ionized~\citep{sutherland_apjs_88_1993}}.

The region of maximum H$\alpha$ surface brightness is located downstream from
the center of the explosion. This happens because the H$\alpha$ emissivity, 
$j_{\rm H\alpha}\propto T^{-0.9}$, is very faint in the region of hot shocked
ejecta (Fig.~\ref{fig:snr_maps_4020}a-c). The simulation with the slowly moving
$40\, \rm M_{\odot}$ progenitor has an emissivity peak along the walls of the wind
cavity (Fig.~\ref{fig:snr_maps_4020}b) because effective cooling of the gas
makes the the post-shock region cool ($T \gtrsim \,10^{4}\, \rm K$) and dense (up
to $n \approx 50\, \rm cm^{-3}$). In the other simulations, the emission
originates from the outer layers of the bow shocks because the walls of their
less massive bow shocks allow a faster and deeper penetration of the shock wave into the
shocked wind material (Fig.~\ref{fig:snr_maps_4070}). Our predicted
maximum H$\alpha$ emission is above the diffuse emission sensitivity limit of
the SuperCOSMOS H-alpha Survey~\citep[SHS][]{parker_mnras_362_2005} of
$1.1$$-$$2.8 \times 10^{-17}\, \mathrm{erg}\, \mathrm{s}^{-1}\,
\mathrm{cm}^{-2}\, \mathrm{arcsec}^{-2}$, e.g. our model OSNR4040 has a maximum
emission larger than $9\times 10^{-16}\, \mathrm{erg}\, \mathrm{s}^{-1}\,
\mathrm{cm}^{-2}\, \mathrm{arcsec}^{-2}$ (Fig.~\ref{fig:snr_maps_4020}b), and
could be compared with data from these surveys.

\textcolor{black}{ 
The [O{\sc iii}] maximum surface brightness of our models
originates from the dense ($n \approx 2$$-$$10\, \rm cm^{-3}$) and warm
($T\le10^{6}\, \rm K$) post-shock region behind the shock wave. It is located at
the walls of the cavity and produces a ringed/tubular structure
(Fig.~\ref{fig:snr_maps_4020}a-c and~\ref{fig:snr_maps_4070}a).} It is generally
not coincident with the projected H$\alpha$ emission because the [O{\sc iii}]
emissivity has a different dependence on the temperature, i.e. $j_{\rm{ [O{{\sc
III}}] }}\,\propto\,e^{-1/T}/T^{1/2}$. However, the simulation OSNR4020 with a
heavy bow shock have its [O{\sc iii}] maximum surface brightness all along the
walls of the wind tunnel and it is coincident with the region of maximum
H$\alpha$ emission, i.e. the behaviour of the emissivities with respect to the
large compression of the gas in the walls ($\propto n^{2}$) overwhelms that of
the gas temperature (Fig.~\ref{fig:snr_maps_4020}b).

The X-ray emission originates from the hot gas with $10^{5} \le T \le \,10^{8}
\rm K$, i.e. from near the shock wave expanding into the unperturbed ISM. The
maximum surface brightness comes from the hot region of doubly shocked gas
upstream from the center of the explosion, and from the post-shock region at the
channelled shock wave
(Figs.~\ref{fig:snr_maps_4020}d-f,~\ref{fig:snr_maps_4070}b). The hard X-ray
surface brightness is several orders of magnitude fainter than the projected
soft X-ray emission, because the gas is not hot enough at $t_{\rm osnr}$
(Fig.~\ref{fig:curves}). The emitting region in soft X-rays is peaked in the
post-shock region at the shock wave whereas the hard X-ray come from a broader
region of shocked gas which outer border is the shock wave
(Fig.~\ref{fig:snr_maps_4020}f). Note also the anti-correlation between the
surface brightness in X-ray (hot region from near the forward shock) and in
[O{\sc iii}] (colder and denser walls) in our remnants generated by a
fast-moving ($v_{\star} \ge 40\, \rm km\, \rm s^{-1}$) progenitor
(Fig.~\ref{fig:snr_maps_4020}b,e;c,f).


\section{Discussion}
\label{sect:discussion}

We here discuss our results in the light of precedent studies and pronounce
on the best manner to observe aspherical supernova remnants generated by massive 
Galactic runaway stars. Finally, we examine our models in the context of recent 
observations.

\subsection{Comparison with previous works}
\label{subsect:previous_works}

We tested that the code {\sc
pluto}~\citep{mignone_apj_170_2007,migmone_apjs_198_2012} reproduced the
one-dimensional models of core-collapse supernovae interacting with their
surroundings in~\citet{whalen_apj_682_2008} and~\citet{vanveelen_aa_50._2009}
using a uniform, spherically symmetric grid. Our numerical method (Paper~I) is
different from that in~\citet{whalen_apj_682_2008} because (i) they utilise a
finite-difference scheme coupled to a network of chemical reactions following
the non-equilibrium rates of the species composing the gas and (ii) their
algorithm includes artificial viscosity~\citep[{\sc zeus} code, see][]{stone_apjs_80_1992}.

We ran tests with their cooling curve~\citep{macdomald_mnras_197_1981} and with
a cooling curve for collisional ionization equilibrium medium (see details in
section 2.4 of Paper~I). We find no notable differences, mostly because they are
similar in the high temperature regime ($T \ge 10^{5}\, \rm K$) that is relevant
for the supernova-wind interaction (Fig.~\ref{fig:sncsm}). We extend this method
to two-dimensional, cylindrically symmetric tests of a supernova shock wave
expanding into a homogeneous ISM to ensure that the sphericity of the shock wave
is conserved throughout its expansion. We notice that the solution behaves well
with respect to the symmetry axis $Oz$.

Models of an off-centered explosion in a wind-driven cavity are available
in~\citet{rozyczka_mnras_261_1993}. Their model produces a parsec-scale
jet-like feature as do our aspherical models 
(Figs.~\ref{fig:snr_large_scale_2020}c to~\ref{fig:snr_large_scale_4070}c) but they
neglect the progenitor's stellar evolution, assume a different microphysics, 
and correspond to a totally different point of the parameter space ($n =
10^{6}\, \rm cm^{-3}$). Our description of the supernova shock wave interacting
with its pre-shaped bow shock is consistent with the works tailored to the
Kepler's supernova remnant; see section 5.2 of~\citet{borkowski_apj_400_1992}
but also~\citet{velazquez_apj_649_2006,toledo_mnras_442_2014}.

The tunnel of unshocked wind that channels the shock wave, e.g.~in our model
OSNR4020 (Fig.~\ref{fig:snr_large_scale_2020}a), is morphologically and
structurally consistent with the model of the red supergiant progenitor of the Crab
nebula in~\citet{cox_mnras_250_1991}, assuming a diluted ambient medium and a
larger space velocity ($n = 0.25\, \rm cm^{-3}$, $v_{\star}=69.5\, \rm km\, \rm
s^{-1}$). Notice that the simulations with fast-moving stars or
post-main-sequence high-mass-loss events have a tunnel with clumpy walls
(Fig.~\ref{fig:snr_large_scale_2020}a and~\ref{fig:snr_large_scale_4070}a) which
do not prevent the tunneling of the shock, in contrast to the suggestion
by~\citet{cox_mnras_250_1991}.

The growth and overall shape of our aspherical remnants are in accordance
with~\citet{brighenti_mnras_270_1994}. Our model OSNR2020 is morphologically
consistent with their model 1 ($v_{\star}=17\, \rm km\, \rm s^{-1}$, $n=1\, \rm
cm^{-3}$). They assumed a comparable mass loss ($\dot{M} = 10^{-5}\, \rm M_{\odot}\,
\rm yr^{-1}$) but a twice larger wind velocity ($v_{\rm w}=20\, \rm km\, \rm
s^{-1}$) during the red supergiant phase which lastes about $7 \times 10^{5}\,
\rm yr$. Identical remarks arise comparing their model 3 and our simulation
OSNR4070 (Fig.~\ref{fig:snr_large_scale_4070}a-c). Because our models include 
thermal conduction, the region of shocked wind in the bow shock from the 
main-sequence phase is larger and the tunnel in our simulation OSNR4020
narrower than in their model 4 and allows a more efficient channeling of the shock 
wave (Fig.~\ref{fig:snr_large_scale_4020}a-c). \textcolor{black}{More details about 
the effects of heat conduction in our remnants is given in Appendix~\ref{section:tc}. }

As a conclusion, for overlapping parameters our results agree well with previous
models of supernova remnants produced by runaway progenitors. We extend the
parameter space with a representative sample of models tailored to the Galactic
plane, whose progenitor's wind properties are taken from self-consistently
pre-calculated stellar evolution models~\citep{brott_aa_530_2011a}.


\subsection{Comparison with observations}
\label{subsect:observations}

\subsubsection{\textcolor{black}{General remarks and comparison approach}}
\label{subsubsect:gal}

Comparing our simulations with observations is not a straightforward task. 
\textcolor{black}{Even though} this paper explores a representative sample of Galactic,
massive, runaway stars, our remnants can only develop asymmetries when the
isotropic shock wave interacts in reality with a dense bow shock, whereas other
mechanisms can also induce asymmetries. They can originate from an intrinsically
anisotropic explosion~\citep{blondin_apj_472_1996}, the rotation of the
progenitor~\textcolor{black}{\citep{langer_ApJ_520_1999}}, the magnetization of the ISM~\citep{rozyczka_274_MNRAS_1995} or the
presence of a neighbouring circumstellar structure~\citep{ferreira_478_aa_2008}.
Consequently, we hereby simply attempt to establish a qualitative \textcolor{black}{discussion}
between \textcolor{black}{the density fields in} our models and observations available in the literature.

\textcolor{black}{Supernova remnants are often observed in radio
wavelengths~\citep{green_cat_2009}}, e.g. at about $325$, $843$ or $1400\, \rm
Mhz$. \textcolor{black}{This emission arises from bremsstrahlung (proportional to
$n_e^2$, where $n_e$ is electron number density ), synchrotron radiation
(proportional to $n_e B$, where $B$ is the field strength), and maser emission
(tracing very dense ISM regions)}. Our discussion is limited to a simple
comparison between such published measures and the density fields in our
hydrodynamical simulations. Additionally, because Galactic SNRs usually have
poorly \textcolor{black}{constrained distances (and hence sizes and masses), we
discuss morphological similarities even though it may not be clear whether or
not our model parameters are appropriate to a given observed remnant}.

\subsubsection{\textcolor{black}{Observations from maser emission such as in 3C391: explosions 
in a dense medium, runaway progenitors or both ?}}
\label{subsubsect:maser}

~\citet{brighenti_mnras_270_1994} and~\citet{eldridge_mnras_414_2011} already
justified the relevance of studying runaway O stars in the understanding of
supernova remnants and gamma-rays bursts. Particularly, they underline the
difficulty of interpreting the shape of incomplete and/or inhomogeneous arc-like
supernova remnants because the overdensities upstream from the center of the
explosion can also arise from pre-existing dense
regions~\citep[see, e.g.][]{orlando_apj_678_2008}. The presence of OH maser 
emission originating from the shock front can discriminate between
\textcolor{black}{these} scenario\textcolor{black}{s}~\citep{frail_aj_111_1996,yusefzadeh_apj_585_2003}, and led to the
classification of about $20$ Galactic arced remnants such as G31.9+0.0 or
G189.1+3.0 as running into a dense cloud.

Examining the necessary conditions to produce OH $1720\, \rm \textcolor{black}{Mhz}$ 
($T\approx50$-$125\, \rm K$, $n_{\rm medium}\approx10^{5}\, \rm cm^{-3}$, 
OH column density $N_{\rm OH}\approx10^{16}\, \rm cm^{-2}$,
see~\citealt{lockett_apj_511_1999}) one finds that such high densities are not
reached in our models. However, according to the Rankine-Hugoniot jump
relations it may be the case in the post-shock region if the shock propagates 
in a medium of pre-shock density $n_{\rm medium}/4\approx 2.5 \times 10^{4}\, \rm cm^{-3}$ 
such as a starless dense core, a molecular cloud or a contracting ISM
filament~\citep{kaufman_EAS2009}. Assuming a temperature of $T\approx50\, \rm
K$, the medium sound speed is $c_{\rm s}\approx 0.65\, \rm km\, \rm s^{-1}$ so
any star moving with velocity $v_{\star}\approx5c_{\rm s}\approx 3.25\, \rm km\,
\rm s^{-1}$ would have an hypersonic motion and produce a bow shock.
Consequently, if OH maser emission is not likely to be detected from bow shocks
in the field, it naturally arises from the surroundings of runaway stars in a
dense medium. Moreover, if the star explodes in such an environment,
additional constraints may be necessary in order to distinguish between maser
emission produced because of the runaway nature of the progenitor, emission
originating from an expanding shock wave, or both.

Particularly, 3C391 was originally believed to be the archetypal remnant
from a moving progenitor~\citep{brighenti_mnras_270_1994} but is now known to be associated with
OH maser emission~\citep{frail_aj_111_1996} together with molecular line
emission. The Infrared Spatial Observatory (ISO) revealed, among
other, CO, HC$\rm O^{+}$ line emission~\citep[see][and references
therein]{reach_apj_511_1999} whereas~\citet{neufeld_apj_781_2014} reported
recent infrared analysis of the $\rm H_{2}$ and $\rm H_{2}$O line emission.
However, as mentioned above, nothing indicates whether this emission
originates from the shock wave colliding with a wind bubble produced by the
progenitor or from the shock wave running into its dense surrounding medium.
Modelling the explosion of a supernova from a runaway progenitor  
with $n_{\rm medium} \gg 1\, \rm cm^{-3}$ is conceivable but beyond the scope 
of the present investigation and left for a follow-up study. 
Considering the past evolution of supernova progenitors in the modelling of
their remnants will help to discriminate between these two situations. Further
investigations, e.g. tailored to 3C391, may provide more severe constraints on
its circumstellar medium at the pre-supernova phase and \textcolor{black}{help to} 
understand its formation scenario.

\subsubsection{Bilateral supernova remnants}
\label{subsubsect:bilateral}

\textcolor{black}{Our simulations PSN4020 and PSN4040 produce approximately cylindrical wind-blown
cavities \textcolor{black}{with} dense walls. The interaction of the supernova with this
circumstellar medium, in our models OSNR4020 and OSNR4040
(Figs.~\ref{fig:snr_large_scale_4020} and~\ref{fig:snr_large_scale_4040})
produces the conditions required for a bilateral supernova remnant, i.e. a
cylindrical cavity with dense, shocked gas on the boundaries.} 
\textcolor{black}{A} previous study identified the material composing 
the bilateral structures of the remnant G296.5+10.0 as shocked pre-supernova 
wind~\citep{manchester_aa_171_1987} and
the presence of a neutron star in between the opposed arcs is reported
in~\citet{zavlin_aa_331_1998}. It constrains the stellar remnant to be a $1.4\,
\rm M_{\odot}$ neutron star, so its progenitor had initial mass below about
$25\, \rm M_{\odot}$~\citep{woosley_rvmp_74_2002}. It is similar to our slowly moving
model of an initially $40\, \rm M_{\odot}$ at time $11400\, \rm yr$ after the
explosion, when the shock wave hits the very dense walls of the wind tunnel 
(Fig.~\ref{fig:snr_large_scale_4020}b). Our understanding of 
the bilateral character of this remnant as a result of \textcolor{black}{its} 
progenitor's supersonic motion only is valid if one considers a lower
initial mass progenitor moving in a medium dense enough to form heavy cavity walls 
resistant to the shock wave. Note that this result is consistent with fig.~6D 
of~\citet{orlando_aa_470_2007}.

However, alternative explanations have been proposed for the fomation of
bilateral remnants like G327.6+14.6 or G3.8-0.3. A strong axisymmetric
background ISM magnetic field \textcolor{black}{(not included in our simulations)} 
has also been suggested to be responsible for the
bilateral character of some Galactic supernova
remnants~\citep{gaensler_apj_493_1998}. It would indeed make their shape more
elongated as long \textcolor{black}{as the field} is strong enough~\citep{rozyczka_274_MNRAS_1995} 
and produce X-ray and/or radio synchrotron emission from the
opposed arcs~\citep{velazquez_apj_601_2004,petruk_393_mnras_2009, 
schneiter_mnras_408_2010}.

\subsubsection{The jet\textcolor{black}{/tubular}-like extension shaped by the motion of the progenitor}
\label{subsubsect:jet}

Our Galactic, slowly moving, initially $20\, \rm M_{\odot}$ progenitor produces
\textcolor{black}{a supernova remnant} whose outer region strongly emits in [O\,{\sc iii}]. The
remnant generated by our slowly moving, initially $40\, \rm M_{\odot}$
progenitor has an [O\,{\sc iii}] jet-like feature that has a H$\alpha$
counterpart (Fig.~\ref{fig:snr_maps_4020}) generated by ejecta channelled with
velocity about $1000\, \rm km\, \rm s^{-1}$ into the wind tunnel of the bow
shock (Fig.~\ref{fig:expansion}b). Those tubular/jet-like features
(Figs.~\ref{fig:snr_maps_4020}-\ref{fig:snr_maps_4070}) are \textcolor{black}{reminiscent of} the
chimney discovered in [O{\sc iii}] in the Crab
nebula~\citep{blandford_301_natur_1983} modeled in~\citet{cox_mnras_250_1991}. 
This morphological resemblance mostly arises because of the
similar stellar evolution history and space velocity $v_{\star}\approx30\, \rm
km\, \rm s^{-1}$ in both simulations.

Nevertheless, further investigations and detailed comparison would have to take
into account the youth of the Crab nebula ($\approx 1000\, \rm yr$) and its
location in the high latitude, low-density ISM of the Galaxy. In our
simulations with larger $v_{\star}$, those jet-like extensions become an
[O\,{\sc iii}] tubular structure that is thinner and closer to the throttling
separating the surroundings of the center of the explosion from the trail of the
bow shock. Supernovae \textcolor{black}{exploding} in a wind cavity could hence form tunnels or
barrel-like shapes, however, alternative quite convincing
demonstration, e.g. based on asymmetric explosions have
\textcolor{black}{already} been
proposed~\citep[WB49,][]{gonzalezcasanova_apj_781_2014}.

\subsubsection{An alternative  explanation for the Cygnus loop nebula ?}
\label{subsubsect:cygloop}

G074.0-8.5, also called the Cygnus loop, is a supernova \textcolor{black}{remnant for} which X-ray
observations, e.g. with ROSAT~\citep{aschenbach_aa_341_1999} reveal a
characteristic overall drop-like morphology. An early interpretation of this
shape is a champagne blowout from the edge of a molecular
cloud~\citep{tenoriotagle_aa_148_1985}. Recent X-ray data 
favours a model with an explosion into a pre-shaped cavity and \textcolor{black}{whose ejecta have}
recently impacted the imperfect walls. ~\citet{uchida_pasj_61_2009} supports
this model and derives the remnant's age to be about $10000\, \rm yr$, its
radius to be $12$-$17\, \rm pc$ and its initial progenitor's mass to be
$12$-$15\, \rm M_{\odot}$, which is consistent with our model OSNR2020
(Fig.~\ref{fig:snr_maps_4020}d).

Moreover, some southern optical filaments have a measured expansion velocity of
a few hundred $\rm km\, \rm s^{-1}$~\citep{medina_791_apj_2014} as in our
simulation OSNR2020 (Fig.~\ref{fig:expansion}a). In our picture, the south
blowout region of the remnant is the low-density wake behind the progenitor in
which the gas is the hottest ($> 10^{6}\, \rm K$), as noted
in~\citet{aschenbach_aa_341_1999}. Nevertheless, our models predict that the
soft and hard X-rays emission  \textcolor{black}{should be limb-brightened, whereas the
observations have a filled morphology}. This may be explained by the presence of
a neutron star~\citep{katsuda_apj_754_2012} that is not taken into account in
our models. \textcolor{black}{Future simulations, investigating the plerionic
nature of the Cygnus loop nebula will \textcolor{black}{allow us to assess this interpretation}.}

\subsubsection{\textcolor{black}{Distinguishing remnants from moving progenitors and remnants with ISM-induced anisotropy.}}
\label{subsubsect:discussion}

\textcolor{black}{  
Distinguishing whether a one-sided supernova remnant is produced thanks to the motion
of its progenitor or because of its surroundings' inhomogeneity is difficult
since both situations can happen together.  As discussed above, the OH emission
can trace the presence of a molecular cloud, but is not sufficient to exclude 
the presence of an hypothetical bow shock. A couple of additional comments can
be given. 
}

\begin{enumerate}

\item \textcolor{black}{  
In the situation of a runaway supernova progenitor, the cavity that channels the subsequent 
shock wave left behind its bow shock may produce a more collimated remnant than in the 
case of an interaction with a dense cloud. This phenomenon could be enhanced by the 
presence of a background ISM magnetic field which elongates the pre-shaped tunnel, reducing the 
instabilities affecting its walls and consequently favoring the channeling. 
}

\item \textcolor{black}{  
If a runaway massive star sheds (an) heavy envelope(s) throughout its evolution, e.g. via a 
luminous blue variable event or a Wolf-Rayet phase, the passage of the shock wave through 
these expelled shell(s) will fragment them. Their temperature will increase by shock heating before 
cooling down, inducing an X-ray rebrightning of the formed foculli. Such a mixed-morphology supernova 
remnant generated by a blown bow shock would have an incomplete arc-like radio envelope around a 
central region containing X-ray-emitting clumps. 
}

\item \textcolor{black}{ 
The situation becomes more complicated when the explosion happens at the 
interface between two different phases of the ISM. If a runaway progenitor leaves a molecular cloud, 
the tubular extension produced by the channeled shock wave will be directed towards 
the center of the cloud. If the progenitor enters the cloud, the imprint of the last mass-loss 
events of the moving star onto the border of the cloud will produce a characteristic hole. 
A famous (nonetheless extragalactic) example of such a phenomenon is the structure that 
overhangs the remnant of SN1987a in the Large Magellanic Cloud~\citep{wang_MNRAS_261_1993}. 
}

\end{enumerate}


\section{Conclusion}
\label{section:cc}

In this paper, we present a grid of hydrodynamical models of asymmetric
supernova remnants generated by a representative sample of Galactic runaway
massive stars whose circumstellar medium \textcolor{black}{during} the main-sequence and red
supergiant phases is studied in~\citet{meyer}. We compute the bow shocks
generated by our progenitors from near the pre-supernova phase and model the
collision between the supernova shock waves and the circumstellar medium which result in
the generation of supernova remnants. The progenitors' initial masses
range from $10$ to $40\, \rm M_{\odot}$ and they move with space velocities
ranging from $20$ to $70\, \rm km\, \rm s^{-1}$. Our models include both
optically-thin cooling and photoheating of the gas. Electronic thermal
conduction is included in the calculations of the circumstellar medium and in
the simulations of the supernova remnants.

We stress that the stellar motion of a core-collapse supernova progenitor 
can be responsible for the deviations from sphericity of its subsequent remnant.
We show that the bow shocks trapping at least $1.5\, \rm M_{\odot}$ of ISM gas
are likely to generate aspherical supernova remnants. They correspond to high
mass and/or slowly moving stars~\textcolor{black}{\citep{brighenti_mnras_270_1994,meyer}}. At the ISM
number density that we consider, they are produced either by our initially $20\,
M_{\odot}$ star moving with space velocity of about $20\, \rm km\, \rm s^{-1} $
or by our initially $40\, \rm M_{\odot}$ runaway star. These mass-accumulating bow
shocks generate a dense bulge of shocked ISM gas upstream from the direction of
motion of the star whereas a cavity of low-density wind material forms in the
opposite direction and extends as a tunnel of unshocked wind material into the
trail of the bow shock.

After the supernova explosion, the shock wave expansion is strongly influenced
by the anisotropy of its circumstellar medium. It collides with the overdense
part of the bow shock whereas it expands \textcolor{black}{freely} at velocities of the order of $1000\,
\rm km\, \rm s^{-1}$ in the opposite direction, channelled by the pre-shaped
tunnel of unshocked wind material~\citep[cf. observations of RCW86
in][]{vink_aa_328_1997}. The mass of shocked ISM trapped in the bow shock
decelerates the shock wave, which continues to penetrate the
unperturbed ISM after the collision with velocity of the order of $100\, \rm
km\, \rm s^{-1}$.

As the shock wave evolves in a non-uniform medium, it is partly reflected towards
the center of the explosion~\citep{ferreira_478_aa_2008} after the collision
with a dense bow shock. It induces mixing of supernova ejecta, stellar wind and
ISM gas that is particularly important for fast-moving progenitors. This \textcolor{black}{reflected} wave
shocks the zone \textcolor{black}{around} the center of the explosion, \textcolor{black}{reheating the gas, which subsequently} 
cools below $10^{7}\, \rm K$ because of the adiabatic 
expansion of the blastwave. Its
luminosity increases, dominated by thermal Bremsstrahlung and soft X-ray
emission originating from the shocked ISM that is upstream from the center of
the explosion. The emission from the ejecta or from the progenitor's wind
material does not contribute significantly to the remnants' total luminosity
once the bow shock is overtaken by the \textcolor{black}{supernova} shock wave.

Our Galactic aspherical supernova remnants have an [O{\sc iii}]
$\lambda \, 5007$ surface brightness larger than their projected H$\alpha$
emission, i.e. the [O{\sc iii}] is more appropriate line to search for
Galactic supernova remnants. Their [O{\sc iii}] surface brightness is maximum in
the post-shock region of the shock wave. It is concentrated along the walls of
the tunnel of wind material. The region of maximum H$\alpha$ emission is
downstream from the direction of motion of the progenitor. It originates from
the outer part of shocked ISM material in the trail of the progenitor's bow
shock. In the case of our slowly moving initially $40\, \rm M_{\odot}$ progenitor,
it mainly comes from the region where the shock wave interact\textcolor{black}{s} with the walls
of the tunnel, i.e. the ejecta forms an [O\,{\sc iii}] \textcolor{black}{tubular}-like feature that has
an H$\alpha$ counterpart. Moreover, we find that our remnants are 
\textcolor{black}{brighter} in soft X-ray emission originating from near the shock wave than in
hard X-ray emission coming from the post-shock region at the shock wave.

Supernova remnants generated by runaway progenitors naturally \textcolor{black}{show} 
structures highly deviating from sphericity. Particularly, our models
of  remnants generated by high-mass, slowly moving progenitors have morphologies
consistent, e.g. with the bilateral character of observed barrel-like Galactic
supernova remnants such as G296.5+10.0 or with the morphology of the Cygnus loop 
nebula. However, other mechanisms are at work in the shaping of
supernova remnants. Forthcoming work will investigate the effects of an ISM
magnetic field on the evolution of our remnants, in order to quantitatively
appreciate its consequences on the remnants\textcolor{black}{'} dynamics and 
emission signatures.


\section*{Acknowledgements}

\textcolor{black}{We thank the anonymous referee for useful comments regarding 
to the effects of thermal conduction and concerning the discussion of our results.}
\textcolor{black}{We} thank Richard Stancliffe who carefully read the manuscript. \textcolor{black}{We} also thank
Thomas Tauris for useful advice on neutron stars, as well as Rolf G\"{u}sten,
John Eldridge, Dave Green, Kazik Borkowski, Daniel Whalen and Allard Jan van
Marle for their help. This work was supported by the Deutsche
Forschungsgemeinschaft priority program 1573, "Physics of the Interstellar
Medium". PFV acknowledges finantial support from CONACyT grant 167611. 
\textcolor{black}{The authors gratefully acknowledge the computing time granted by the John von
Neumann Institute for Computing (NIC) and provided on the supercomputer JUROPA
at J\" ulich Supercomputing Centre (JSC).}


\bibliographystyle{mn2e}

\footnotesize{
\bibliography{grid_snr}
}


\appendix

\section{The effects of thermal conduction}
\label{section:tc}

In this Appendix we discuss the effects of thermal conduction on the shape and internal 
structures of our circumstellar nebulae, before and after the supernova explosion.

\subsection{The effects of thermal conduction on the pre-shaped cirumstellar medium}
\label{section:subtc1}

In the situation of our slowly-moving stars ($v_{\star}=20\, \rm km\, \rm
s^{-1}$), we have an off-centered explosion {\it inside} the wind-bubble
generated during the main-sequence phase of the moving
star~\citep{brighenti_mnras_270_1994}. The effects of thermal conduction mainly
arise from the structural differences it induces in the bow shock, which consist
in the circumstellar medium of the star at the pre-supernova phase (see our model PSN4020 in
Fig.~\ref{fig:snr_large_scale_2020}a). The high effective temperatures of our
main-sequence, massive stars produce a heat
flux which timescale $\propto T^{-7/2}$~\citep[][Paper~I]{orlando_aa_444_2005} is
much faster than both the dynamical advection timescale of the stellar wind
and ISM gas in the bow shock and cooling by optically-thin
radiative processes timescale. It transports large amount of the gas internal energy
from the reverse shock of the bow shock towards the contact discontinuity that
slightly enlarges the bow shock in the direction of motion of the star. 
Moreover, it damps the instabilities developing at the reverse shock, i.e. the
walls of the tunnel, and reorganises its internal structure (see discussion in
section 3.3 of Paper~I).

In the case of our fast-moving stars ($v_{\star}\ge40\, \rm km\, \rm s^{-1}$), we have an off-centered 
explosion {\it outside} the wind-bubble generated during the main-sequence phase of the moving 
star and the supernova explosion consequently takes place in a pre-shaped circumstellar medium where any information 
relative to the main-sequence wind bubble is located behind the star~\citep{brighenti_mnras_270_1994}. 
The effects of thermal conduction mainly arise from the instabilities that develop in the bow shock 
produced during the post-main sequence phase of the stellar evolution. 
Since the seeds and the growth of the Kelvin-Helmholtz and Rayleigh-Taylor instabilities 
partly depend on density distribution at the end of the main-sequence phase, the effects of 
thermal conduction consist in changing the shape of the instabilities at the apex of these bow shocks 
(see our model PSN4070 in Fig.~\ref{fig:snr_large_scale_2020}b). 
Note that only the morphology of the eddies at a given time differ, the developing instability 
remains of the same kind, i.e. a thin-shell-related instability~\citep{vishniac_apj_428_1994,blondin_na_57_1998}.

\begin{figure}
	\centering
	\begin{minipage}[b]{ 0.42\textwidth}
\includegraphics[width=1.0\textwidth]{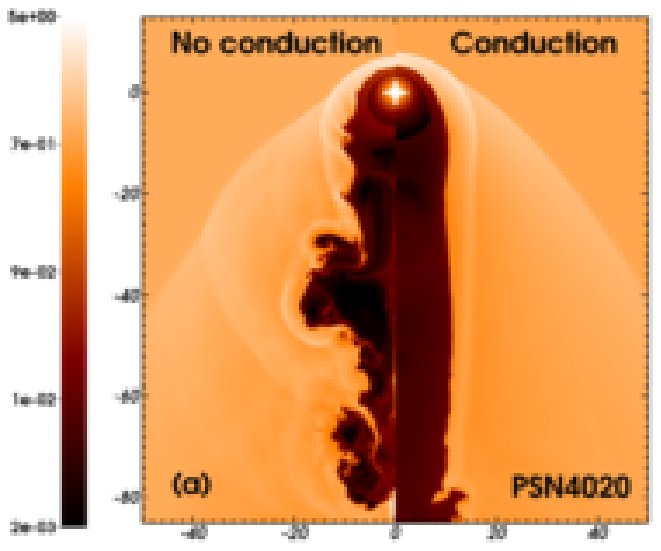}
	\end{minipage} 
	\centering
	\begin{minipage}[b]{ 0.42\textwidth}
\includegraphics[width=1.0\textwidth]{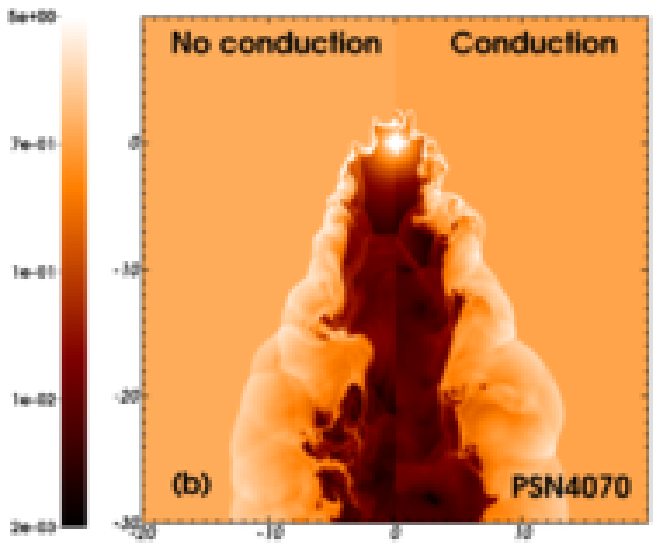}
	\end{minipage} 	
	\caption{ 
\textcolor{black}{ Effects of heat conduction on the pre-supernova circumstellar medium of our 
initially $40\, \rm M_{\odot}$ progenitor moving with velocity $20\, \rm km\, \rm s^{-1}$ (a)
and $70\, \rm km\, \rm s^{-1}$ (b). The figure shows the gas number density field of simulations 
carried out without (left-hand part of the panels) and with heat conduction 
(right-hand part of the panels) with a density range from about 
$10^{-3}$ to $5.0\, \mathrm{cm}^{-3}$ on the logarithmic scale, at a time $t_{\rm osnr}$.
The white cross marks the position of the runaway star. }
		}
	\label{fig:snr_large_scale_2020}  
\end{figure}

\subsection{The effects of thermal conduction on the old supernova remnants}
\label{section:subtc2}

\begin{figure}
	\centering
	\begin{minipage}[b]{ 0.42\textwidth}
	\centering
\includegraphics[width=1.0\textwidth]{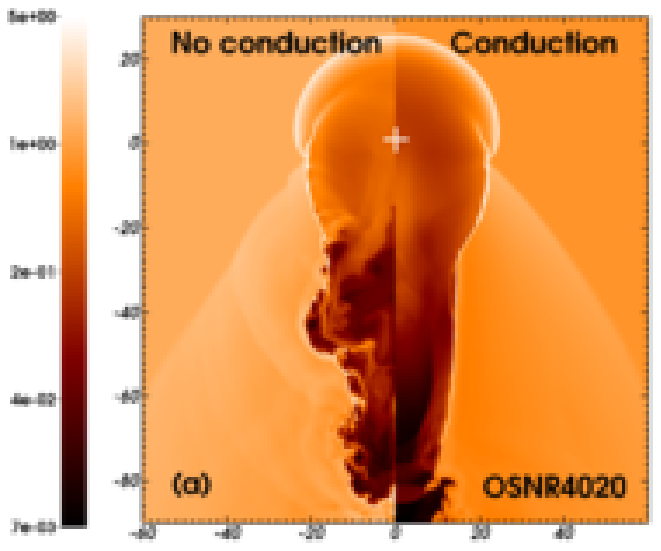}
	\end{minipage} 
	\centering
	\begin{minipage}[b]{ 0.42\textwidth}
\includegraphics[width=1.0\textwidth]{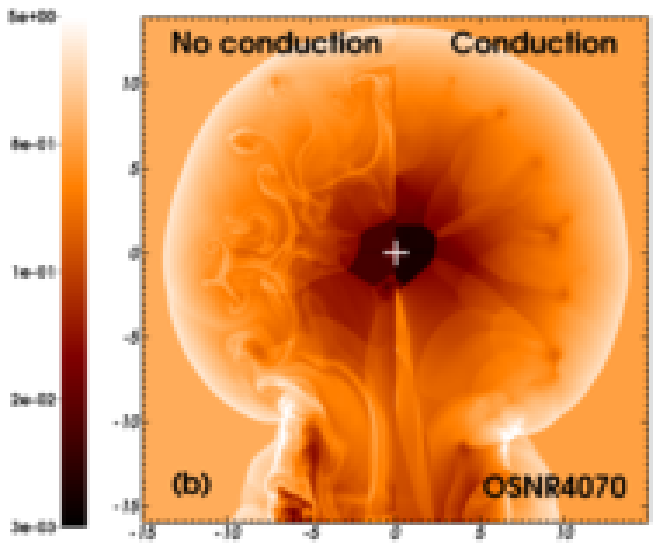}
	\end{minipage} 
	\caption{ 
\textcolor{black}{ Effects of heat conduction on the old supernova remnant generated by our 
initially $40\, \rm M_{\odot}$ progenitor moving with velocity $20\, \rm km\, \rm s^{-1}$ (a)
and $70\, \rm km\, \rm s^{-1}$ (b).
The figure shows the gas number density field of simulations carried out without (left-hand part of the panels) 
and with heat conduction (right-hand part of the panels) with a density range from about $10^{-3}$ to 
$5.0\, \mathrm{cm}^{-3}$ on the logarithmic scale, at a time $t_{\rm osnr}$. 
The white cross marks the center of the explosion. }
		}
	\label{fig:snr_large_scale_2070}  
\end{figure}

In the situation of our slowly-moving supernova progenitors, the remnant is affected by thermal 
conduction in the sense that the channeled supernova shock wave interacting with the walls 
of the cavity encounters a more irregular and clumpy medium (Fig.~\ref{fig:snr_large_scale_2070}a). 
It changes neither the shape of the outflow along the direction of the progenitor nor the 
global remnant' morphology, however, it may generate ring-like emission artefacts, 
e.g. by projection effect. Note that heat conduction is in its turn sensitive to the ISM magnetisation
which makes it anisotropic~\citep{balsara_MNRAS_386_2008b} as it has been shown in several studies 
devoted to old supernova remnants~\citep{velazquez_apj_601_2004,balsara_MNRAS_386_2008a}.

In the case of our fast-moving supernova progenitors, the absence of thermal transfers (i) 
slightly modify, as discussed above, the shape of the eddies constituting the pre-supernova 
circumstellar medium distribution and (ii) allows the ejecta interacting with the bow shock to 
cool independently of the rest of the gas during the radiative phase of the remnant. 
Consequently, the fingers of the developing Rayleigh-Taylor instabilities become denser 
(Fig.~\ref{fig:snr_large_scale_2070}b).

\end{document}